\documentstyle[emulateapj]{article}

\lefthead{Nagao, Murayama, \& Taniguchi}
\righthead{Seyfert-Type Dependences of Narrow Emission-Line Ratios}

\begin{document}

\submitted{Publication of the Astronomical Society of Japan, 53, in press}
\title{Seyfert-Type Dependences of Narrow Emission-Line Ratios and\\
       Physical Properties of High-Ionization Nuclear
       Emission-Line Regions \\
       in Seyfert Galaxies}
\author{Tohru NAGAO, Takashi MURAYAMA, and Yoshiaki TANIGUCHI}
\affil{Astronomical Institute, Graduate School of Science, 
       Tohoku University, Aramaki, Aoba, Sendai 980-8578, Japan\\
       tohru@astr.tohoku.ac.jp, 
       murayama@astr.tohoku.ac.jp, 
       tani@astr.tohoku.ac.jp}

\begin{abstract}
            
In order to examine how narrow emission-line flux ratios depend on the 
Seyfert type, we compiled 
various narrow emission-line flux ratios of 355 Seyfert galaxies
from the literature. 
We present in this paper that the intensity of the high-ionization 
emission lines, [Fe {\sc vii}]$\lambda$6087, [Fe {\sc x}]$\lambda$6374
and [Ne {\sc v}]$\lambda$3426, tend to be stronger in Seyfert 1 galaxies
than in Seyfert 2 galaxies.
In addition to these lines, [O {\sc iii}]$\lambda$4363 and
[Ne {\sc iii}]$\lambda$3869, whose ionization potentials are not high
($<$ 100 eV), but whose critical densities are significantly high
($\gtrsim 10^7$ cm$^{-3}$), also exhibit the same tendency.
On the other hand, the emission-line flux ratios among low-ionization
emission lines do not show such a tendency.
We point out that the most plausible interpretation of these results is 
that the high-ionization emission lines arise mainly from highly-ionized,
dense gas clouds, which are located very close to nuclei, and thus 
can be hidden by dusty tori. 
To examine the physical properties of these highly-ionized dense gas clouds,
photoionization model calculations were performed.
As a result, we find that the hydrogen density and the ionization
parameter of these highly-ionized dense gas clouds are constrained to be
$n_{\rm H} > 10^6$ cm$^{-3}$ and $U > 10^{-2}$, respectively.
These lower limits are almost independent both from the metallicity 
of gas clouds and from the spectral energy distribution of 
the nuclear ionizing radiation.

\end{abstract}
       

   

\section{Introduction}
            
Seyfert nuclei are typical active galactic nuclei (AGNs) in the nearby
universe. They have been broadly classified into two types based on
the presence or absence of broad (typically $\gtrsim$ 2000 km s$^{-1}$)
permitted emission lines in their optical spectra
(Khachikian, Weedman 1974); Seyfert nuclei with broad lines are type 1 
(hereafter S1), while those without broad lines are type 2 (S2).
This difference is thought to be due to a dependence of
the visibility of broad-line regions (BLRs) on the viewing angle.
Accordingly, the BLR is thought to be located in a very inner region
(a typical radial distance from the central black hole is
$r \sim 0.01$ pc; see, e.g., Peterson 1993) and surrounded by a
geometrically- and optically-thick dusty torus
(AGN unified model; see Antonucci 1993 for a review).

Contrary to the BLR emission, narrow (typically $\lesssim$ 1000 km s$^{-1}$)
permitted and forbidden emission lines, arising from narrow-line regions
(NLRs), are seen in the spectra of both S1s and S2s.
Therefore, the NLR is believed to be located far from the nucleus, and 
thus not to be hidden by dusty tori.
If this is the case, the observed physical properties of ionized gas in NLRs
do not depend on the viewing angle toward dusty tori.
However, several studies have statistically shown that some 
high-ionization emission lines, such as
[Fe {\sc vii}]$\lambda$6087, [Fe {\sc x}]$\lambda$6374, and
[Ne {\sc v}]$\lambda$3426, are stronger in spectra of 
S1s than in those of S2s (e.g., Shuder, Osterbrock 1981; Cohen 1983;
Murayama, Taniguchi 1998a; Schmitt 1998; 
Nagao et al. 2000).
Some emission lines whose ionization potential is not so high 
($<$ 100 eV), but whose
critical density is high, such as [O {\sc iii}]$\lambda$4363 and
[Ne {\sc iii}]$\lambda$3869, also show the same Seyfert-type dependence 
(e.g., Osterbrock et al. 1976; Heckman, Balick 1979;
Shuder, Osterbrock 1981; Schmitt 1998; Nagao et al. 2001b).
These Seyfert-type dependences in some emission-line intensities seem to
conflict with the framework of the AGN unified model.
Therefore, some possible models have been proposed to explain such
Seyfert-type dependences in the narrow emission-line intensities. 
One of the proposed ideas is that these 
Seyfert-type dependences in the strength of the high-ionization emission 
lines are caused by a viewing-angle dependence of the 
visibility of highly-ionized dense gas clouds, which are located very 
close to the nucleus, and thus can be hidden by dusty tori 
(Murayama, Taniguchi 1998a, b; Nagao et al. 2000, 2001b).
On the contrary, Schmitt (1998) pointed out that the Seyfert-type 
dependences in the intensities of the high-critical-density transition
can be understood if the intrinsic difference in the NLR size between S1s 
and S2s (see, e.g., Schmitt, Kinney 1996) is taken into account.

Which is the origin of the Seyfert-type dependence of those emission lines,
the inclination effect (i.e., obscuration by dusty tori) 
or the intrinsic difference in the size of NLRs? 
To investigate this issue, it should be examined 
how emission-line strengths depend on the Seyfert-type.
Thus, we have compiled emission-line flux ratios of many Seyfert galaxies
from the literature. In this paper, we show the Seyfert-type dependence
of various emission-line flux ratios based on this compiled database
and discuss the origin of the Seyfert-type dependence.

   

\section{The Data}
            
\subsection{Data}

In this section, we briefly describe the properties of the database of
emission-line flux ratios. See Nagao (2001) for details of this database.
The database contains various emission-line flux ratios of 355 Seyfert
nuclei in total; 33 narrow-line S1s (NLS1s; see, e.g., 
Osterbrock, Pogge 1985), 48 broad-line S1s (BLS1s), 
48 Seyfert 1.2 galaxies (S1.2s), 78 Seyfert 1.5 galaxies (S1.5s), 
9 Seyfert 1.8 galaxies (S1.8s), 28 Seyfert 1.9 galaxies (S1.9s),
6 S2s with broad emission in near-infrared spectra (S2$_{\rm NIR-BLR}$s),
12 S2s with broad polarized emission (S2$_{\rm HBLR}$s),
and 93 S2s without any broad emission in their spectra (S2$^-$s).
In this paper, S1.2s are basically included in BLS1s, and
S1.8s, S1.9s, and S2$_{\rm NIR-BLR}$s are gathered into the class of 
``S2 with reddened BLR (S2$_{\rm RBLR}$)''.
When necessary, NLS1s and BLS1s are gathered into the class of
``S1$_{\rm total}$'', and S2$_{\rm RBLR}$, S2$_{\rm HBLR}$s, and S2$^-$s
are gathered into the class of ``S2$_{\rm total}$''.
We adopt the classification of the Seyfert types by
V\'{e}ron-Cetty and V\'{e}ron (2000).

Since it is often difficult to measure the flux of narrow Balmer components
accurately for S1s (and S1.5s), the correction for dust extinction 
adopting the Balmer
decrement method (see, e.g., Osterbrock 1989) would cause possible
systematic errors. Therefore, we did not correct any dust extinction effect
on the emission-line flux ratios in the database.
The effects of dust extinction on the following discussion are mentioned in 
subsection 3.2.

\subsection{Selection Biases}
            
The sample used here is not a complete one in any sense.
Therefore, it is necessary to check whether or not the sample is appropriate
for the statistical and comparative investigations carried out in the
following sections. Since some possible biases may be caused
if there are large systematic differences in their redshift or intrinsic
luminosity distributions, we check their distributions below.

First, we check the frequency distribution of redshift for each
type of Seyfert galaxies.
The histogram of the frequency distribution, the median, the average, and 
the 1 $\sigma$ deviation of the redshift for each type of Seyfert galaxies 
are presented in figure 1 and table 1.
It appears that the S1s and the S1.5s tend to have higher redshifts than
the S2s. In order to check whether or not the frequency distributions
of the redshift are statistically different among the various types of Seyfert
galaxies, we apply the Kolmogorov--Smirnov (KS) statistical test
(see, e.g., Press et al. 1988).
The null hypothesis is that the redshift distributions of two types 
of Seyfert galaxies (in any combination) come from 
the same underlying population.
The derived KS probabilities (i.e., the probabilities of two samples 
drawn from the same parent population) are summarized in table 2.
The KS test leads to the following results:
(1) the redshift distributions of the NLS1s, the BLS1s,
and the S1.5s are statistically indistinguishable, (2) those of 
the S2$_{\rm RBLR}$, S2$_{\rm HBLR}$, and S2$^-$ are also statistically 
indistinguishable, while (3) the NLS1s, the BLS1s, and the S1.5s have
systematically higher redshifts than 
the S2$_{\rm RBLR}$, S2$_{\rm HBLR}$, and S2$^-$. 
Therefore, we must keep in mind that some
properties investigated in the following sections may be affected
by the redshift bias. We will check this possibility when we
carry out statistical investigations of the
emission-line intensity ratios (see subsection 3.2).

Second, we check the intrinsic luminosity distribution of each type of 
Seyfert galaxies.
Following the AGN unified model, the nuclear nonthermal continuum radiation
of S2s is absorbed by dusty tori and cannot be observed directly.
This results in weaker continuum emission in the sample of S2s 
than in the sample of S1s, if the distribution of the intrinsic luminosity
is the same between the samples of S1s and S2s. In other words,
we may pick up intrinsically luminous S2s compared to S1s by survey
observations with a certain limiting flux.
Since such biases may affect the statistical properties of our samples, 
we have to check the intrinsic luminosity distribution of 
each type of Seyfert galaxy using some isotropic radiation.
Here we investigate the distributions of the mid- and far-infrared 
luminosity, i.e., IRAS 25 $\mu$m and 60 $\mu$m luminosities.\footnote{
In this paper, a cosmology of $H_0$ = 50 km s$^{-1}$ Mpc$^{-1}$ and
$q_0$ = 0 is assumed. The data of the IRAS 25 $\mu$m and 60 $\mu$m are
taken from Moshir et al. (1992).}
These luminosities are thought to scale the nuclear continuum luminosity
which is absorbed and re-radiated by dusty tori, and 
to have little viewing-angle dependence (e.g., Pier, Krolik 1992; 
Efstathiou, Rowan-Robinson 1995; Fadda et al. 1998). 
Note that the effect of star formation at 25 $\mu$m is less than at 60 $\mu$m
though active starburst may contribute to the 25 $\mu$m luminosity,
not only the 60 $\mu$m luminosity.
The histograms of the frequency distribution, the median, the average, 
and the 1 $\sigma$ 
deviation of the IRAS 25 $\mu$m and 60 $\mu$m luminosities are
given in figures 2 and 3, and table 2. 
The KS test leads to the result that the samples of S1s 
and S1.5s have higher IRAS 25 $\mu$m luminosity than that of the S2s,
though the samples are statistically indistinguishable in the 
IRAS 60 $\mu$m luminosity. Since the intrinsic luminosity of Seyfert
nuclei is more accurately represented by the IRAS 25 $\mu$m luminosity,
this may mean that the samples of S1s and S1.5s are more luminous than 
that of S2s, statistically. We check whether or not this difference
in the intrinsic luminosity distribution affect the later discussions
in subsection 3.2.

   

\section{Results}

 \subsection{Emission-Line Ratios with Balmer lines}

Historically, the diagnostic diagrams proposed by Veilleux and Osterbrock
(1987; hereafter VO87) have often been used to examine the physical
properties of gas clouds in NLRs. 
The VO87 diagrams are made from the emission-line flux ratios of a forbidden
line to a narrow component of a hydrogen Balmer line;
e.g., [O {\sc iii}]$\lambda$5007/H$\beta$, 
[N {\sc ii}]$\lambda$6583/H$\alpha$, and so on. 
However, since it is often difficult to measure 
the narrow Balmer components for S1s (and S1.5s) accurately, 
it is unclear whether or not these flux ratios of S1s (and S1.5s) can be 
used to study the properties of NLRs and can be compared with those of S2s.
Therefore, prior to comparing the frequency distributions of these
emission-line flux ratios among various types of Seyfert galaxies,
we examine whether or not the VO87 diagrams can work even for S1s
and S1.5s.

In figure 4, we show the diagnostic diagrams proposed by VO87, 
i.e., the diagrams of the 
emission-line flux ratio of [O {\sc iii}]$\lambda$5007/H$\beta$ versus
that of [N {\sc ii}]$\lambda$6583/H$\alpha$, 
[S {\sc ii}]$\lambda \lambda$6717,6731/H$\alpha$, and
[O {\sc i}]$\lambda$6300/H$\alpha$.
The compiled data of Seyfert galaxies are plotted in these diagrams
with the data of extragalactic H {\sc ii} systems for references.
Note that these emission-line ratios are little influenced by
dust extinction, because the wavelength separations between the 
concerned two lines are small.
It appears that the data of the S1s and the S1.5s show larger scatters
than those of the S2s in each diagram,
especially in the diagram of [O {\sc iii}]$\lambda$5007/H$\beta$ versus
[S {\sc ii}]$\lambda$6717,6731/H$\alpha$ (figure 4b).
Does this suggest that there is a systematic difference in the physical
properties of NLR gas clouds between the S2s and the others, or that
the measurements of the narrow components of Balmer lines are not 
well determined for the S1s and the S1.5s?
To examine this issue, we investigate the relationship of the emission-line 
flux ratio between H$\gamma$/H$\beta$ and H$\alpha$/H$\beta$ (figure 5). 
When we assume
the case B approximation, their theoretically predicted ratios are
H$\alpha$/H$\beta$ = 2.9 and H$\gamma$/H$\beta$ = 0.47 
(see, e.g., Osterbrock 1989).\footnote{
These predictions are insensitive to gas temperature and
hydrogen density except for a very high-dense gas clouds as existing 
in BLRs. In the range of 5000 K $<$ $T$ $<$ 20000 K
and 10$^2$ cm$^{-3}$ $<$ $n_{\rm H}$ $<$ 10$^4$ cm$^{-3}$,
these predicted flux ratios vary within 10 \% (see Osterbrock 1989).
Note that, in NLRs of Seyfert galaxies, the H$\alpha$ emission may
be enhanced by collisional excitation
(e.g., Ferland, Netzer 1983; Halpern, Steiner 1983).}
In figure 5, although most of the S2s can be well described by the case B
prediction taking the effects of dust extinction into account (approximately
0 mag $\lesssim A_V \lesssim$ 3 mag), most of the S1s cannot be explained 
by the case B prediction with dust extinction. 
This suggests that the fluxes of the narrow components of Balmer lines 
are not well determined for the S1s (and S1.5s) although other 
possibilities (e.g., contribution of optically-thin gas clouds) 
cannot be ruled out. In any case, it is safe to avoid the
fluxes of the narrow components of the Balmer lines in order to investigate
the physical properties of NLR gas clouds in S1s and in S1.5s and in order to 
study any systematic differences of the NLRs between S1s and S2s.

 \subsection{Emission-Line Ratios without Balmer Lines}

Some other diagnostic diagrams in which the narrow Balmer emission is 
not used have been proposed to discuss the physical properties of  
gas clouds in Seyfert nuclei
(e.g., Baldwin et al. 1981; Ohyama 1996;
Nagao et al. 2001a; see also Halpern, Steiner 1983).
Such diagnostic diagrams seem to be highly useful in investigating both
the physical properties of NLRs of S1s and S1.5s and any systematic 
difference from those of S2s.
Therefore, we consider statistical properties of the NLRs only by 
using various forbidden emission-line flux ratios 
(see also Nagao et al. 2001a).

In figures 6a--p, we show the frequency distributions of the compiled
emission-line ratios, 
[O {\sc i}]$\lambda$6300/[O {\sc iii}]$\lambda$5007,
[O {\sc ii}]$\lambda$3727/[O {\sc iii}]$\lambda$5007,
[O {\sc i}]$\lambda$6300/[O {\sc ii}]$\lambda$3727,
[O {\sc iii}]$\lambda$4363/[O {\sc iii}]$\lambda$5007,
[S {\sc ii}]$\lambda$6717/[S {\sc ii}]$\lambda$6731,
[O {\sc i}]$\lambda$6300/[S {\sc ii}]$\lambda \lambda$6717,6731,
[O {\sc ii}]$\lambda$3727/[S {\sc ii}]$\lambda \lambda$6717,6731,
[S {\sc ii}]$\lambda \lambda$6717,6731/[O {\sc iii}]$\lambda$5007,
[O {\sc i}]$\lambda$6300/[N {\sc ii}]$\lambda$6583,
[O {\sc ii}]$\lambda$3727/[N {\sc ii}]$\lambda$6583,
[N {\sc ii}]$\lambda$6583/[O {\sc iii}]$\lambda$5007,
[S {\sc ii}]$\lambda \lambda$6717,6731/[N {\sc ii}]$\lambda$6583,
[Ne {\sc iii}]$\lambda$3869/[O {\sc iii}]$\lambda$5007,
[Ne {\sc iii}]$\lambda$3869/[O {\sc ii}]$\lambda$3727,
[Ne {\sc v}]$\lambda$3426/[O {\sc ii}]$\lambda$3727, and
[Fe {\sc vii}]$\lambda$6087/[O {\sc iii}]$\lambda$5007 
for the S1s, the S1.5s and the S2s.
In these figures, the distributions of the flux ratios 
for the NLS1s, the BLS1s, 
the S2$_{\rm RBLR}$s, the S2$_{\rm HBLR}$s, and the S2$^-$s 
are also shown in the right-hand panels.
The frequency distributions of 
[O {\sc ii}]$\lambda$7325/[O {\sc ii}]$\lambda$3727,
[S {\sc iii}]$\lambda$9069/[S {\sc ii}]$\lambda \lambda$6717,6731,
[N {\sc i}]$\lambda$5199/[N {\sc ii}]$\lambda$6583,
[N {\sc ii}]$\lambda$5755/[N {\sc ii}]$\lambda$6583,
[Ar {\sc iii}]$\lambda$7136/[O {\sc iii}]$\lambda$5007,
[Fe {\sc x}]$\lambda$6374/[O {\sc iii}]$\lambda$5007, and
[Fe {\sc xi}]$\lambda$7892/[O {\sc iii}]$\lambda$5007
for the S1s, the S1.5s and the S2s are shown in figure 7.
Since these line ratios have been measured for a subset of
the samples, we do not show their histograms for each subclass of
S1s and S2s in this figure.
In table 3, the median, the average, and the 1$\sigma$ deviation of
each emission-line flux ratio for the S1s, the S1.5s and the S2s are given.

In order to check whether or not the frequency distributions of
these emission-line flux ratios are statistically different among the
S1s, the S1.5s and the S2s, we apply the KS test.
The resultant KS probabilities are given in table 4.
These results can be summarized as follows.
(1) As for the emission-line flux ratios of
[O {\sc i}]$\lambda$6300/[O {\sc iii}]$\lambda$5007,
[O {\sc ii}]$\lambda$3727/[O {\sc iii}]$\lambda$5007,
[O {\sc i}]$\lambda$6300/[O {\sc ii}]$\lambda$3727,
[S {\sc ii}]$\lambda$6717/[S {\sc ii}]$\lambda$6731,
[S {\sc iii}]$\lambda$9069/[S {\sc ii}]$\lambda \lambda$6717,6731,
[O {\sc i}]$\lambda$6300/[S {\sc ii}]$\lambda \lambda$6717,6731,
[O {\sc ii}]$\lambda$3717/[S {\sc ii}]$\lambda \lambda$6717,6731,
[N {\sc i}]$\lambda$5199/[N {\sc ii}]$\lambda$6583,
[O {\sc i}]$\lambda$6300/[N {\sc ii}]$\lambda$6583,
[O {\sc ii}]$\lambda$3727/[N {\sc ii}]$\lambda$6583,
[N {\sc ii}]$\lambda$6583/[O {\sc iii}]$\lambda$5007,
[S {\sc ii}]$\lambda \lambda$6717,6731/[N {\sc ii}]$\lambda$6583 and
[Ar {\sc iii}]$\lambda$7136/[O {\sc iii}]$\lambda$5007,
there are little or no systematic differences among the S1s, the S1.5s
and the S2s.
(2) The emission-line flux ratios of
[O {\sc ii}]$\lambda$7325/[O {\sc ii}]$\lambda$3727,
[N {\sc ii}]$\lambda$5755/[N {\sc ii}]$\lambda$6583 and
[Fe {\sc xi}]$\lambda$7892/[O {\sc iii}]$\lambda$5007 appear to be slightly
higher in the S1s than those in the S2s, although
these differences are statistically insignificant.
(3) It is statistically significant that the 
emission-line flux ratios of
[O {\sc iii}]$\lambda$4363/[O {\sc iii}]$\lambda$5007,
[S {\sc ii}]$\lambda \lambda$6717,6731/[O {\sc iii}]$\lambda$5007,
[Ne {\sc iii}]$\lambda$3869/[O {\sc iii}]$\lambda$5007,
[Ne {\sc iii}]$\lambda$3869/[O {\sc ii}]$\lambda$3727,
[Ne {\sc v}]$\lambda$3426/[O {\sc ii}]$\lambda$3727, 
[Fe {\sc vii}]$\lambda$6087/[O {\sc iii}]$\lambda$5007 and
[Fe {\sc x}]$\lambda$6374/[O {\sc iii}]$\lambda$5007 are different
between the S2s and the other types of Seyfert galaxies
(i.e., the S1s and the S1.5s).
Since these 23 emission-line flux ratios are not correlated with the redshift
and IRAS 25 $\mu$m luminosity as shown in figure 8,
the above results seem to be almost free from the redshift and 
luminosity biases mentioned in subsection 2.2.

As mentioned in subsection 2.1, no reddening correction has been made
for all of the compiled emission-line flux ratios.
Since it is known that the dust extinction is larger on average
in S2s than that in S1s (e.g., Dahari, De Robertis 1988), the
above results may be caused by the difference in the amount of
extinction between the S1s and the S2s.
In order to check whether or not this is the case,
we examine the effect of the extinction correction for each
emission-line flux ratio adopting the Cardelli's extinction curve
(Cardelli et al. 1989).
In table 5, we summarize the correction factors, 
by which the emission-line flux ratios should be multiplied to be
converted into the extinction-corrected values for the case of 
$A_V$ = 1 mag.
Note that the mean difference of the amount of dust extinction between
S1s and S2s is $\sim$1 mag (Dahari, De Robertis 1988; see also
De Zotti, Gaskell 1985).
Since the effect of the extinction correction is too small for the cases of
[O {\sc iii}]$\lambda$4363/[O {\sc iii}]$\lambda$5007,
[Ne {\sc iii}]$\lambda$3869/[O {\sc iii}]$\lambda$5007
and [Ne {\sc v}]$\lambda$3426/[O {\sc ii}]$\lambda$3727,
the differences in these emission-line flux ratios between the S1s and the S2s
cannot be attributed only to the effect of the dust extinction. 
Furthermore, the differences in the flux ratios of
[Ne {\sc iii}]$\lambda$3869/[O {\sc ii}]$\lambda$3727,
[Fe {\sc vii}]$\lambda$6087/[O {\sc iii}]$\lambda$5007 and
[Fe {\sc x}]$\lambda$6374/[O {\sc iii}]$\lambda$5007
cannot be also interpreted by the effect of the dust extinction.
Therefore, we conclude that 
the AGN-type dependence of these emission-line flux ratios
is due not to the difference in the amounts of
extinction, but to some other factors, as discussed later.

Here, we mention that these results are consistent with the previous 
studies. The excess of the flux ratio of
[O {\sc iii}]$\lambda$4363/[O {\sc iii}]$\lambda$5007 
in S1s (and in S1.5s) has been reported by 
Osterbrock et al. (1976), Heckman and Balick (1979),
Shuder and Osterbrock (1981), Cohen (1983), and Nagao et al. (2001b).
The excess of the intensities of high-ionization iron emission lines of S1s 
has been mentioned by Shuder and Osterbrock (1981), Cohen (1983), 
Murayama and Taniguchi (1998a), and Nagao et al. (2000).
Schmitt (1998) noted that S1s exhibit higher ratios of
[Ne {\sc iii}]$\lambda$3869/[O {\sc ii}]$\lambda$3727 and
[Ne {\sc v}]$\lambda$3426/[O {\sc ii}]$\lambda$3727 than S2s.

   

\section{Discussion}

 \subsection{The Origin of the Seyfert-Type Dependence of 
             the Emission-Line Flux Ratios}

The results presented in the last section can be summarized as follows:
(1) Most of the emission-line flux ratios which show the Seyfert-type
dependence contain a high critical-density ($\gtrsim 10^7$ cm$^{-3}$)
and/or high ionization-potential ($\gtrsim$ 100 eV) emission lines, except
for the ratio of 
[S {\sc ii}]$\lambda \lambda$6717,6731/[O {\sc iii}]$\lambda$5007.
(2) On the other hand, the flux ratios which do not exhibit the
Seyfert-type dependence do not contain high-ionization emission lines.
In table 6, we summarize the ionization potential and the critical density
of each emission line used in our analysis. Since the 
[Fe {\sc xi}]$\lambda$7892 emission has a very high ionization potential,
its relative intensity could be different between S1s and S2s. However,
it is unclear in our sample whether or not the [Fe {\sc xi}]$\lambda$7892
emission is stronger in the S1s than in the S2s. 
One of the reasons for this may be the small number of 
[Fe {\sc xi}]-detected objects. 
Further observations will be necessary to confirm the
difference in the frequency distribution of the [Fe {\sc xi}]$\lambda$7892
intensity between S1s and S2s.

We now consider the origin of the Seyfert-type 
dependence of those emission-line flux ratios.
There are two possible alternatives.
One is that the high-ionization emission lines arise mainly
from dense gas clouds which are located very close to nuclei
(Torus HINER\footnote{
The term ``HINER'' means high-ionization nuclear emission-line region
(Binette 1985; Murayama et al. 1998)}; see
Murayama, Taniguchi 1998a, b).
Since such a region can be hidden by dusty tori, the visibility
of the dense gas clouds may depend on a viewing angle toward the tori.
This component may correspond to highly-ionized, dense 
($\sim 10^{7-8}$ cm$^{-3}$) gas beside the inner wall of dusty tori
(see Pier, Voit 1995).
This idea was proposed by Murayama and Taniguchi (1998a) to explain
the stronger [Fe {\sc vii}]$\lambda$6087 emission in S1s compared to S2s.
Taking this Torus-HINER component into account, Murayama and Taniguchi (1998b)
showed that the difference in the [Fe {\sc vii}]$\lambda$6087 intensity 
between S1s and S2s can be successfully interpreted by a dual-component 
photoionization model proposed by them (see also Nagao et al. 2001b).
The other idea is that the difference in the emission-line flux ratios
reflects the intrinsic difference of NLR properties, e.g., physical size,
density, temperature, and so on.
Osterbrock (1978) mentioned that the systematic difference in
the flux ratio of [O {\sc iii}]$\lambda$4363/[O {\sc iii}]$\lambda$5007
can be understood assuming $n_{\rm H} \sim 10^{6-7}$ cm$^{-3}$ for S1s and
$n_{\rm H} < 10^5$ cm$^{-3}$ for S2s. On the contrary,
Heckman and Balick (1979) and Cohen (1983) claimed that the origin of the
difference in [O {\sc iii}]$\lambda$4363/[O {\sc iii}]$\lambda$5007 
between S1s and S2s is attributed
not to the density difference, but to the temperature difference; i.e., 
$T_{\rm e} > 2 \times 10^4$ K for S1s while $T_{\rm e} \sim 10^4$ K for S2s.
Here, it must be noted that
these situations may occur when high-density or high-temperature regions
are hidden by any obscuring matter in S2s. Namely, these two scenarios do
not necessarily mean the intrinsic difference in the NLR properties.
On the other hand, Schmitt (1998) reported that the systematic differences
in the ratios of [Ne {\sc iii}]$\lambda$3869/[O {\sc ii}]$\lambda$3727 and
[Ne {\sc v}]$\lambda$3426/[O {\sc ii}]$\lambda$3727 
between S1s and S2s can be explained by taking
the physically (i.e., not projected) smaller NLR size of S1s
compared to that of S2s, as suggested by Schmitt and Kinney (1996),
into account. 
Since smaller NLRs contain fewer ionization-bounded clouds,
which radiate low-ionization emission lines selectively,
this results in weaker [O {\sc ii}]$\lambda$3727 emission
compared to [Ne {\sc iii}]$\lambda$3869 and [Ne {\sc v}]$\lambda$3426
in S1s (hereafter ``smaller NLR model'').

Which scheme is more realistic, the obscuration of the Torus HINER in S2s
or the smaller NLR model? 
In either case, both high-ionization gas clouds and low-ionization ones
are necessary to explain the observations, and a certain systematic 
difference in the relative contribution between the two components can
be regarded as the origin of the Seyfert-type dependence of the
emission-line flux ratios.
However, the two ideas make different predictions.
The Torus-HINER model predicts that the intrinsic luminosity of
high-ionization lines are brighter in S1s than in S2s because the Torus
HINER is assumed to be hidden in the S2s (Murayama, Taniguchi 1998a).
On the other hand, the smaller NLR model predicts that the
intrinsic luminosities of low-ionization lines are lower in the S1s than 
in the S2s because the low-ionization lines arise mostly from 
ionization-bounded gas clouds at outer NLRs. 
Therefore, the Torus-HINER model predicts similar [O {\sc iii}]$\lambda$5007
luminosity among the Seyfert types and higher luminosities of 
[O {\sc iii}]$\lambda$4363, [Ne {\sc iii}]$\lambda$3869, and 
[Fe {\sc vii}]$\lambda$6087 in S1s than in S2s, 
while the ``smaller NLR model'' predicts lower [O {\sc iii}]$\lambda$5007
luminosity in S1s than that in S2s, and similar luminosities of 
[O {\sc iii}]$\lambda$4363, [Ne {\sc iii}]$\lambda$3869, and 
[Fe {\sc vii}]$\lambda$6087 among the Seyfert types.

We now discuss observational tests for the two models.
Since the frequency distribution of the intrinsic luminosity
is different among the S1s, the S1.5s and S2s in our sample, as mentioned
in subsection 2.2, these emission-line luminosities should be normalized by
IRAS 25 $\mu$m luminosity to test the above predictions.
In figure 9, we show the frequency distributions of the 
[O {\sc iii}]$\lambda$5007, [O {\sc iii}]$\lambda$4363, 
[Ne {\sc iii}]$\lambda$3869, and [Fe {\sc vii}]$\lambda$6087 luminosities 
of the S1s, the S1.5s, and the S2s, which are normalized by the IRAS 25 $\mu$m
luminosity. The median, the average, and the 1$\sigma$ deviation of each 
relative luminosity for the S1s, the S1.5s, and the S2s are given in table 7.
In order to examine whether or not the distribution of the relative strength
of the emission lines depends on the Seyfert type, we apply the KS test. 
The resultant KS probabilities are given in table 8. 
The KS test leads to the following results:

\begin{itemize}
\item The [O {\sc iii}]$\lambda$5007 luminosity normalized by IRAS 25 $\mu$m 
      luminosity is statistically indistinguishable among the Seyfert types.
      This is consistent with the Torus-HINER model, but in conflict with 
      the prediction of the ``smaller NLR model''.
\item The [O {\sc iii}]$\lambda$4363 and the [Fe {\sc vii}]$\lambda$6087 
      luminosity normalized by IRAS 25 $\mu$m luminosity appear to be higher 
      in the S1s and in the S1.5s than in the S2s, although the statistical 
      significance is low. This agrees with the 
      prediction of the Torus-HINER model, but cannot be understood 
      in terms of the ``smaller NLR model''.
\item The [Ne {\sc iii}]$\lambda$3869 luminosity normalized by IRAS 25 $\mu$m 
      luminosity is statistically indistinguishable among the Seyfert types.
      This seems to disagree with the prediction of the Torus-HINER model.
\end{itemize}
The first two results seem to support the Torus-HINER model. As for 
the last result, we have to take into account that the critical density of 
the [Ne {\sc iii}]$\lambda$3869 transition is smaller than those of 
the [O {\sc iii}]$\lambda$4363 and [Fe {\sc vii}]$\lambda$6087 
transitions. This may be the reason why the frequency distribution of 
the relative luminosity of the [Ne {\sc iii}]$\lambda$3869 emission 
is indistinguishable among the Seyfert types. 
We therefore conclude that the Torus-HINER model appears to be much more
consistent with the observation than the smaller NLR model.
We thus adopt the Torus-HINER model in the following discussion.

One of our interesting results is that the emission-line flux ratio of 
[S {\sc ii}]$\lambda \lambda$6717,6731/[O {\sc iii}]$\lambda$5007
exhibits a systematic difference in its frequency distribution between
the S1.5s and the S2s.
If this difference is also attributed to different contributions of the
Torus HINER between the two types, it is suggested that the visibility of the
[O {\sc iii}]$\lambda$5007 emitting region depends on the viewing angle 
toward dusty tori.
Indeed, the isotropy of the [O {\sc iii}]$\lambda$5007 emission
has sometimes been called into question, especially in radio-loud AGNs.
Jackson and Browne (1990) reported that the [O {\sc iii}]$\lambda$5007 
luminosity of the narrow-line radio galaxies is lower by 5--10 times
than that of the broad-line quasars, matched in redshift and extended
radio luminosity.
On the other hand, Hes et al. (1993) found that these two
kinds of radio-loud AGNs show no difference in the [O {\sc ii}]$\lambda$3727
luminosity. These results suggest that the [O {\sc ii}]$\lambda$3727 
emission is isotropic, but the [O {\sc iii}]$\lambda$5007 emission is not
(see also Baker, Hunstead 1995; Baker 1997; cf., Simpson 1998).
Note that this conclusion is based on an assumption that
broad-line quasars and (powerful) narrow-line radio galaxies are
intrinsically similar, but different only in the viewing angle 
(see, e.g., Barthel 1989).
Polarized [O {\sc iii}]$\lambda$5007 emission has been detected in 
some radio galaxies (di Serego Alighieri et al. 1997), supporting
the scenario that a part of the [O {\sc iii}]$\lambda$5007 flux is
hidden by the tori.
Because such polarized [O {\sc iii}]$\lambda$5007 emission has also been 
detected in a Seyfert 2 galaxy, NGC 4258 (Wilkes et al. 1995;
Barth et al. 1999), it is interesting to investigate whether or not
the [O {\sc iii}]$\lambda$5007 emission has an anisotropic property also
in the sample of Seyfert galaxies, although Mulchaey et al. (1994) reported
a negative result for this possibility.

If the [O {\sc iii}]$\lambda$5007 emission of our sample is affected by
an orientation-dependent dust obscuration, 
the frequency distribution of $L$([O {\sc iii}]$\lambda$5007) would show
a larger scatter than that of $L$([O {\sc ii}]$\lambda$3727) 
as a result of the obscuration of 
[O {\sc iii}]$\lambda$5007 emission at a large inclination angle 
(see Kuraszkiewicz et al. 2000). 
Here, the [O {\sc ii}]$\lambda$3727 emission is thought to have no 
viewing-angle dependence.
In figure 9, we show a diagram of $L$([O {\sc iii}]$\lambda$5007) versus
$L$([O {\sc ii}]$\lambda$3727), in which no obscuration effect is found.
Recently, Kuraszkiewicz et al. (2000) reported that radio-quiet quasars
do not show evidence for an anisotropic property of 
the [O {\sc iii}]$\lambda$5007 emission.
Therefore, taking the results of both this study and 
Kuraszkiewicz et al. (2000) into account, 
radio-quiet AGNs including quasars and 
Seyfert galaxies may not have the anisotropic property of 
the [O {\sc iii}]$\lambda$5007 emission, contrary to radio-loud AGNs
(see also Mulchaey et al. 1994).
Then, why is the frequency distribution of the flux ratio of
[S {\sc ii}]$\lambda \lambda$6717,6731/[O {\sc iii}]$\lambda$5007
different between the S1.5s and the S2s?
A possible reason may be contamination of 
lower-ionization emission-line fluxes arising from
circumnuclear star-forming regions
associated in S2s, because it is known that 
circumnuclear star-forming activity tends to associate with S2s more
frequently than with S1s (e.g., Thuan 1984; Heckman et al. 1995, 1997).
Indeed, there is a weak negative correlation between the flux ratio of
[S {\sc ii}]$\lambda \lambda$6717,6731/[O {\sc iii}]$\lambda$5007 and that 
of $\nu F_{\nu}$(25 $\mu$m)/$\nu F_{\nu}$(60 $\mu$m), as shown in figure 11. 
Since the ratio of $\nu F_{\nu}$(25$\mu$m)/$\nu F_{\nu}$(60$\mu$m)
becomes smaller when the star-formation activity contributes much more
to the infrared
continuum radiation, this correlation suggests that the star-formation 
activity may enhance the flux ratio of 
[S {\sc ii}]$\lambda \lambda$6717,6731/[O {\sc iii}]$\lambda$5007.

 \subsection{Physical Properties of the Torus HINERs}

Although the Torus HINER is regarded as a ``highly-ionized and dense'' 
component, there are some remaining issues to be investigated:
How dense is the Torus HINER?  And, how high is the ionization
parameter of ionized gas in the Torus HINER? To examine these issues,
we carry out single cloud photoionization model calculations 
using the spectral synthesis code $Cloudy$ version 90.04 (Ferland 1996),
which solves the equations of statistical and thermal equilibrium and
produces a self-consistent model of the run of temperature as a function 
of depth into a nebula. Here, we assume that a uniform density, dust-free
gas cloud with plane-parallel geometry is ionized by a power-law
continuum source. The parameters for the calculations are: 
(1) the spectral energy distribution (SED) of the input radiation, 
(2) the chemical composition, 
(3) the hydrogen density of the cloud ($n_{\rm H}$),
(4) the ionization parameter ($U$), and 
(5) the hydrogen column density. 
We adopt the following input continuum spectrum: 
(i) $\alpha$ = 2.5 for $\lambda > 10 \mu$m,
(ii) $\alpha$ = --2.0, --1.5, and --1.0, between 10 $\mu$m and 50 keV, and 
(iii) $\alpha$ = --2.0 for $h \nu >$ 50 keV, where $\alpha$ is a
spectral index ($f_{\nu} \propto \nu^{\alpha}$).
Koski (1978) reported that the optical continua of S2s can be
approximated by a stellar contribution diluted by a featureless
continuum, with the latter component described by a power law with
$\alpha = -1.5 \pm 0.5$ (see also Storchi-Bergmann, Pastoriza 1989,
1990; Kinney et al. 1991). We set the gas-phase elemental abundances to
be the solar ones taken from Grevesse and Anders (1989) with extensions by
Grevesse and Noels (1993). To see the effects of metallicity, we also
calculated for the case of $Z$ = 0.5 and 2.0 in addition to $Z = 1.0$.
We performed several model runs covering 
$10^2$ ${\rm cm}^{-3} \leq n_{\rm H} \leq 10^8$ ${\rm cm}^{-3}$ and
$10^{-3.5} \leq U \leq 10^{-1.5}$ for three kinds of SED.
Since an unusually strong [O {\sc i}]$\lambda$6300 
emission with respect to the
[O III]$\lambda$5007 emission is predicted by the model
assuming high-density, ionization-bounded gas clouds, 
we assumed truncated (i.e., matter-bounded) clouds
(see, e.g., Murayama, Taniguchi 1998b).
To make the gas clouds matter-bounded ones, 
the calculations were stopped at a hydrogen column density when 
a Lyman limit optical depth ($\tau_{912}$) reached up to 0.1.
In this condition, more than 96 \% of hydrogen is ionized
at the most outer region of a cloud.

In figure 12, the calculated emission-line flux ratios of
[O {\sc iii}]$\lambda$4363/[O {\sc iii}]$\lambda$5007,
[Ne {\sc iii}]$\lambda$3869/[O {\sc iii}]$\lambda$5007 and 
[Fe {\sc vii}]$\lambda$6087/[O {\sc iii}]$\lambda$5007 are plotted
as a function of $n_{\rm H}$. 
In order to explain the observed flux ratio of
[O {\sc iii}]$\lambda$4363/[O {\sc iii}]$\lambda$5007 
(= 0.122 $\pm$ 0.116 for the S1s), rather high density 
($n_{\rm H} \sim 10^6$ cm$^{-3}$) is necessary, 
regardless of $U$, $\alpha$,
and the metallicity. 
To examine the ionization parameter which can consistently explain
the observed emission-line flux ratios of 
[Ne {\sc iii}]$\lambda$3869/[O {\sc iii}]$\lambda$5007 and 
[Fe {\sc vii}]$\lambda$6087/[O {\sc iii}]$\lambda$5007 
(0.246 $\pm$ 0.145 and 0.122 $\pm$ 0.116 for the S1s, respectively) 
adopting $n_{\rm H} \sim 10^6$ cm$^{-3}$, 
we show the calculated flux ratios of
[Ne {\sc iii}]$\lambda$3869/[O {\sc iii}]$\lambda$5007 and 
[Fe {\sc vii}]$\lambda$6087/[O {\sc iii}]$\lambda$5007
as a function of $U$ for the case of $n_{\rm H} \sim 10^6$ cm$^{-3}$
in figure 13. 
This figure suggests that the observed emission-line 
flux ratios can be described by the models with $U = 10^{-2}$.
This value is almost independent 
of the input SEDs and metallicities.

However, these estimates ($n_{\rm H} \sim 10^6$ cm$^{-3}$ and 
$U = 10^{-2}$) are based on the assumption that
all of the observed emission-line fluxes arise from the Torus HINER.
In fact, it is evident that a part of the emission-line flux arises
from the outer low-density regions.
Since the gas clouds in such regions are expected
to emit relatively low-ionization emission-line spectra compared to
those in the Torus HINER, the Torus-HINER emission must be diluted by
such low-ionization emission-line spectra.
Therefore, we should remind that the derived properties of Torus HINERs
are lower limits. i.e., $n_{\rm H} > 10^6$ cm$^{-3}$ and $U > 10^{-2}$. 

These results are not modified significantly
if we take a smaller value of $\tau_{912}$.
In figure 14, we show the calculated emission-line flux ratios of
[O {\sc iii}]$\lambda$4363/[O {\sc iii}]$\lambda$5007,
[Ne {\sc iii}]$\lambda$3869/[O {\sc iii}]$\lambda$5007, and 
[Fe {\sc vii}]$\lambda$6087/[O {\sc iii}]$\lambda$5007 in the case of
$\tau_{912} = 0.01$ as a function of $n_{\rm H}$
(only the models with solar abundances are shown).
The results of the calculations are nearly the same as those for the
case of $\tau_{912} = 0.1$.
On the contrary, relatively smaller flux ratios of 
[Fe {\sc vii}]$\lambda$6087/[O {\sc iii}]$\lambda$5007 are predicted
if we take a larger value of $\tau_{912}$.
For reference we show the calculation results for the case of 
$\tau_{912} = 1.0$ in figure 15. Accordingly, larger gas density and/or
ionization parameters than those estimated for the case of
$\tau_{912} = 0.1$ are necessary 
to explain the observations in the cases of $\tau_{912} > 0.1$.
In any case, we conclude that the {\it lower limits} of the gas density
and the ionization parameter of the Torus-HINER are 
$n_{\rm H} \sim 10^{6}$ cm$^{-3}$ and $U \sim 10^{-2}$.

   

\section{Summary}

Based on a compilation of the optical emission-line spectra of Seyfert 
galaxies, we have investigated the
Seyfert-type dependences of various emission-line flux ratios.
Our analysis was made using only forbidden emission lines.
This method has enabled us to compare the physical properties of the NLR
among the various types of Seyfert nuclei (e.g., S1, S1.5, and S2).

In consequence of the statistical comparisons of various forbidden
emission-line flux
ratios among the Seyfert types, we have obtained the following results:
\begin{itemize}
\item
As for the emission-line flux ratios of
[O {\sc i}]$\lambda$6300/[O {\sc iii}]$\lambda$5007,
[O {\sc ii}]$\lambda$3727/[O {\sc iii}]$\lambda$5007,
[O {\sc i}]$\lambda$6300/[O {\sc ii}]$\lambda$3727,
[S {\sc ii}]$\lambda$6717/[S {\sc ii}]$\lambda$6731,
[S {\sc iii}]$\lambda$9069/[S {\sc ii}]$\lambda \lambda$6717,6731,
[O {\sc i}]$\lambda$6300/[S {\sc ii}]$\lambda \lambda$6717,6731,
[O {\sc ii}]$\lambda$3717/[S {\sc ii}]$\lambda \lambda$6717,6731,
[N {\sc i}]$\lambda$5199/[N {\sc ii}]$\lambda$6583,
[O {\sc i}]$\lambda$6300/[N {\sc ii}]$\lambda$6583,
[O {\sc ii}]$\lambda$3727/[N {\sc ii}]$\lambda$6583,
[N {\sc ii}]$\lambda$6583/[O {\sc iii}]$\lambda$5007,
[S {\sc ii}]$\lambda \lambda$6717,6731/[N {\sc ii}]$\lambda$6583 and
[Ar {\sc iii}]$\lambda$7136/[O {\sc iii}]$\lambda$5007,
there are little or no systematic differences among the S1s, the S1.5s
and the S2s.
\item
On the other hand, it is statistically significant that the 
emission-line flux ratios of
[O {\sc iii}]$\lambda$4363/[O {\sc iii}]$\lambda$5007,
[S {\sc ii}]$\lambda \lambda$6717,6731/[O {\sc iii}]$\lambda$5007
[Ne {\sc iii}]$\lambda$3869/[O {\sc iii}]$\lambda$5007,
[Ne {\sc iii}]$\lambda$3869/[O {\sc ii}]$\lambda$3727,
[Ne {\sc v}]$\lambda$3426/[O {\sc ii}]$\lambda$3727, 
[Fe {\sc vii}]$\lambda$6087/[O {\sc iii}]$\lambda$5007 and
[Fe {\sc x}]$\lambda$6374/[O {\sc iii}]$\lambda$5007 are systematically
higher in the S1s and the S1.5s than in the S2s.
\end{itemize}
These results can be interpreted as that the flux ratios of
a high-ionization and a high-critical-density emission line
to a low-ionization emission line
are systematically higher in the S1s than in S2s.

There are two ideas which can possibly explain the Seyfert-type dependences
of the relative strength of high-ionization emission lines;
i.e., the viewing-angle dependence of visibility of Torus HINER and
the intrinsic difference of the NLR size between S1s and S2s.
To discriminate these two possibilities, we have examined the Seyfert-type
dependences of some emission-line luminosities. The results are consistent 
with the prediction of the Torus-HINER model, but disagree with the 
``smaller NLR model''.
We have concluded that the difference of visibility of Torus HINER between
S1s and S2s causes the Seyfert-type dependences of the relative strength of 
the high-ionization emission lines.

In order to investigate the properties of the Torus HINER,
we have compared the relative strengths of high-ionization emission lines
with the results of the photoionization model calculations.
The estimated lower limit of the density and the ionization parameter
are $n_{\rm H} \sim 10^6$ cm$^{-3}$ and $U \sim 10^{-2}$.
These constraints are almost independent of input SEDs and metallicities
of the ionized gas.

   

\vspace{1pc}

We would like to thank Gary Ferland for making his code $Cloudy$ 
available to the public. 
We also acknowledge the referee, Neil Trentham,
for useful comments and suggestions.
Alberto Rodr\'{\i}guez-Ardila kindly provided information about
the emission-line ratios of some Seyfert galaxies.
We thank Martin Gaskell for useful comments.
This research has made use of the NED (NASA extragalactic database),
which is operated by the Jet Propulsion Laboratory,
California Institute of Technology, under constructed with the 
National Aeronautics and Science Administration.
This work was financially supported in part by Grant-in-Aids for the
Scientific Research (Nos. 10044052 and 10304013) of the Japanese
Ministry of Education, Culture, Sports, Science, and Technology.




\vspace{5pc}

{\it Note added in proof.} --- C. M. Gaskell (Ap Letts, 24, 43 [1984])
also pointed out that there is a high-density component whose visibility
depends on a viewing angle and which is seen only in S1s.
This result was derived by the consideration of the emission-line flux ratio 
of [S {\sc ii}]$\lambda$4074/[S {\sc ii}]$\lambda\lambda$6717,6731.


\clearpage

\begin{table*}
\begin{center}
Table~1.\hspace{4pt}
	Medians, averages, and 1 $\sigma$ deviations of 
	redshift, $\nu L_{\nu}$(25 $\mu$m), and $\nu L_{\nu}$(60 $\mu$m) for 
	each type of Seyfert galaxies.\\
\end{center}
\vspace{6pt}
\begin{tabular*}{\textwidth}{@{\hspace{\tabcolsep}
\extracolsep{\fill}}lcccccc}
\hline\hline\\[-6pt]
     & \multicolumn{2}{c}{Redshift} 
     & \multicolumn{2}{c}{log $\nu L_{\nu}$(25 $\mu$m)}
     & \multicolumn{2}{c}{log $\nu L_{\nu}$(60 $\mu$m)}\\
Type & Median & Average and 1 $\sigma$  & Median & Average and 1 $\sigma$ & Median & Average and 1 $\sigma$ \\
[4pt]\hline\\[-6pt]
S1$_{\rm total}$        \dotfill & 0.0502 & 0.1023 $\pm$ 0.1306 & 42.296 & 42.315 $\pm$ 0.555 & 42.179 & 42.203 $\pm$ 0.629 \\
\ \ \ \ NLS1            \dotfill & 0.0532 & 0.0687 $\pm$ 0.0633 & 42.207 & 42.166 $\pm$ 0.672 & 42.065 & 42.235 $\pm$ 0.632 \\
\ \ \ \ BLS1            \dotfill & 0.0500 & 0.1138 $\pm$ 0.1452 & 42.297 & 42.383 $\pm$ 0.490 & 42.189 & 42.190 $\pm$ 0.636 \\
S1.5                    \dotfill & 0.0384 & 0.0985 $\pm$ 0.1514 & 42.553 & 42.339 $\pm$ 0.639 & 42.240 & 42.231 $\pm$ 0.606 \\
S2$_{\rm total}$        \dotfill & 0.0202 & 0.0373 $\pm$ 0.0660 & 41.933 & 41.795 $\pm$ 0.765 & 42.000 & 41.979 $\pm$ 0.653 \\
\ \ \ \ S2$_{\rm RBLR}$ \dotfill & 0.0133 & 0.0437 $\pm$ 0.0967 & 41.545 & 41.432 $\pm$ 0.820 & 41.735 & 41.728 $\pm$ 0.617 \\
\ \ \ \ S2$_{\rm HBLR}$ \dotfill & 0.0136 & 0.0214 $\pm$ 0.0185 & 42.430 & 42.447 $\pm$ 0.322 & 42.176 & 42.213 $\pm$ 0.523 \\
\ \ \ \ S2$^-$          \dotfill & 0.0240 & 0.0364 $\pm$ 0.0509 & 41.933 & 41.883 $\pm$ 0.693 & 42.093 & 42.061 $\pm$ 0.661 \\
\hline
\end{tabular*}
\end{table*}


\begin{table*}
\small
\begin{center}
Table~2.\hspace{4pt}
	Resultant KS probabilities$^*$ concerning redshift,
	$\nu L_{\nu}$(25 $\mu$m), and $\nu L_{\nu}$(60 $\mu$m).\\
\end{center}
\vspace{6pt}
\begin{tabular*}{\textwidth}{@{\hspace{\tabcolsep}
\extracolsep{\fill}}lcccccccc}
\hline\hline\\[-6pt]
Type & S1$_{\rm total}$ & NLS1 & BLS1 & S1.5 & 
S2$_{\rm total}$ & S2$_{\rm RBLR}$ & S2$_{\rm HBLR}$ & S2$^-$ \\
[4pt]\hline\\[-6pt]
\hline
\multicolumn{9}{c}{Redshift}\\
\hline
S1$_{\rm total}$  &$\cdots$&$\cdots$& $\cdots$              &1.232$\times$10$^{-1}$&1.291$\times$10$^{-12}$&7.673$\times$10$^{-10}$&5.925$\times$10$^{-4}$&2.343$\times$10$^{-7}$\\
NLS1              &$\cdots$&$\cdots$& 6.382$\times$10$^{-1}$&6.153$\times$10$^{-1}$&4.563$\times$10$^{-5}$ &1.237$\times$10$^{-5}$ &3.103$\times$10$^{-3}$&1.688$\times$10$^{-3}$\\
BLS1              &$\cdots$&$\cdots$& $\cdots$              &1.466$\times$10$^{-1}$&4.865$\times$10$^{-11}$&2.515$\times$10$^{-9}$ &6.994$\times$10$^{-4}$&1.068$\times$10$^{-6}$\\
S1.5              &$\cdots$&$\cdots$& $\cdots$              &$\cdots$              &1.458$\times$10$^{-7}$ &5.421$\times$10$^{-7}$ &8.856$\times$10$^{-3}$&1.248$\times$10$^{-4}$\\
S2$_{\rm total}$  &$\cdots$&$\cdots$& $\cdots$              &$\cdots$              &$\cdots$               &$\cdots$               &$\cdots$              &$\cdots$              \\
S2$_{\rm RBLR}$   &$\cdots$&$\cdots$& $\cdots$              &$\cdots$              &$\cdots$               &$\cdots$               &5.264$\times$10$^{-1}$&1.656$\times$10$^{-2}$\\
S2$_{\rm HBLR}$   &$\cdots$&$\cdots$& $\cdots$              &$\cdots$              &$\cdots$               &$\cdots$               &$\cdots$              &8.871$\times$10$^{-2}$\\
S2$^-$            &$\cdots$&$\cdots$& $\cdots$              &$\cdots$              &$\cdots$               &$\cdots$               &$\cdots$              &$\cdots$              \\
\hline
\multicolumn{9}{c}{$\nu L_{\nu}$(25 $\mu$m)}\\
\hline
S1$_{\rm total}$ \dotfill &$\cdots$&$\cdots$& $\cdots$              &6.952$\times$10$^{-2}$&5.904$\times$10$^{-4}$ &1.816$\times$10$^{-4}$ &5.321$\times$10$^{-1}$&2.534$\times$10$^{-3}$\\
NLS1             \dotfill &$\cdots$&$\cdots$& 3.223$\times$10$^{-1}$&2.371$\times$10$^{-1}$&1.724$\times$10$^{-1}$ &2.475$\times$10$^{-2}$ &3.242$\times$10$^{-1}$&2.987$\times$10$^{-1}$\\
BLS1             \dotfill &$\cdots$&$\cdots$& $\cdots$              &1.238$\times$10$^{-1}$&1.679$\times$10$^{-4}$ &4.898$\times$10$^{-5}$ &6.685$\times$10$^{-1}$&6.470$\times$10$^{-4}$\\
S1.5             \dotfill &$\cdots$&$\cdots$& $\cdots$              &$\cdots$              &4.121$\times$10$^{-4}$ &2.000$\times$10$^{-4}$ &7.709$\times$10$^{-1}$&1.483$\times$10$^{-3}$\\
S2$_{\rm total}$ \dotfill &$\cdots$&$\cdots$& $\cdots$              &$\cdots$              &$\cdots$               &$\cdots$               &$\cdots$              &$\cdots$              \\
S2$_{\rm RBLR}$  \dotfill &$\cdots$&$\cdots$& $\cdots$              &$\cdots$              &$\cdots$               &$\cdots$               &1.184$\times$10$^{-3}$&1.642$\times$10$^{-1}$\\
S2$_{\rm HBLR}$  \dotfill &$\cdots$&$\cdots$& $\cdots$              &$\cdots$              &$\cdots$               &$\cdots$               &$\cdots$              &1.037$\times$10$^{-2}$\\
S2$^-$           \dotfill &$\cdots$&$\cdots$& $\cdots$              &$\cdots$              &$\cdots$               &$\cdots$               &$\cdots$              &$\cdots$              \\
\hline
\multicolumn{9}{c}{$\nu L_{\nu}$(60 $\mu$m)}\\
\hline
S1$_{\rm total}$ \dotfill &$\cdots$&$\cdots$& $\cdots$              &7.619$\times$10$^{-1}$&5.383$\times$10$^{-2}$ &2.389$\times$10$^{-3}$ &9.811$\times$10$^{-1}$&3.852$\times$10$^{-1}$\\
NLS1             \dotfill &$\cdots$&$\cdots$& 8.215$\times$10$^{-1}$&9.358$\times$10$^{-1}$&6.148$\times$10$^{-1}$ &1.230$\times$10$^{-1}$ &9.585$\times$10$^{-1}$&9.072$\times$10$^{-1}$\\
BLS1             \dotfill &$\cdots$&$\cdots$& $\cdots$              &5.805$\times$10$^{-1}$&2.455$\times$10$^{-2}$ &1.834$\times$10$^{-3}$ &9.958$\times$10$^{-1}$&1.911$\times$10$^{-1}$\\
S1.5             \dotfill &$\cdots$&$\cdots$& $\cdots$              &$\cdots$              &1.044$\times$10$^{-1}$ &9.109$\times$10$^{-3}$ &8.899$\times$10$^{-1}$&4.809$\times$10$^{-1}$\\
S2$_{\rm total}$ \dotfill &$\cdots$&$\cdots$& $\cdots$              &$\cdots$              &$\cdots$               &$\cdots$               &$\cdots$              &$\cdots$              \\
S2$_{\rm RBLR}$  \dotfill &$\cdots$&$\cdots$& $\cdots$              &$\cdots$              &$\cdots$               &$\cdots$               &1.252$\times$10$^{-1}$&8.458$\times$10$^{-2}$\\
S2$_{\rm HBLR}$  \dotfill &$\cdots$&$\cdots$& $\cdots$              &$\cdots$              &$\cdots$               &$\cdots$               &$\cdots$              &7.279$\times$10$^{-1}$\\
S2$^-$           \dotfill &$\cdots$&$\cdots$& $\cdots$              &$\cdots$              &$\cdots$               &$\cdots$               &$\cdots$              &$\cdots$              \\
\hline
\end{tabular*}
\vspace{6pt}\par\noindent
$^*$ These values means the probabilities of two sample drawn from the
     same parent population.
\end{table*}


\begin{table*}
\small
\begin{center}
Table~3.\hspace{4pt}
	Median, average, and 1 $\sigma$ deviation of 
        each line ratio.\\
\end{center}
\vspace{6pt}
\begin{tabular*}{\textwidth}{@{\hspace{\tabcolsep}
\extracolsep{\fill}}lcccccc}
\hline\hline\\[-6pt]
Line Ratio & \multicolumn{2}{c}{Seyfert 1 Galaxies} & 
\multicolumn{2}{c}{Seyfert 1.5 Galaxies} & \multicolumn{2}{c}{Seyfert 2 Galaxies} \\
 & Median & Average and 1 $\sigma$ & Median & Average and 1 $\sigma$ & Median & Average and 1 $\sigma$ \\
[4pt]\hline\\[-6pt]
{}[O {\sc i}]$\lambda$6300/[O {\sc iii}]$\lambda$5007              \dotfill & 
   0.074 & 0.129 $\pm$ 0.175 & 0.064 & 0.110 $\pm$ 0.165 & 0.091 & 0.135 $\pm$ 0.149 \\
{}[O {\sc ii}]$\lambda$3727/[O {\sc iii}]$\lambda$5007             \dotfill & 
   0.201 & 0.280 $\pm$ 0.200 & 0.165 & 0.223 $\pm$ 0.206 & 0.233 & 0.341 $\pm$ 0.396 \\
{}[O {\sc i}] $\lambda$6300/[O {\sc iii}]$\lambda$3727	           \dotfill &
   0.322 & 0.447 $\pm$ 0.378 & 0.397 & 0.522 $\pm$ 0.402 & 0.331 & 0.514 $\pm$ 1.048 \\
{}[O {\sc iii}]$\lambda$4363/[O {\sc iii}]$\lambda$5007            \dotfill & 
   0.073 & 0.101 $\pm$ 0.101 & 0.063 & 0.104 $\pm$ 0.200 & 0.025 & 0.038 $\pm$ 0.043 \\
{}[O {\sc ii}]$\lambda$7325/[O {\sc ii}]$\lambda$3727              \dotfill & 
   0.340 & 0.496 $\pm$ 0.639 & 0.195 & 0.231 $\pm$ 0.099 & 0.239 & 0.244 $\pm$ 0.112 \\
{}[S {\sc ii}]$\lambda$6717/[S {\sc ii}]$\lambda$6731              \dotfill & 
   1.025 & 1.083 $\pm$ 0.307 & 1.000 & 1.001 $\pm$ 0.167 & 1.072 & 1.086 $\pm$ 0.187 \\
{}[S {\sc iii}]$\lambda$9069/[S {\sc ii}]$\lambda \lambda$6717,6731\dotfill & 
   0.692 & 0.617 $\pm$ 0.315 & 0.481 & 0.538 $\pm$ 0.209 & 0.567 & 0.605 $\pm$ 0.208 \\
{}[O {\sc i}]$\lambda$6300/[S {\sc ii}]$\lambda \lambda$6717,6731  \dotfill & 
   0.302 & 0.446 $\pm$ 0.392 & 0.359 & 0.483 $\pm$ 0.352 & 0.267 & 0.333 $\pm$ 0.274 \\
{}[O {\sc ii}]$\lambda$3727/[S {\sc ii}]$\lambda \lambda$6717,6731 \dotfill & 
   0.903 & 1.211 $\pm$ 1.103 & 0.891 & 0.990 $\pm$ 0.545 & 0.853 & 0.947 $\pm$ 0.485 \\
{}[S {\sc ii}]$\lambda \lambda$6717,6731/[O {\sc iii}]$\lambda$5007\dotfill & 
   0.209 & 0.330 $\pm$ 0.285 & 0.180 & 0.242 $\pm$ 0.238 & 0.332 & 0.555 $\pm$ 0.704 \\
{}[N {\sc i}]$\lambda$5199/[N {\sc ii}]$\lambda$6583               \dotfill & 
   0.037 & 0.045 $\pm$ 0.025 & 0.035 & 0.051 $\pm$ 0.047 & 0.030 & 0.033 $\pm$ 0.017 \\
{}[N {\sc ii}]$\lambda$5755/[N {\sc ii}]$\lambda$6583              \dotfill & 
   0.028 & 0.048 $\pm$ 0.055 & 0.022 & 0.027 $\pm$ 0.017 & 0.012 & 0.013 $\pm$ 0.009 \\
{}[O {\sc i}]$\lambda$6300/[N {\sc ii}]$\lambda$6583               \dotfill & 
   0.166 & 0.264 $\pm$ 0.303 & 0.271 & 0.395 $\pm$ 0.725 & 0.136 & 0.180 $\pm$ 0.126 \\
{}[O {\sc ii}]$\lambda$3727/[N {\sc ii}]$\lambda$6583              \dotfill & 
   0.432 & 0.635 $\pm$ 0.622 & 0.454 & 0.550 $\pm$ 0.372 & 0.424 & 0.593 $\pm$ 0.459 \\
{}[N {\sc ii}]$\lambda$6583[O {\sc iii}]$\lambda$5007              \dotfill & 
   0.459 & 0.820 $\pm$ 0.967 & 0.311 & 0.486 $\pm$ 0.554 & 0.570 & 1.080 $\pm$ 1.523 \\
{}[S {\sc ii}]$\lambda \lambda$6717,6731/[N {\sc ii}]$\lambda$6583 \dotfill & 
   0.462 & 0.724 $\pm$ 0.665 & 0.550 & 0.708 $\pm$ 0.472 & 0.541 & 0.613 $\pm$ 0.450 \\
{}[Ar {\sc iii}]$\lambda$7136/[O {\sc iii}]$\lambda$5007           \dotfill & 
   0.029 & 0.036 $\pm$ 0.021 & 0.021 & 0.029 $\pm$ 0.021 & 0.041 & 0.054 $\pm$ 0.032 \\
{}[Ne {\sc iii}]$\lambda$3869/[O {\sc iii}]$\lambda$5007           \dotfill & 
   0.210 & 0.246 $\pm$ 0.145 & 0.134 & 0.152 $\pm$ 0.088 & 0.080 & 0.110 $\pm$ 0.118 \\
{}[Ne {\sc iii}]$\lambda$3869/[O {\sc ii}]$\lambda$3727            \dotfill & 
   0.965 & 1.178 $\pm$ 0.951 & 0.674 & 0.873 $\pm$ 0.556 & 0.358 & 0.449 $\pm$ 0.327 \\
{}[Ne {\sc v}]$\lambda$3426/[O {\sc ii}]$\lambda$3727              \dotfill & 
   1.280 & 1.467 $\pm$ 1.031 & 1.006 & 1.519 $\pm$ 1.373 & 0.370 & 0.472 $\pm$ 0.433 \\
{}[Fe {\sc vii}]$\lambda$6087/[O {\sc iii}]$\lambda$5007           \dotfill & 
   0.089 & 0.122 $\pm$ 0.116 & 0.047 & 0.082 $\pm$ 0.092 & 0.017 & 0.022 $\pm$ 0.020 \\
{}[Fe {\sc x}]$\lambda$6374/[O {\sc iii}]$\lambda$5007             \dotfill & 
   0.070 & 0.077 $\pm$ 0.061 & 0.035 & 0.045 $\pm$ 0.034 & 0.009 & 0.021 $\pm$ 0.034 \\
{}[Fe {\sc xi}]$\lambda$7892/[O {\sc iii}]$\lambda$5007            \dotfill & 
   0.067 & 0.073 $\pm$ 0.053 & 0.026 & 0.049 $\pm$ 0.057 & 0.002 & 0.021 $\pm$ 0.036 \\
\hline
\end{tabular*}
\end{table*}

\clearpage

\begin{table*}
\begin{center}
Table~4.\hspace{4pt}
	Resultant KS probabilities$^*$ concerning the forbidden
        emission-line flux ratios.\\
\end{center}
\vspace{6pt}
\begin{tabular*}{\textwidth}{@{\hspace{\tabcolsep}
\extracolsep{\fill}}lccc}
\hline\hline\\[-6pt]
Line Ratio & S1$_{\rm total}$ vs. S1.5s &
S1$_{\rm total}$ vs. S2$_{\rm total}$ & S1.5s vs. S2$_{\rm total}$ \\
[4pt]\hline\\[-6pt]
{}[O {\sc i}]$\lambda$6300/[O {\sc iii}]$\lambda$5007              \dotfill & 
 2.066$\times 10^{-1}$ & 3.140$\times 10^{-1}$ & 2.551$\times 10^{-2}$ \\
{}[O {\sc ii}]$\lambda$3727/[O {\sc iii}]$\lambda$5007             \dotfill & 
 6.928$\times 10^{-2}$ & 8.763$\times 10^{-1}$ & 1.061$\times 10^{-2}$ \\
{}[O {\sc i}]$\lambda$6300/[O {\sc ii}]$\lambda$3727               \dotfill & 
 1.467$\times 10^{-1}$ & 6.653$\times 10^{-1}$ & 5.610$\times 10^{-2}$ \\
{}[O {\sc iii}]$\lambda$4363/[O {\sc iii}]$\lambda$5007            \dotfill & 
 3.120$\times 10^{-1}$ & 8.701$\times 10^{-13}$ & 2.193$\times 10^{-7}$ \\
{}[O {\sc ii}]$\lambda$7325/[O {\sc ii}]$\lambda$3727              \dotfill & 
 2.006$\times 10^{-1}$ & 8.458$\times 10^{-2}$ & 9.894$\times 10^{-1}$ \\
{}[S {\sc ii}]$\lambda$6717/[S {\sc ii}]$\lambda$6731              \dotfill & 
 2.660$\times 10^{-1}$ & 5.359$\times 10^{-1}$ & 8.009$\times 10^{-2}$ \\
{}[S {\sc iii}]$\lambda$9069/[S {\sc ii}]$\lambda \lambda$6717,6731\dotfill & 
 3.009$\times 10^{-1}$ & 4.673$\times 10^{-1}$ & 3.168$\times 10^{-1}$ \\
{}[O {\sc i}]$\lambda$6300/[S {\sc ii}]$\lambda \lambda$6717,6731  \dotfill & 
 3.423$\times 10^{-1}$ & 6.035$\times 10^{-2}$ & 2.670$\times 10^{-3}$ \\
{}[O {\sc ii}]$\lambda$3727/[S {\sc ii}]$\lambda \lambda$6717,6731 \dotfill & 
 6.566$\times 10^{-1}$ & 3.909$\times 10^{-1}$ & 8.521$\times 10^{-1}$ \\
{}[S {\sc ii}]$\lambda \lambda$6717,6731/[O {\sc iii}]$\lambda$5007\dotfill & 
 9.093$\times 10^{-2}$ & 8.885$\times 10^{-3}$ & 5.849$\times 10^{-5}$ \\
{}[N {\sc i}]$\lambda$5199/[N {\sc ii}]$\lambda$6583               \dotfill & 
 8.958$\times 10^{-1}$ & 4.265$\times 10^{-1}$ & 4.116$\times 10^{-1}$ \\
{}[N {\sc ii}]$\lambda$5755/[N {\sc ii}]$\lambda$6583              \dotfill & 
 8.096$\times 10^{-1}$ & 7.785$\times 10^{-2}$ & 1.866$\times 10^{-2}$ \\
{}[O {\sc i}]$\lambda$6300/[N {\sc ii}]$\lambda$6583               \dotfill & 
 1.588$\times 10^{-1}$ & 2.394$\times 10^{-1}$ & 1.839$\times 10^{-3}$ \\
{}[O {\sc ii}]$\lambda$3727/[N {\sc ii}]$\lambda$6583              \dotfill & 
 8.310$\times 10^{-1}$ & 8.702$\times 10^{-1}$ & 9.879$\times 10^{-1}$ \\
{}[N {\sc ii}]$\lambda$6583[O {\sc iii}]$\lambda$5007              \dotfill & 
 7.421$\times 10^{-2}$ & 5.395$\times 10^{-1}$ & 2.386$\times 10^{-3}$ \\
{}[S {\sc ii}]$\lambda \lambda$6717,6731/[N {\sc ii}]$\lambda$6583 \dotfill &
 3.472$\times 10^{-2}$ & 1.658$\times 10^{-1}$ & 4.095$\times 10^{-1}$ \\
{}[Ar {\sc iii}]$\lambda$7136/[O {\sc iii}]$\lambda$5007           \dotfill & 
 2.826$\times 10^{-1}$ & 2.472$\times 10^{-1}$ & 6.642$\times 10^{-3}$ \\
{}[Ne {\sc iii}]$\lambda$3869/[O {\sc iii}]$\lambda$5007           \dotfill & 
 7.646$\times 10^{-5}$ & 1.685$\times 10^{-15}$ & 4.231$\times 10^{-7}$ \\
{}[Ne {\sc iii}]$\lambda$3869/[O {\sc ii}]$\lambda$3727            \dotfill &  
 9.533$\times 10^{-2}$ & 3.494$\times 10^{-13}$ & 1.070$\times 10^{-6}$ \\
{}[Ne {\sc v}]$\lambda$3426/[O {\sc ii}]$\lambda$3727              \dotfill & 
 8.275$\times 10^{-1}$ & 2.068$\times 10^{-4}$ & 1.343$\times 10^{-5}$ \\
{}[Fe {\sc vii}]$\lambda$6087/[O {\sc iii}]$\lambda$5007           \dotfill & 
 1.051$\times 10^{-1}$ & 2.562$\times 10^{-10}$ & 1.172$\times 10^{-5}$ \\
{}[Fe {\sc x}]$\lambda$6374/[O {\sc iii}]$\lambda$5007             \dotfill & 
 2.938$\times 10^{-2}$ & 9.535$\times 10^{-6}$ & 7.356$\times 10^{-4}$ \\
{}[Fe {\sc xi}]$\lambda$7892/[O {\sc iii}]$\lambda$5007            \dotfill & 
 1.963$\times 10^{-1}$ & 1.006$\times 10^{-2}$ & 1.518$\times 10^{-1}$ \\
\hline
\end{tabular*}
\vspace{6pt}\par\noindent
$^*$ These values means the probabilities of two sample drawn from the
     same parent population.
\end{table*}

\clearpage

\begin{table*}
\begin{center}
Table~5.\hspace{4pt}
	Correction factors for the extinction ($A_V$ = 1.0 mag).\\
\end{center}
\vspace{6pt}
\begin{tabular*}{\textwidth}{@{\hspace{\tabcolsep}
\extracolsep{\fill}}lc}
\hline\hline\\[-6pt]
Line Ratio & Correction Factor \\
[4pt]\hline\\[-6pt]
{}[O {\sc i}]$\lambda$6300/[O {\sc iii}]$\lambda$5007              \dotfill & 0.786\\
{}[O {\sc ii}]$\lambda$3727/[O {\sc iii}]$\lambda$5007             \dotfill & 1.471\\
{}[O {\sc i}]$\lambda$6300/[O {\sc ii}]$\lambda$3727               \dotfill & 1.872\\
{}[O {\sc iii}]$\lambda$4363/[O {\sc iii}]$\lambda$5007            \dotfill & 1.222\\
{}[O {\sc ii}]$\lambda$7325/[O {\sc ii}]$\lambda$3727              \dotfill & 0.461\\
{}[S {\sc ii}]$\lambda$6717/[S {\sc ii}]$\lambda$6731              \dotfill & 1.002\\
{}[S {\sc iii}]$\lambda$9069/[S {\sc ii}]$\lambda \lambda$6717,6731\dotfill & 0.745\\
{}[O {\sc i}]$\lambda$6300/[S {\sc ii}]$\lambda \lambda$6717,6731  \dotfill & 1.062\\
{}[O {\sc ii}]$\lambda$3727/[S {\sc ii}]$\lambda \lambda$6717,6731 \dotfill & 1.988\\
{}[S {\sc ii}]$\lambda \lambda$6717,6731/[O {\sc iii}]$\lambda$5007\dotfill & 0.740\\
{}[N {\sc i}]$\lambda$5199/[N {\sc ii}]$\lambda$6583               \dotfill & 1.263\\
{}[N {\sc ii}]$\lambda$5755/[N {\sc ii}]$\lambda$6583              \dotfill & 1.132\\
{}[O {\sc i}]$\lambda$6300/[N {\sc ii}]$\lambda$6583               \dotfill & 1.041\\
{}[O {\sc ii}]$\lambda$3727/[N {\sc ii}]$\lambda$6583              \dotfill & 1.949\\
{}[N {\sc ii}]$\lambda$6583[O {\sc iii}]$\lambda$5007              \dotfill & 0.755\\
{}[S {\sc ii}]$\lambda \lambda$6717,6731/[N {\sc ii}]$\lambda$6583 \dotfill & 0.980\\
{}[Ar {\sc iii}]$\lambda$7136/[O {\sc iii}]$\lambda$5007           \dotfill & 0.697\\
{}[Ne {\sc iii}]$\lambda$3869/[O {\sc iii}]$\lambda$5007           \dotfill & 1.423\\
{}[Ne {\sc iii}]$\lambda$3869/[O {\sc ii}]$\lambda$3727            \dotfill & 0.967\\
{}[Ne {\sc v}]$\lambda$3426/[O {\sc ii}]$\lambda$3727              \dotfill & 1.069\\
{}[Fe {\sc vii}]$\lambda$6087/[O {\sc iii}]$\lambda$5007           \dotfill & 0.811\\
{}[Fe {\sc x}]$\lambda$6374/[O {\sc iii}]$\lambda$5007             \dotfill & 0.778\\
{}[Fe {\sc xi}]$\lambda$7892/[O {\sc iii}]$\lambda$5007            \dotfill & 0.627\\
\hline
\end{tabular*}
\end{table*}

\clearpage

\begin{table*}
\begin{center}
Table~6.\hspace{4pt}
	Ionization potentials and the critical densities
        for forbidden emission lines.\\
\end{center}
\vspace{6pt}
\begin{tabular*}{\textwidth}{@{\hspace{\tabcolsep}
\extracolsep{\fill}}lccc}
\hline\hline\\[-6pt]
Line & \multicolumn{2}{c}{Ionization Potential} & Critical Density \\
     & Lower (eV) & Upper (eV) & (cm$^{-3}$) \\
[4pt]\hline\\[-6pt]
{}[Ne {\sc v}]$\lambda$3426    & 97.1 & 126.2 & 1.6 $\times 10^7$ \\
{}[O {\sc ii}]$\lambda$3726    & 13.6 & 35.1 & 1.4 $\times 10^4$ \\
{}[O {\sc ii}]$\lambda$3729    & 13.6 & 35.1 & 3.3 $\times 10^3$ \\
{}[Ne {\sc iii}]$\lambda$3869  & 41.0 & 63.5 & 9.7 $\times 10^6$ \\
{}[O {\sc iii}]$\lambda$4363   & 35.1 & 54.9 & 3.3 $\times 10^7$ \\
{}[O {\sc iii}]$\lambda$5007   & 35.1 & 54.9 & 7.0 $\times 10^5$ \\
{}[N {\sc i}]$\lambda$5199     & 0.0 & 14.5 & 2.0 $\times 10^3$ \\
{}[N {\sc ii}]$\lambda$5755    & 14.5 & 29.6 & 3.2 $\times 10^7$ \\
{}[Fe {\sc vii}]$\lambda$6087  & 99.1 & 125.0 & 3.6 $\times 10^7$ \\
{}[O {\sc i}]$\lambda$6300     & 0.0 & 13.6 & 1.8 $\times 10^6$ \\
{}[Fe {\sc x}]$\lambda$6374    & 233.6 & 262.1 & 4.8 $\times 10^9$ \\
{}[N {\sc ii}]$\lambda$6583    & 14.5 & 29.6 & 8.7 $\times 10^4$ \\
{}[S {\sc ii}]$\lambda$6717    & 10.4 & 23.3 & 1.5 $\times 10^3$ \\
{}[S {\sc ii}]$\lambda$6731    & 10.4 & 23.3 & 3.9 $\times 10^3$ \\
{}[Ar {\sc iii}]$\lambda$7136  & 27.6 & 40.7 & 4.8 $\times 10^6$ \\
{}[O {\sc ii}]$\lambda$7318.6  & 13.6 & 35.1 & 4.8 $\times 10^6$ \\
{}[O {\sc ii}]$\lambda$7319.9  & 13.6 & 35.1 & 7.4 $\times 10^6$ \\
{}[O {\sc ii}]$\lambda$7329.9  & 13.6 & 35.1 & 8.6 $\times 10^6$ \\
{}[O {\sc ii}]$\lambda$7330.7  & 13.6 & 35.1 & 7.0 $\times 10^6$ \\
{}[Fe {\sc xi}]$\lambda$7892   & 262.1 & 290.1 & $\cdots$ \\
{}[S {\sc iii}]$\lambda$9069   & 23.3 & 34.8 & 1.2 $\times 10^6$ \\
\hline
\end{tabular*}
\end{table*}

\clearpage

\begin{table*}
\begin{center}
Table~7.\hspace{4pt}
	Median, average, and 1 $\sigma$ deviation 
        of the line luminosities normalized by $\nu L_{\nu}$(25 $\mu$m).\\
\end{center}
\vspace{6pt}
\begin{tabular*}{\textwidth}{@{\hspace{\tabcolsep}
\extracolsep{\fill}}lcccccc}
\hline\hline\\[-6pt]
 & \multicolumn{2}{c}{Seyfert 1 Galaxies} &
\multicolumn{2}{c}{Seyfert 1.5 Galaxies} & \multicolumn{2}{c}{Seyfert 2 Galaxies} \\
  & Median & Average and 1 $\sigma$ & Median & Average and 1 $\sigma$ & Median & Average and 1 $\sigma$ \\
[4pt]\hline\\[-6pt]
$L$([O {\sc iii}]$\lambda$5007)/$\nu L_{\nu}$(25$\mu$m)  & 
   0.245 & 0.251 $\pm$ 0.165 & 
   0.216 & 0.536 $\pm$ 1.117 & 
   0.179 & 0.334 $\pm$ 0.403 \\
$L$([O {\sc iii}]$\lambda$4363)/$\nu L_{\nu}$(25$\mu$m)  & 
   0.018 & 0.021 $\pm$ 0.016 & 
   0.013 & 0.020 $\pm$ 0.024 & 
   0.008 & 0.015 $\pm$ 0.018 \\
$L$([Ne {\sc iii}]$\lambda$3869)/$\nu L_{\nu}$(25$\mu$m) & 
   0.039 & 0.060 $\pm$ 0.076 & 
   0.028 & 0.074 $\pm$ 0.108 & 
   0.037 & 0.044 $\pm$ 0.041 \\
$L$([Fe {\sc vii}]$\lambda$6087)/$\nu L_{\nu}$(25$\mu$m) & 
   0.026 & 0.037 $\pm$ 0.034 & 
   0.013 & 0.017 $\pm$ 0.014 & 
   0.007 & 0.011 $\pm$ 0.014 \\
\hline
\end{tabular*}
\end{table*}


\begin{table*}
\begin{center}
Table~8.\hspace{4pt}
	Resultant KS probabilities$^*$ concerning 
        emission-line luminosities normalized by $\nu L_{\nu}$(25 $\mu$m).\\
\end{center}
\vspace{6pt}
\begin{tabular*}{\textwidth}{@{\hspace{\tabcolsep}
\extracolsep{\fill}}lccc}
\hline\hline\\[-6pt]
Line & S1$_{\rm total}$ vs. S1.5s & 
S1$_{\rm total}$ vs. S2$_{\rm total}$ & S1.5s vs. S2$_{\rm total}$ \\
[4pt]\hline\\[-6pt]
$L$([O {\sc iii}]$\lambda$5007)/$\nu L_{\nu}$(25$\mu$m)      & 
   3.513$\times 10^{-1}$  &  1.800$\times 10^{-1}$  &  1.280$\times 10^{-1}$ \\
$L$([O {\sc iii}]$\lambda$4363)/$\nu L_{\nu}$(25$\mu$m)      & 
   2.916$\times 10^{-1}$  &  2.474$\times 10^{-2}$  &  3.038$\times 10^{-1}$ \\
$L$([Ne {\sc iii}]$\lambda$3869)/$\nu L_{\nu}$(25$\mu$m)     & 
   4.313$\times 10^{-1}$  &  2.450$\times 10^{-1}$  &  6.721$\times 10^{-1}$ \\
$L$([Fe {\sc vii}]$\lambda$6087)/$\nu L_{\nu}$(25$\mu$m)     & 
   4.983$\times 10^{-2}$  &  1.801$\times 10^{-3}$  &  4.809$\times 10^{-2}$ \\
\hline
\end{tabular*}
\vspace{6pt}\par\noindent
$^*$ These values means the probabilities of two sample drawn from the
     same parent population.
\end{table*}

\clearpage


\begin{figure*}
\figurenum{1}
\epsscale{0.5}
\plotone{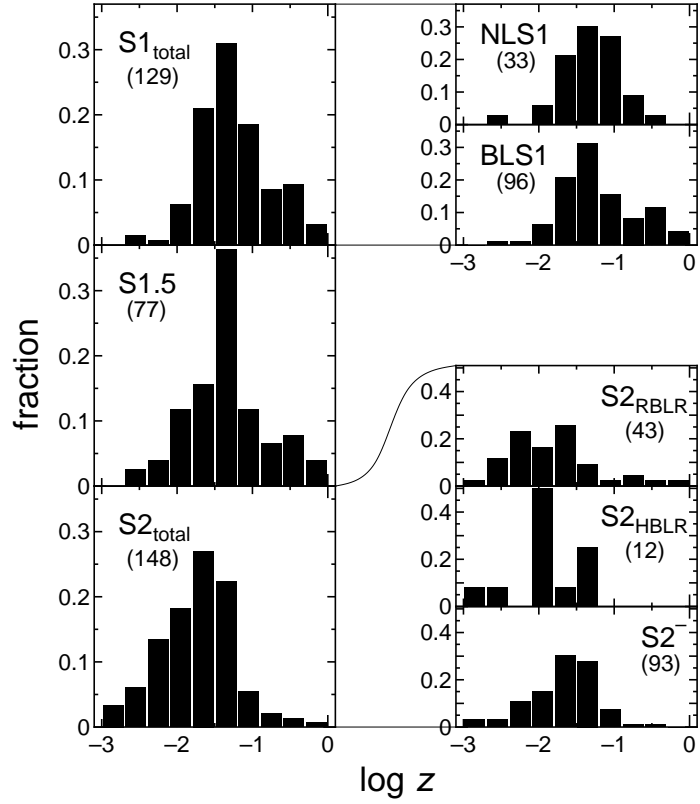}
\caption{Frequency distributions of the redshift for each class of 
         Seyfert nuclei. The number of objects for each types is written
         in parenthesis.
         Note that the data of a nearby S1.5, NGC 3031, is not 
         included in this figure because its redshift is negative.}
\end{figure*}

\begin{figure*}
\figurenum{2}
\epsscale{0.5}
\plotone{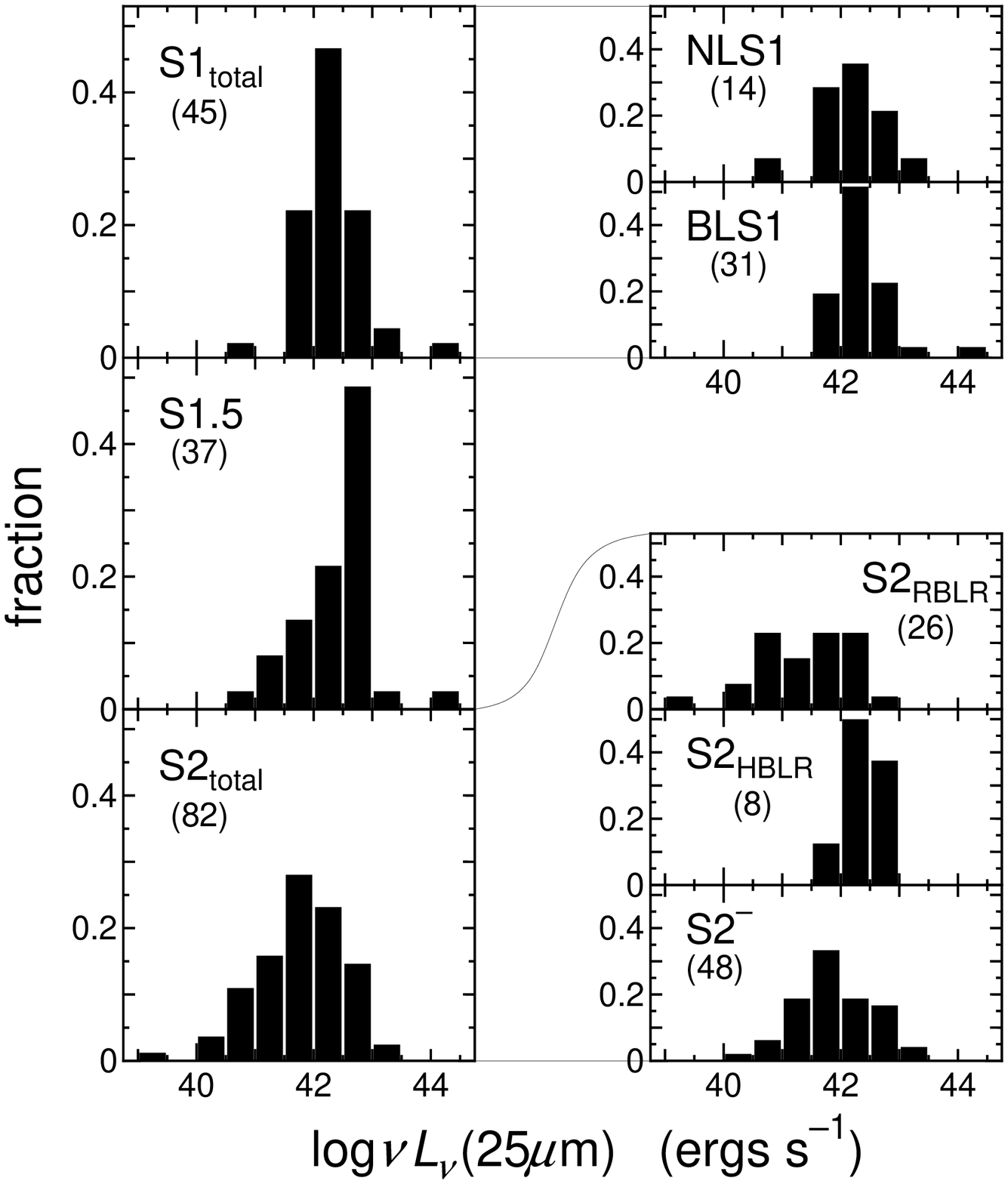}
\caption{Same as figure 1, but for the IRAS 25 $\mu$m luminosity.}
\end{figure*}

\begin{figure*}
\figurenum{3}
\epsscale{0.5}
\plotone{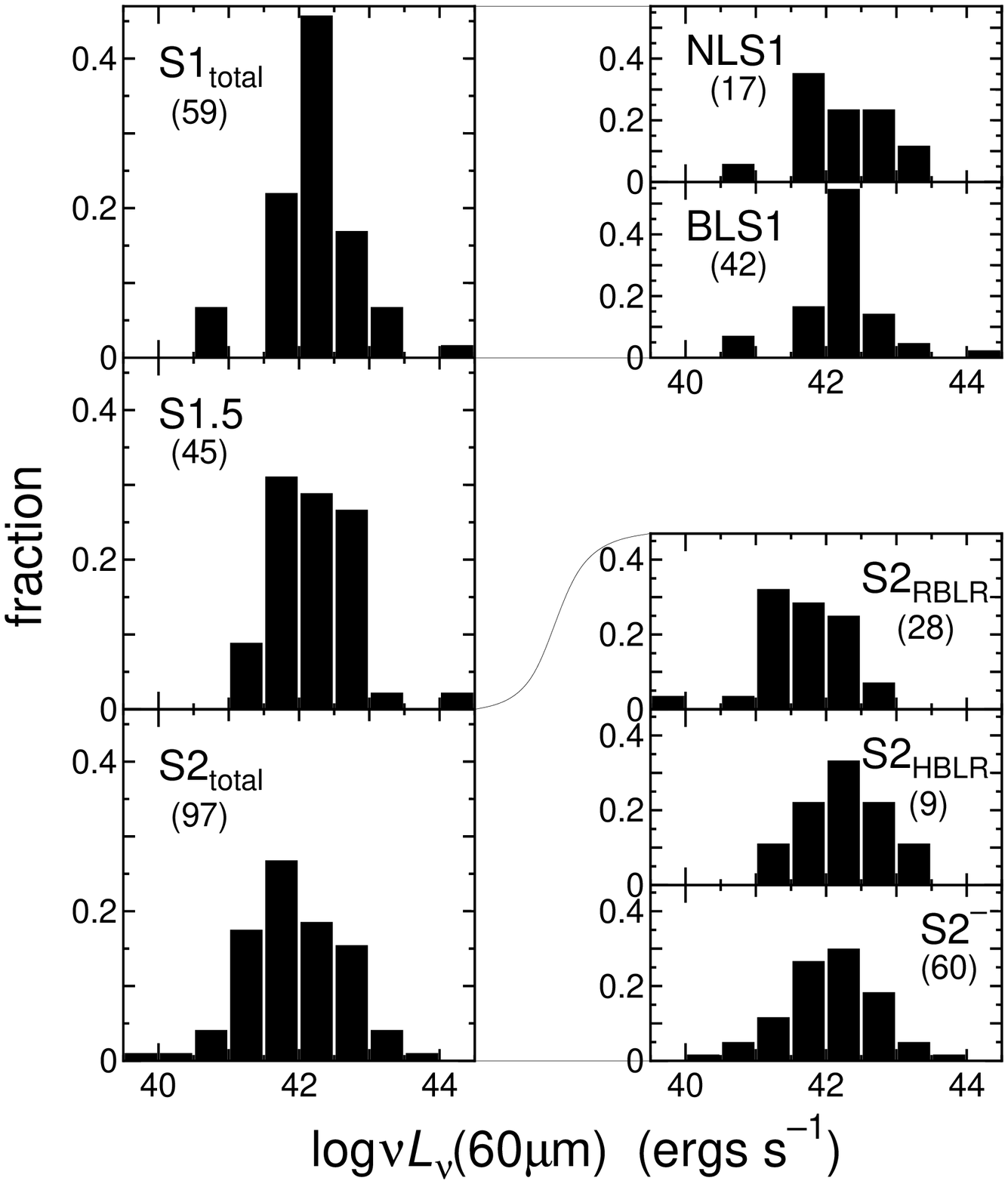}
\caption{Same as figure 1, but for the IRAS 60 $\mu$m luminosity.}
\end{figure*}

\begin{figure*}
\figurenum{4}
\epsscale{1.0}
\plotone{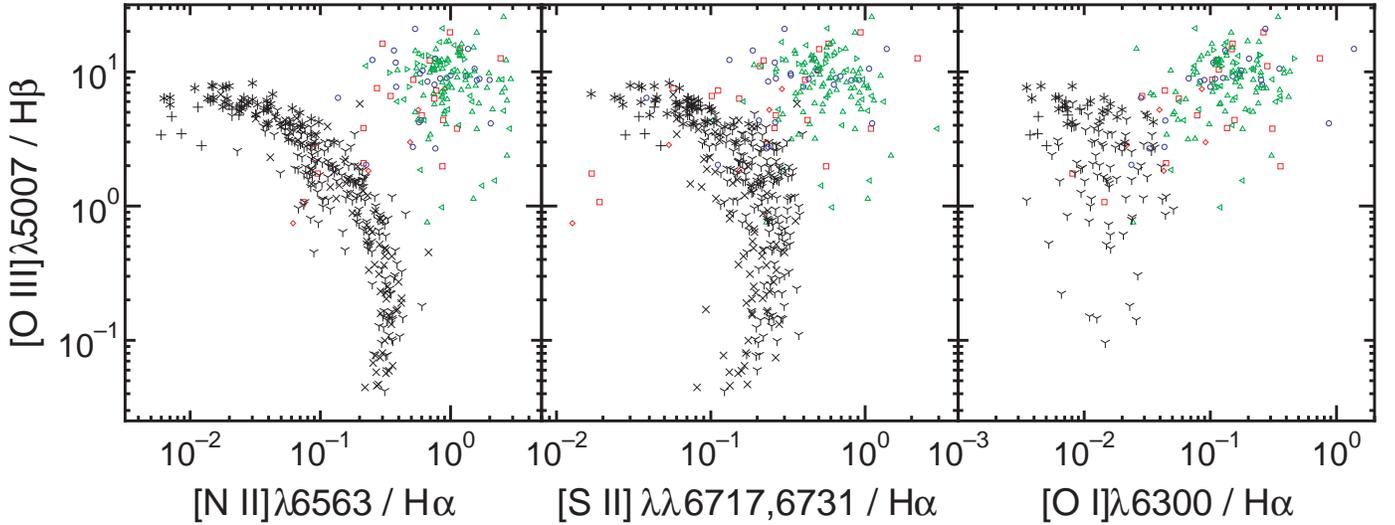}
\caption{Compiled objects are shown in the diagnostic diagrams proposed
         by Veilleux and Osterbrock (1987).
         The diamonds are NLS1s, the squares are BLS1s, the circles are
         S1.5s, the triangles pointing to the left are S2$_{\rm RBLR}$s,
         those pointing to the right are S2$_{\rm HBLR}$s, and
         those pointing to the upper side are S2$^-$.
         Red color means S1s, blue color means S1.5s, and green color
         means S2s.
         For comparison, compiled H {\sc ii} systems are also
         plotted in this figure. 
         The plus signs are blue compact
         galaxies (Izotov et al. 1994), the crosses and
         the ``Y'' signs are extragalactic H {\sc ii} galaxies 
         (McCall et al. 1985; van Zee et al. 1998), and the
         asterisks are H {\sc ii} galaxies (Masegosa et al. 1994).
         (a) Diagram of [O {\sc iii}]$\lambda$5007/H$\beta$ versus 
             [N {\sc ii}]$\lambda$6583/H$\alpha$.
         (b) Diagram of [O {\sc iii}]$\lambda$5007/H$\beta$ versus 
             [S {\sc ii}]$\lambda \lambda$6717,6731/H$\alpha$.
         (c) Diagram of [O {\sc iii}]$\lambda$5007/H$\beta$ versus 
             [O {\sc i}]$\lambda$6300/H$\alpha$.}
\end{figure*}

\begin{figure*}
\figurenum{5}
\epsscale{0.45}
\plotone{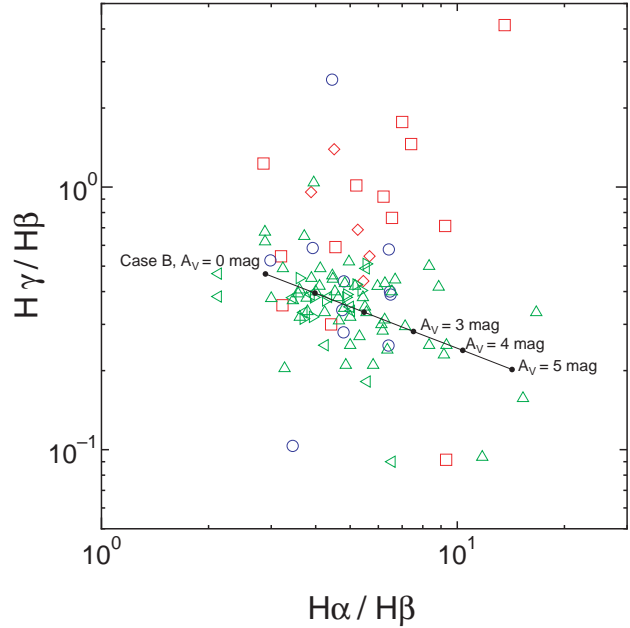}
\caption{Diagram of H$\gamma$/H$\beta$ versus H$\alpha$/H$\beta$.
         The symbols are the same as those in figure 4.
         Theoretically predicted emission-line ratios assuming the case B 
         and dust extinction of 0 mag $ \leq A_V \leq$ 5 mag
         are also plotted by a solid line.}
\end{figure*}

\begin{figure*}
\figurenum{6a}
\epsscale{0.5}
\plotone{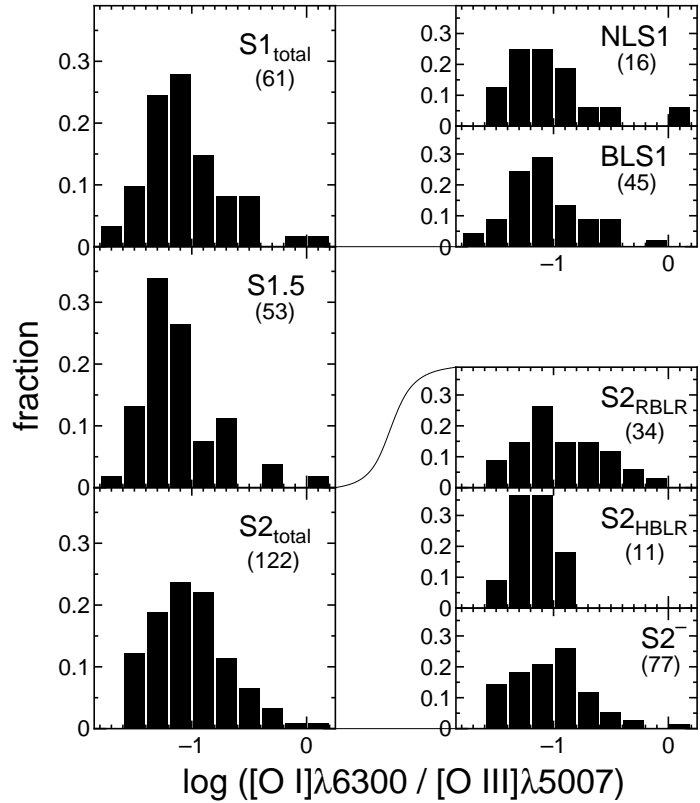}
\caption{Frequency distributions of the various emission-line flux ratios.
         (a) Flux ratio of 
             [O {\sc i}]$\lambda$6300/[O {\sc iii}]$\lambda$5007.
         (b) Flux ratio of 
             [O {\sc ii}]$\lambda$3727/[O {\sc iii}]$\lambda$5007.
         (c) Flux ratio of 
             [O {\sc i}]$\lambda$6300/[O {\sc ii}]$\lambda$3727.
         (d) Flux ratio of 
             [O {\sc iii}]$\lambda$4363/[O {\sc iii}]$\lambda$5007.
         (e) Flux ratio of 
             [S {\sc ii}]$\lambda$6717/[S {\sc ii}]$\lambda$6731.
         (f) Flux ratio of 
             [O {\sc i}]$\lambda$6300/[S {\sc ii}]$\lambda \lambda$6717,6731.
         (g) Flux ratio of 
             [O {\sc ii}]$\lambda$3727/[S {\sc ii}]$\lambda \lambda$6717,6731.
         (h) Flux ratio of 
             [S {\sc ii}]$\lambda \lambda$6717,6731/[O {\sc iii}]$\lambda$5007.
         (i) Flux ratio of 
             [O {\sc i}]$\lambda$6300/[N {\sc ii}]$\lambda$6583.
         (j) Flux ratio of 
             [O {\sc ii}]$\lambda$3727/[N {\sc ii}]$\lambda$6583.
         (k) Flux ratio of 
             [N {\sc ii}]$\lambda$6583/[O {\sc iii}]$\lambda$5007.
         (l) Flux ratio of 
             [S {\sc ii}]$\lambda \lambda$6717,6731/[N {\sc ii}]$\lambda$6583.
         (m) Flux ratio of 
             [Ne {\sc iii}]$\lambda$3869/[O {\sc iii}]$\lambda$5007.
         (n) Flux ratio of 
             [Ne {\sc iii}]$\lambda$3869/[O {\sc ii}]$\lambda$3727.
         (o) Flux ratio of 
             [Ne {\sc v}]$\lambda$3426/[O {\sc ii}]$\lambda$3727.
         (p) Flux ratio of 
             [Fe {\sc vii}]$\lambda$6087/[O {\sc iii}]$\lambda$5007.}
\end{figure*}

\begin{figure*}
\figurenum{6b}
\epsscale{0.5}
\plotone{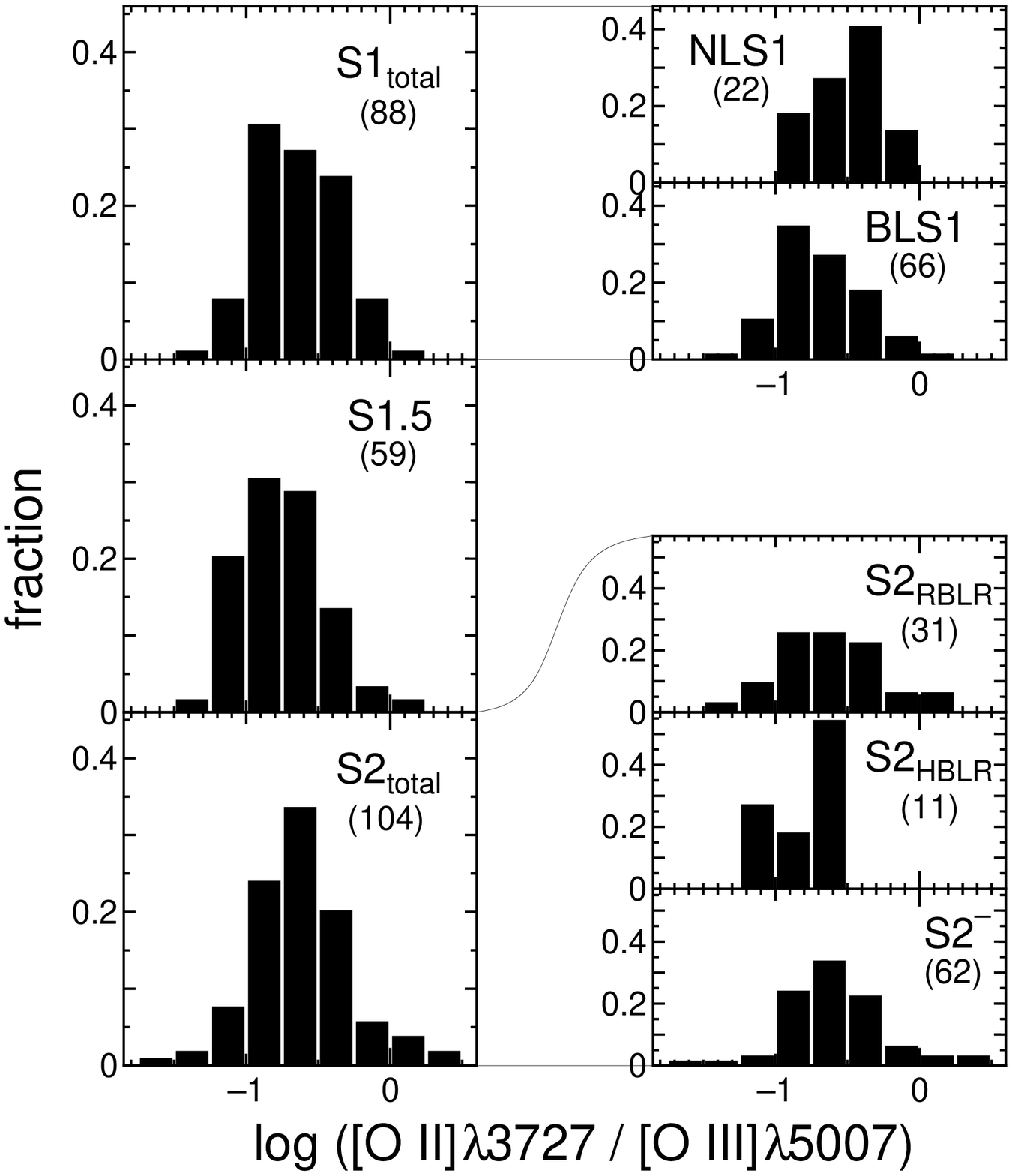}
\caption{continued.}
\end{figure*}

\begin{figure*}
\figurenum{6c}
\epsscale{0.5}
\plotone{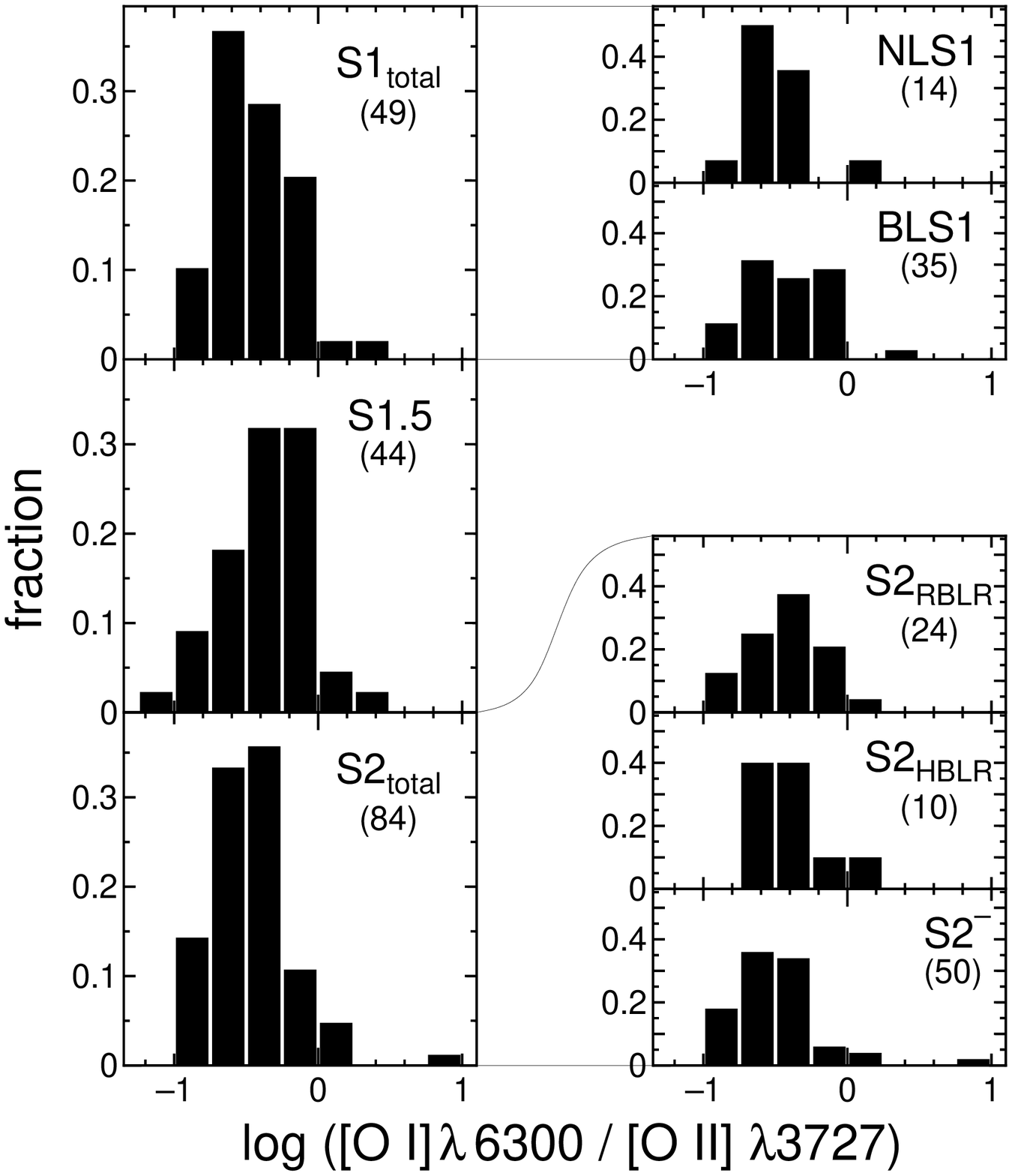}
\caption{continued.}
\end{figure*}

\begin{figure*}
\figurenum{6d}
\epsscale{0.5}
\plotone{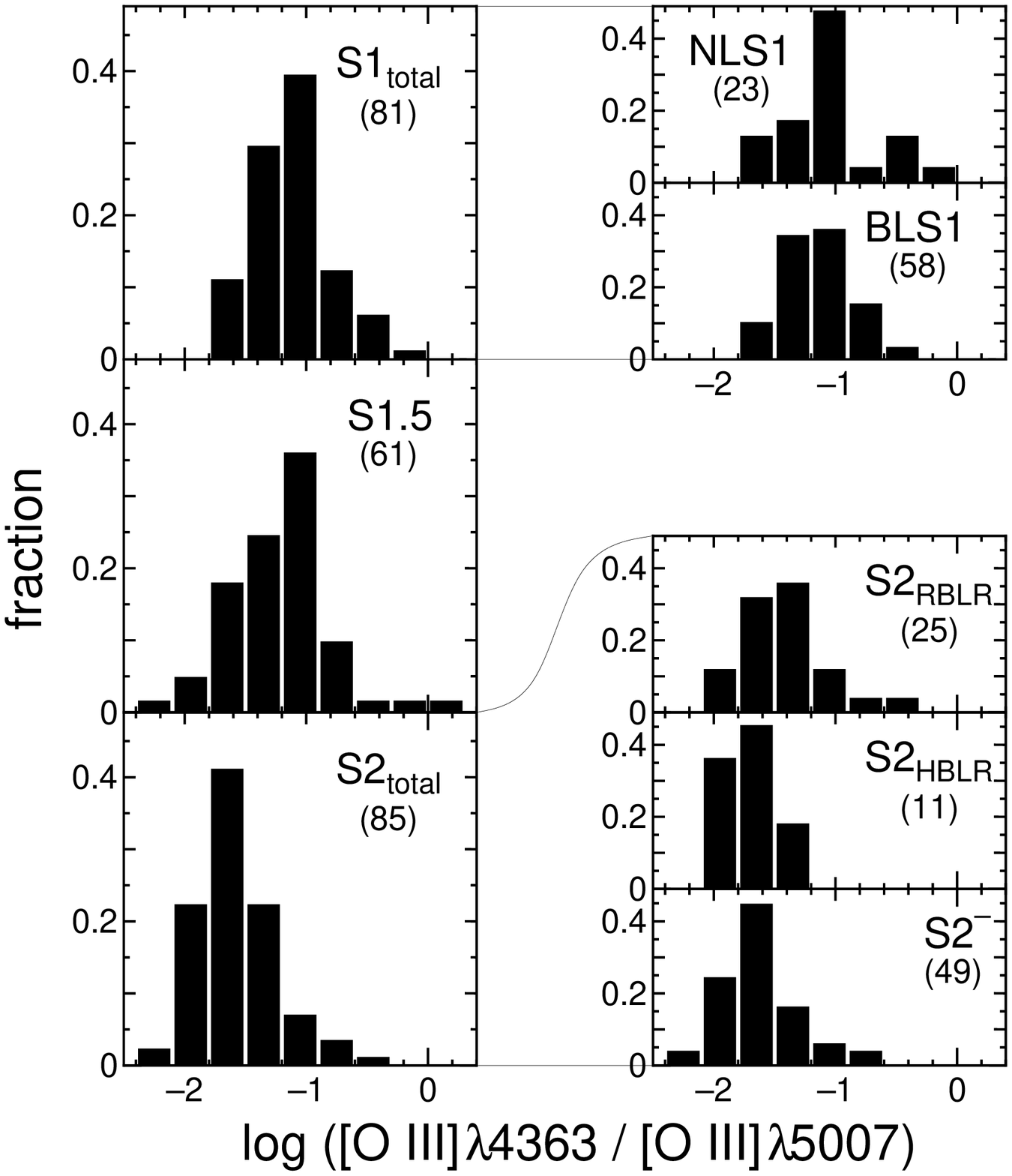}
\caption{continued.}
\end{figure*}

\begin{figure*}
\figurenum{6e}
\epsscale{0.5}
\plotone{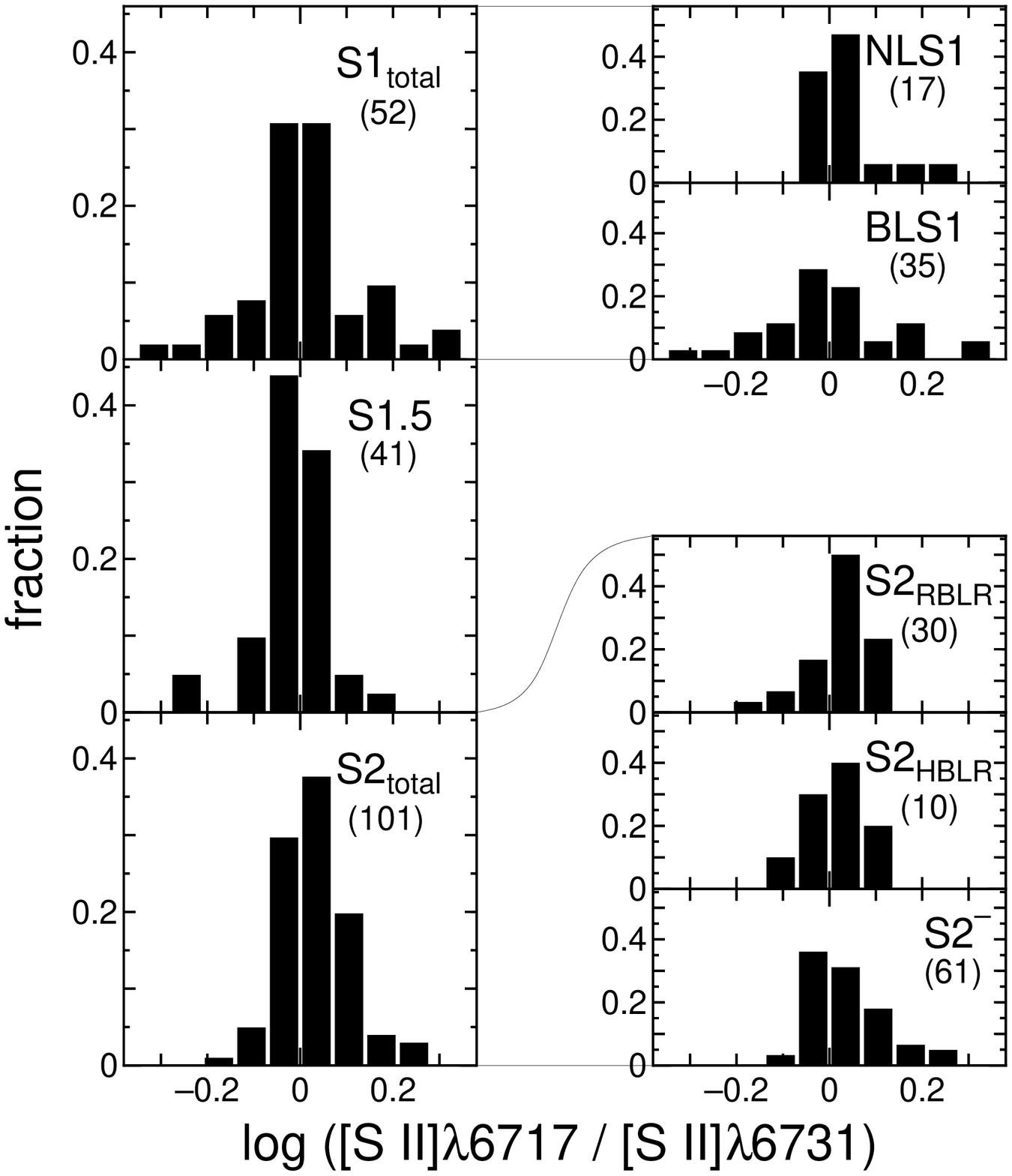}
\caption{continued.}
\end{figure*}

\begin{figure*}
\figurenum{6f}
\epsscale{0.5}
\plotone{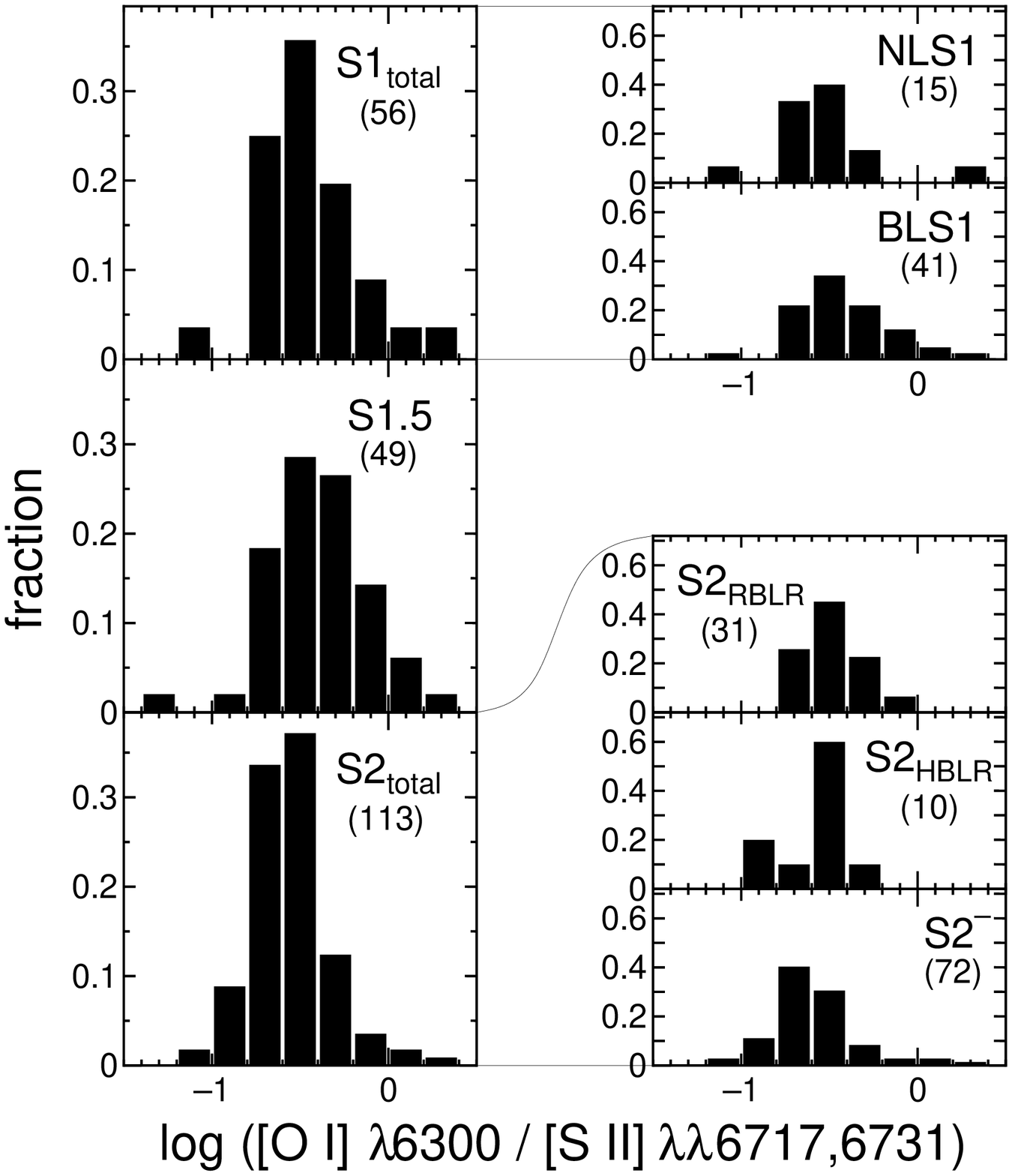}
\caption{continued.}
\end{figure*}

\begin{figure*}
\figurenum{6g}
\epsscale{0.5}
\plotone{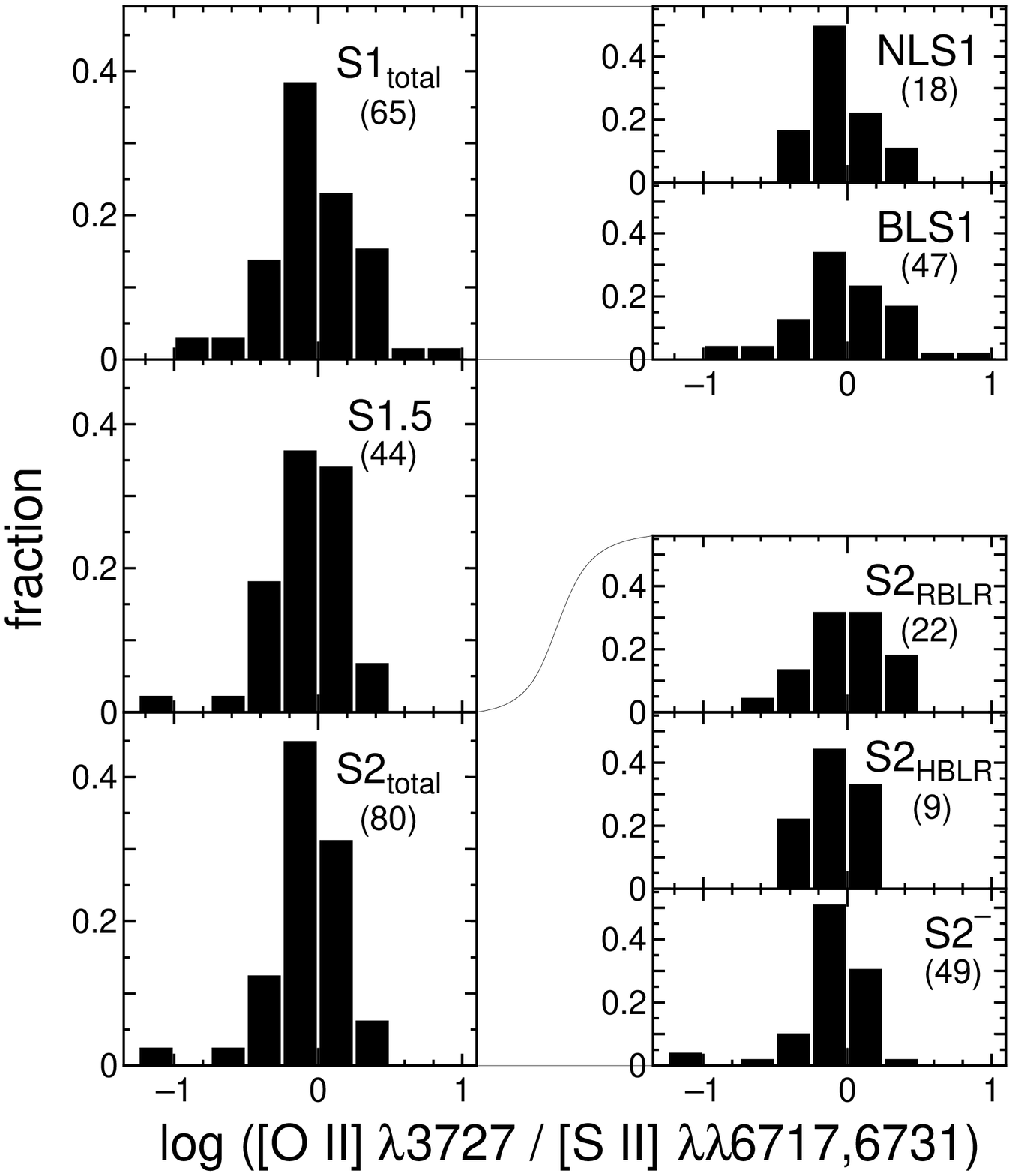}
\caption{continued.}
\end{figure*}

\begin{figure*}
\figurenum{6h}
\epsscale{0.5}
\plotone{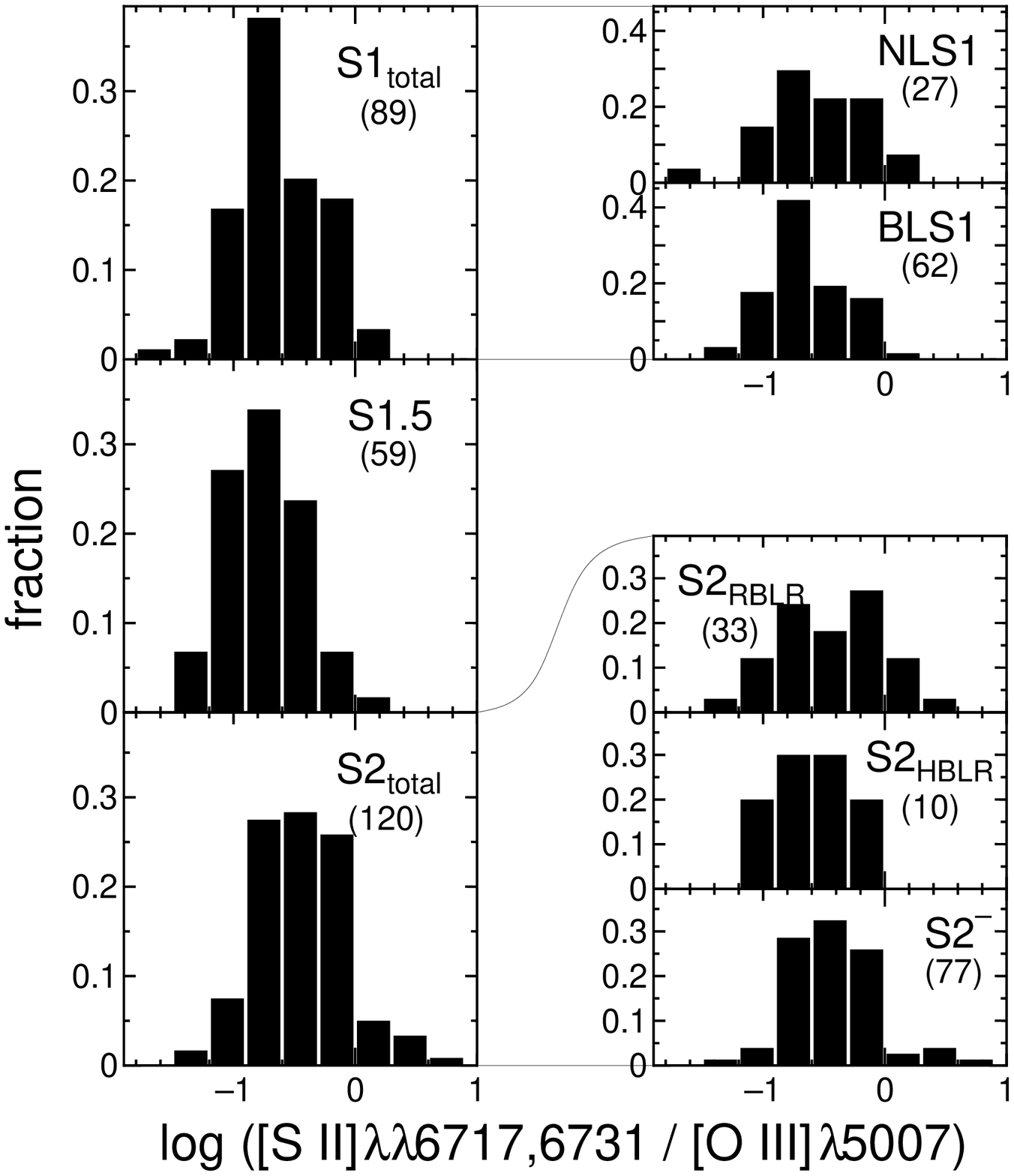}
\caption{continued.}
\end{figure*}

\begin{figure*}
\figurenum{6i}
\epsscale{0.5}
\plotone{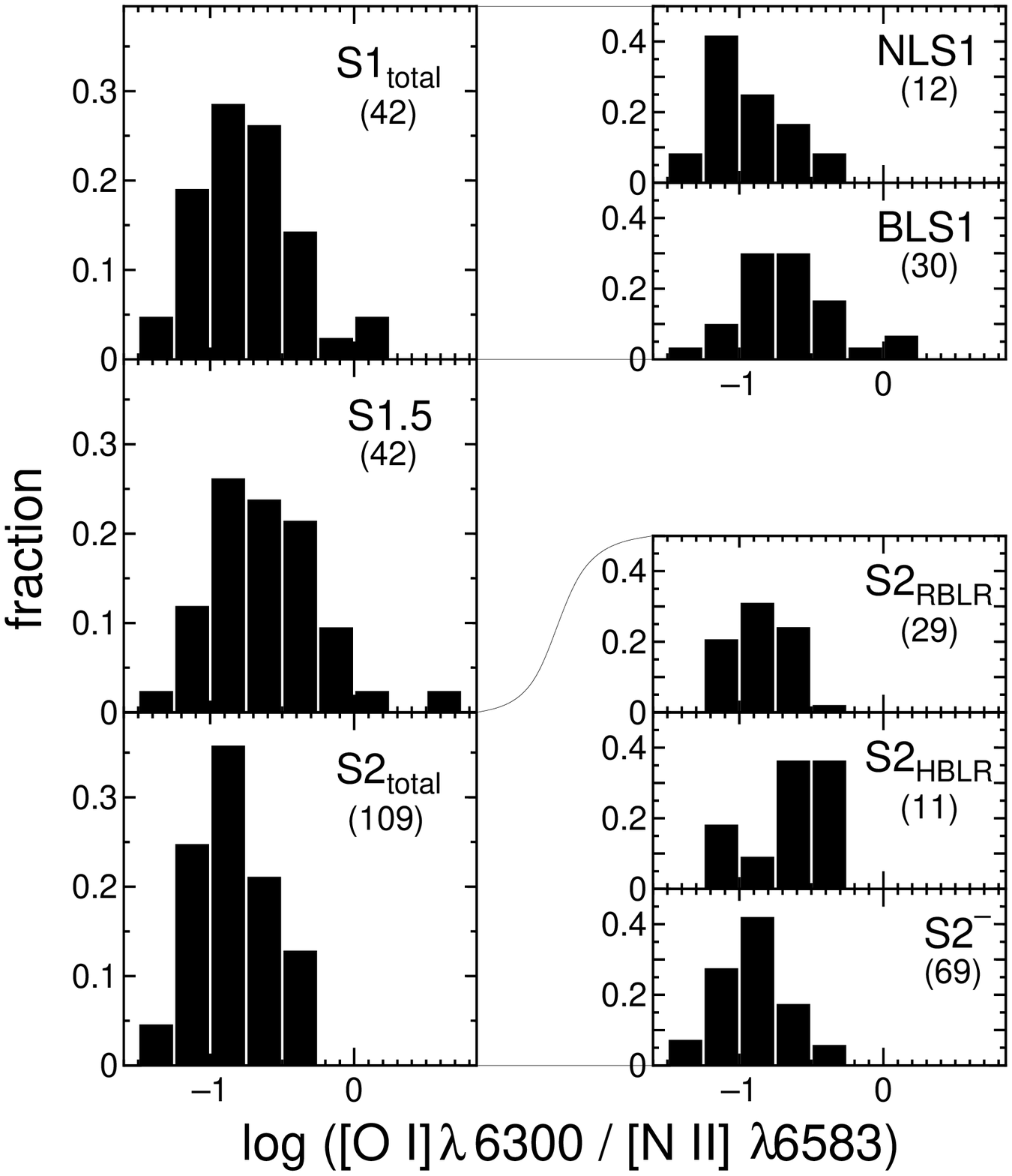}
\caption{continued.}
\end{figure*}

\begin{figure*}
\figurenum{6j}
\epsscale{0.5}
\plotone{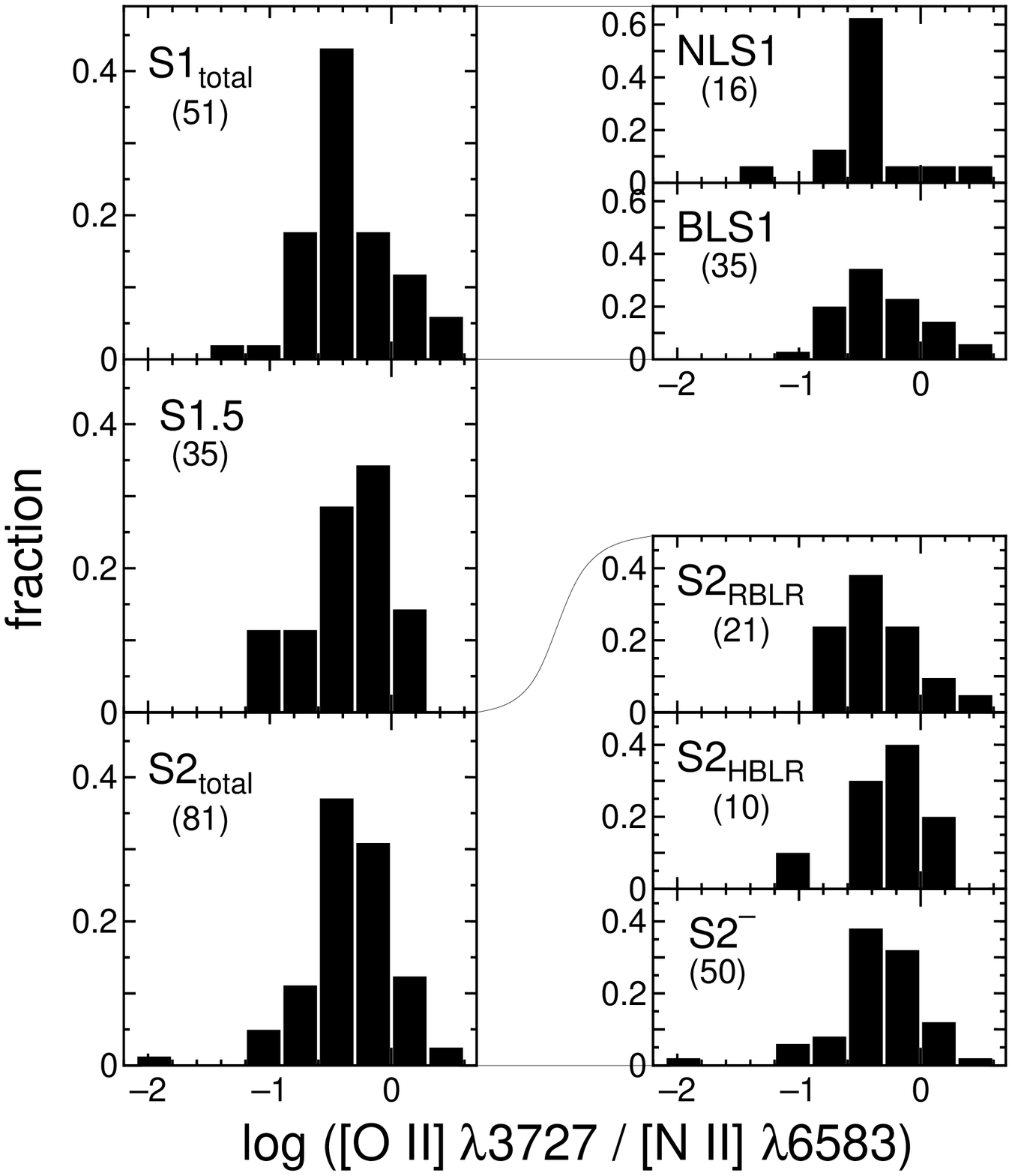}
\caption{continued.}
\end{figure*}

\begin{figure*}
\figurenum{6k}
\epsscale{0.5}
\plotone{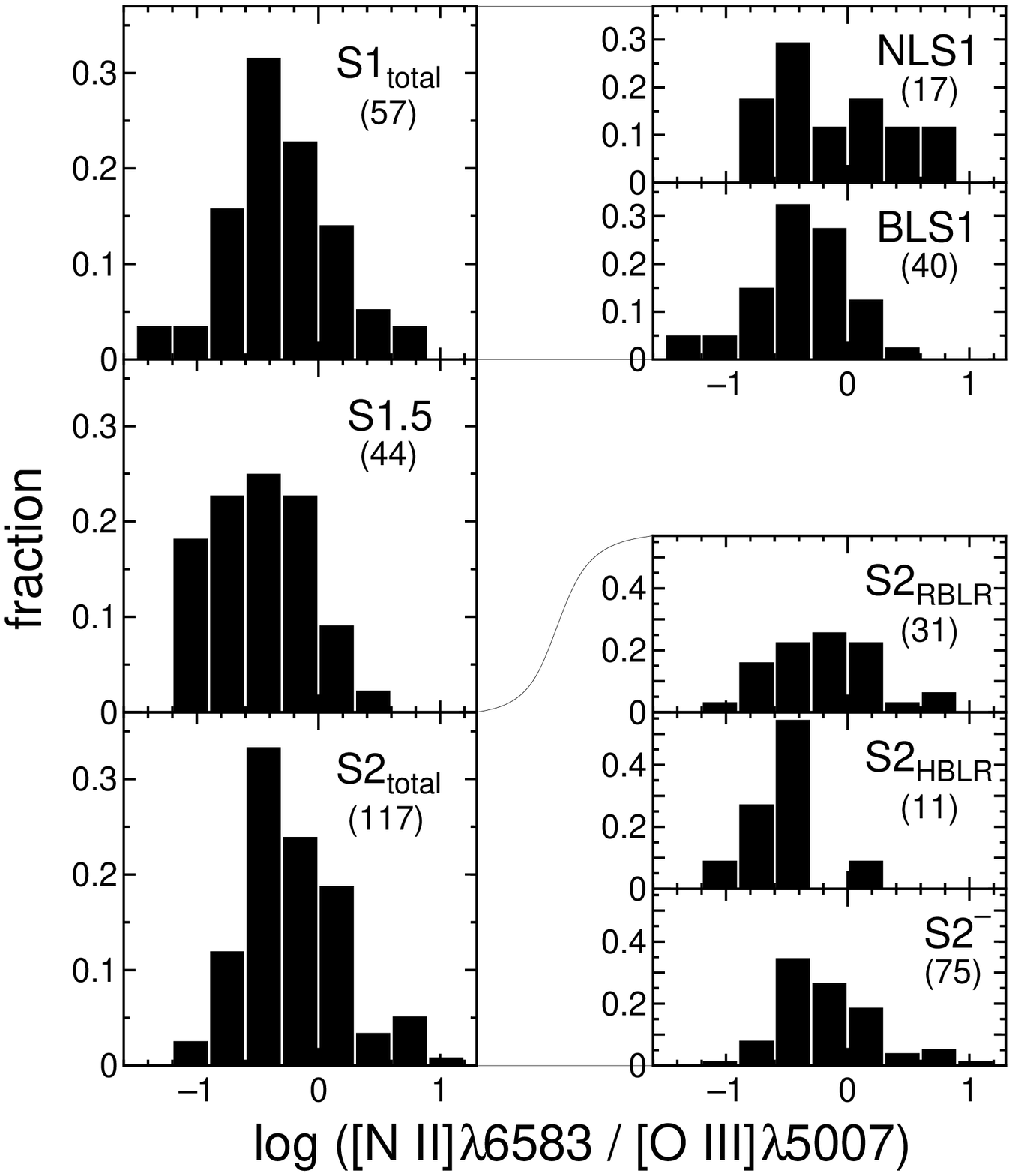}
\caption{continued.}
\end{figure*}

\begin{figure*}
\figurenum{6l}
\epsscale{0.5}
\plotone{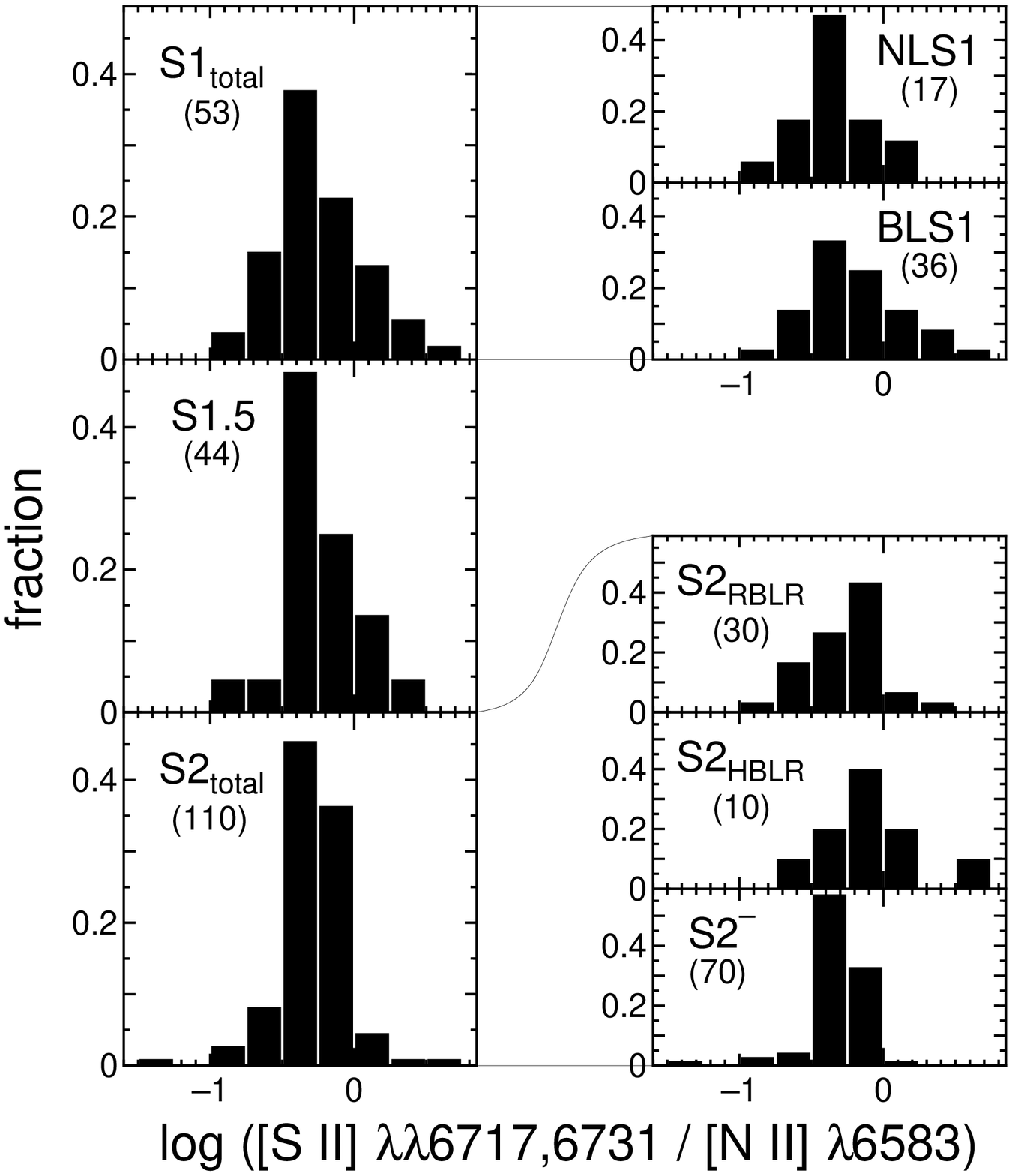}
\caption{continued.}
\end{figure*}

\begin{figure*}
\figurenum{6m}
\epsscale{0.5}
\plotone{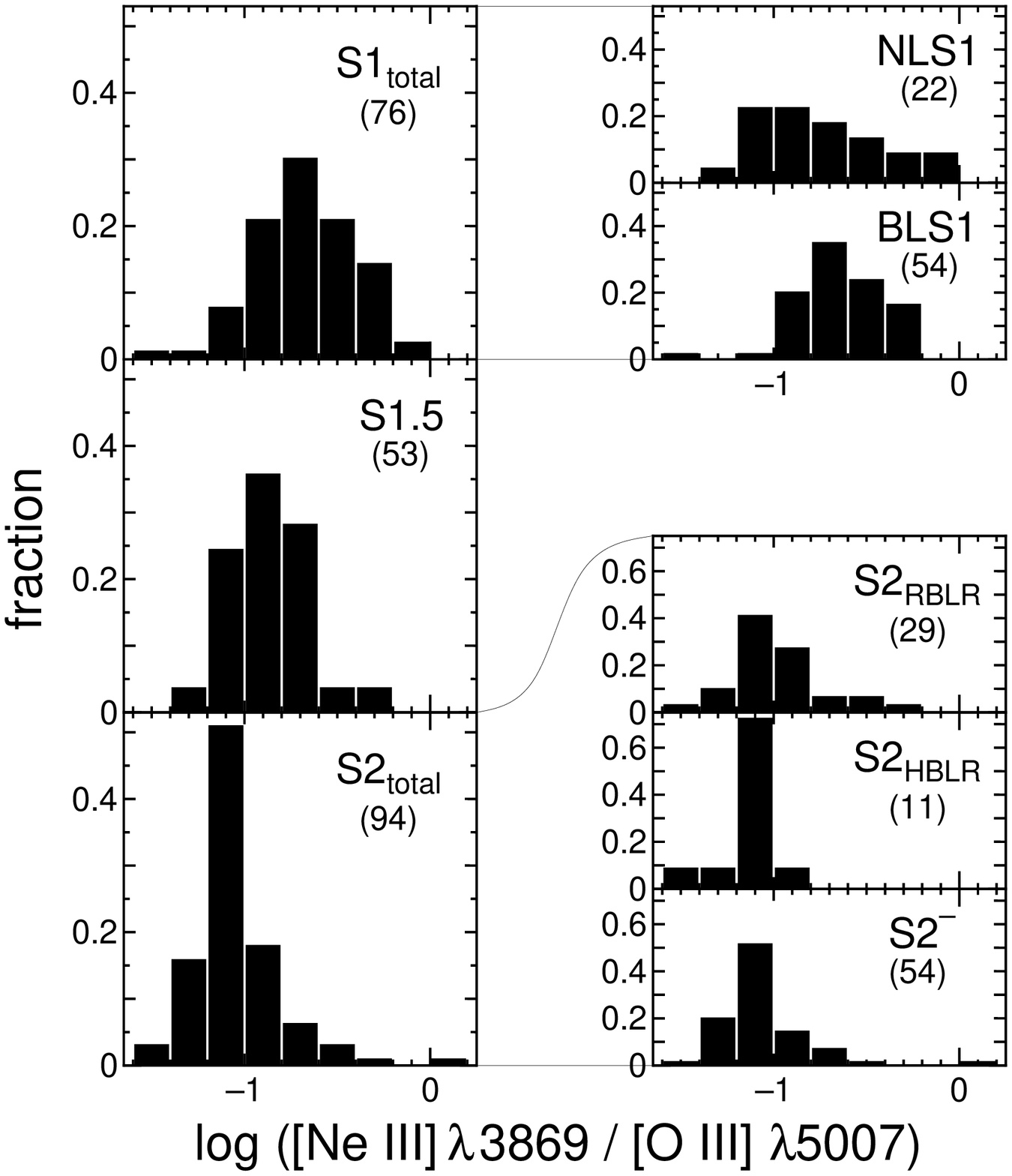}
\caption{continued.}
\end{figure*}

\clearpage

\begin{figure*}
\figurenum{6n}
\epsscale{0.5}
\plotone{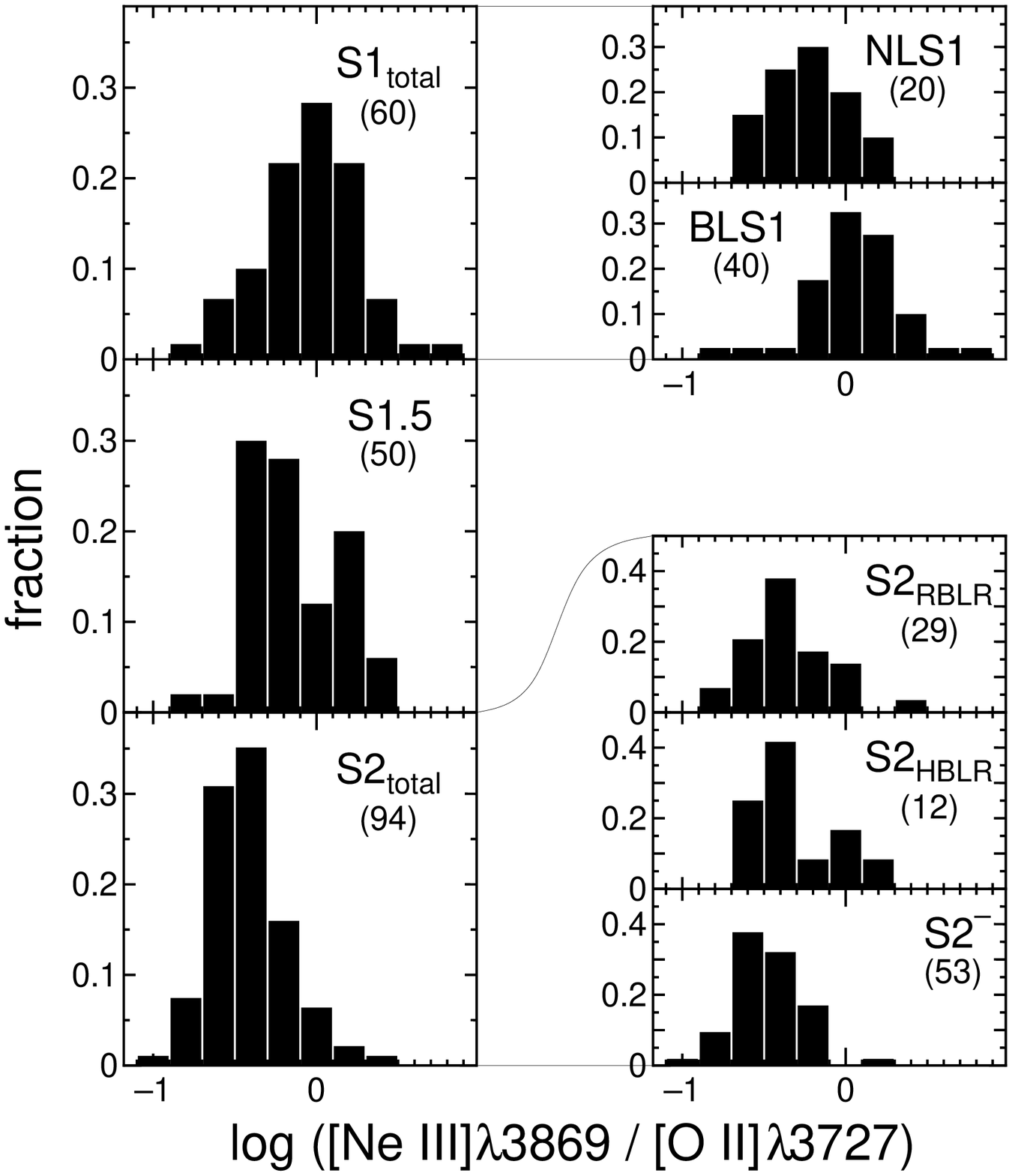}
\caption{continued.}
\end{figure*}

\begin{figure*}
\figurenum{6o}
\epsscale{0.5}
\plotone{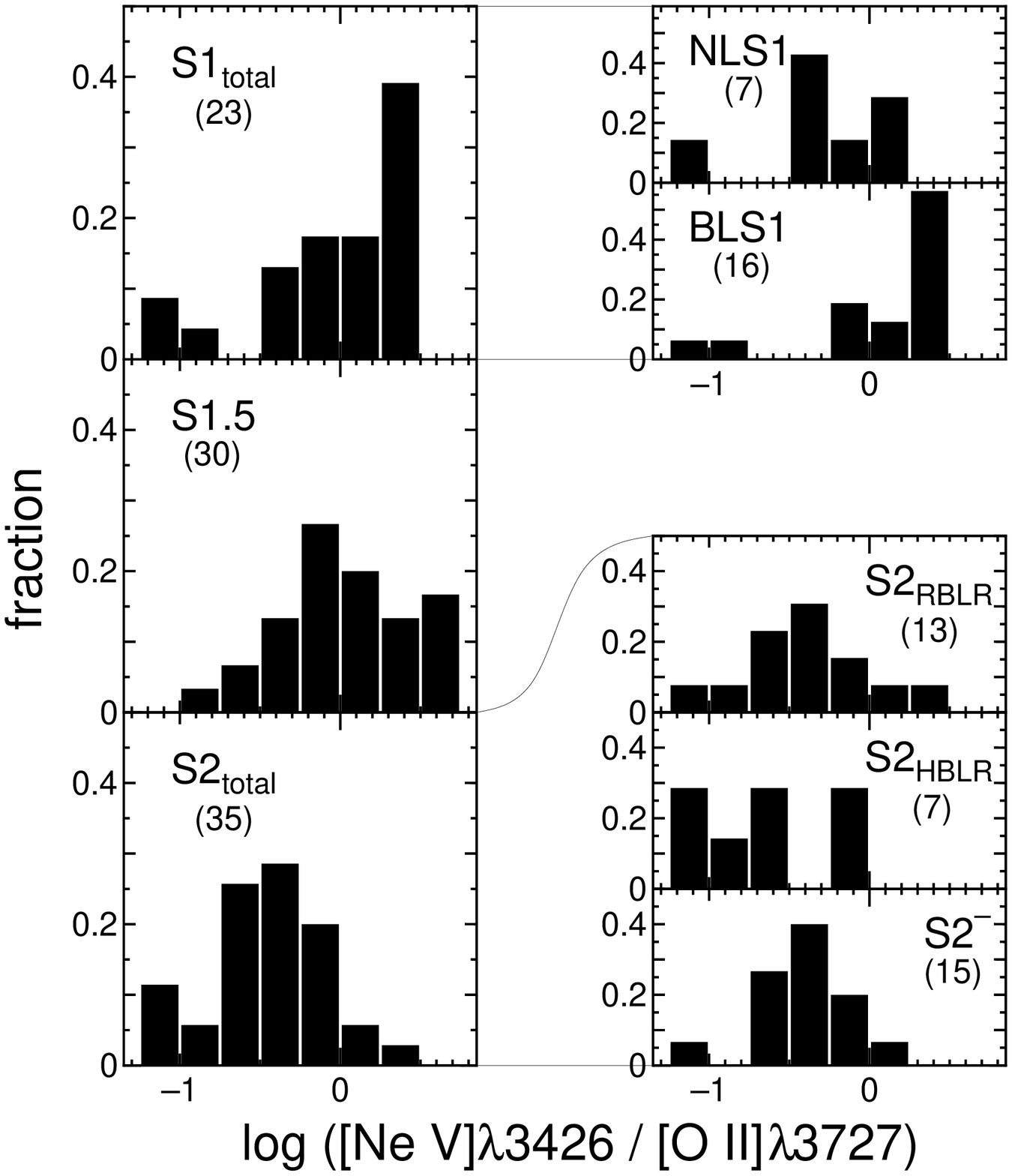}
\caption{continued.}
\end{figure*}

\begin{figure*}
\figurenum{6p}
\epsscale{0.5}
\plotone{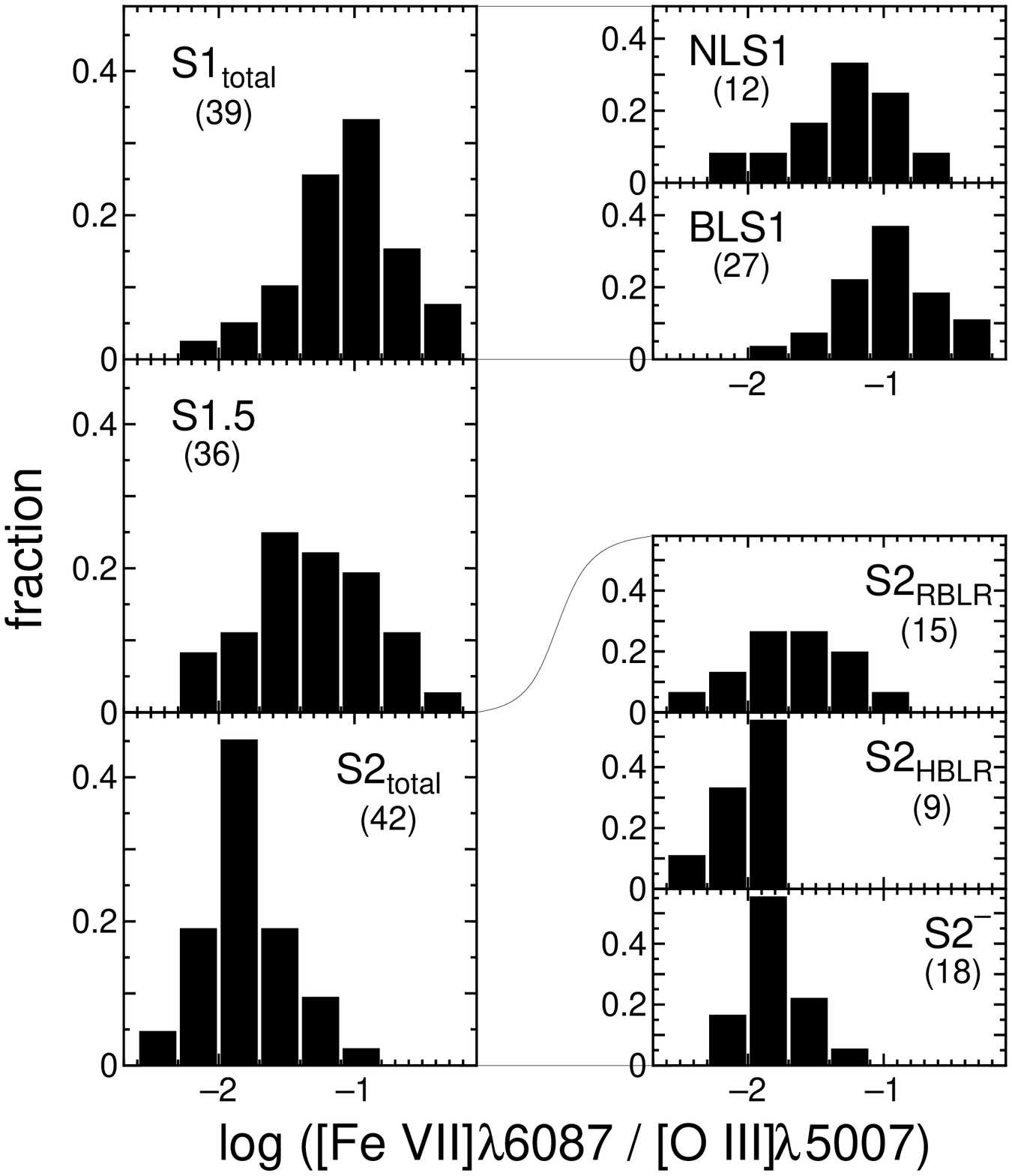}
\caption{continued.}
\end{figure*}

\begin{figure*}
\figurenum{7}
\epsscale{1.0}
\plotone{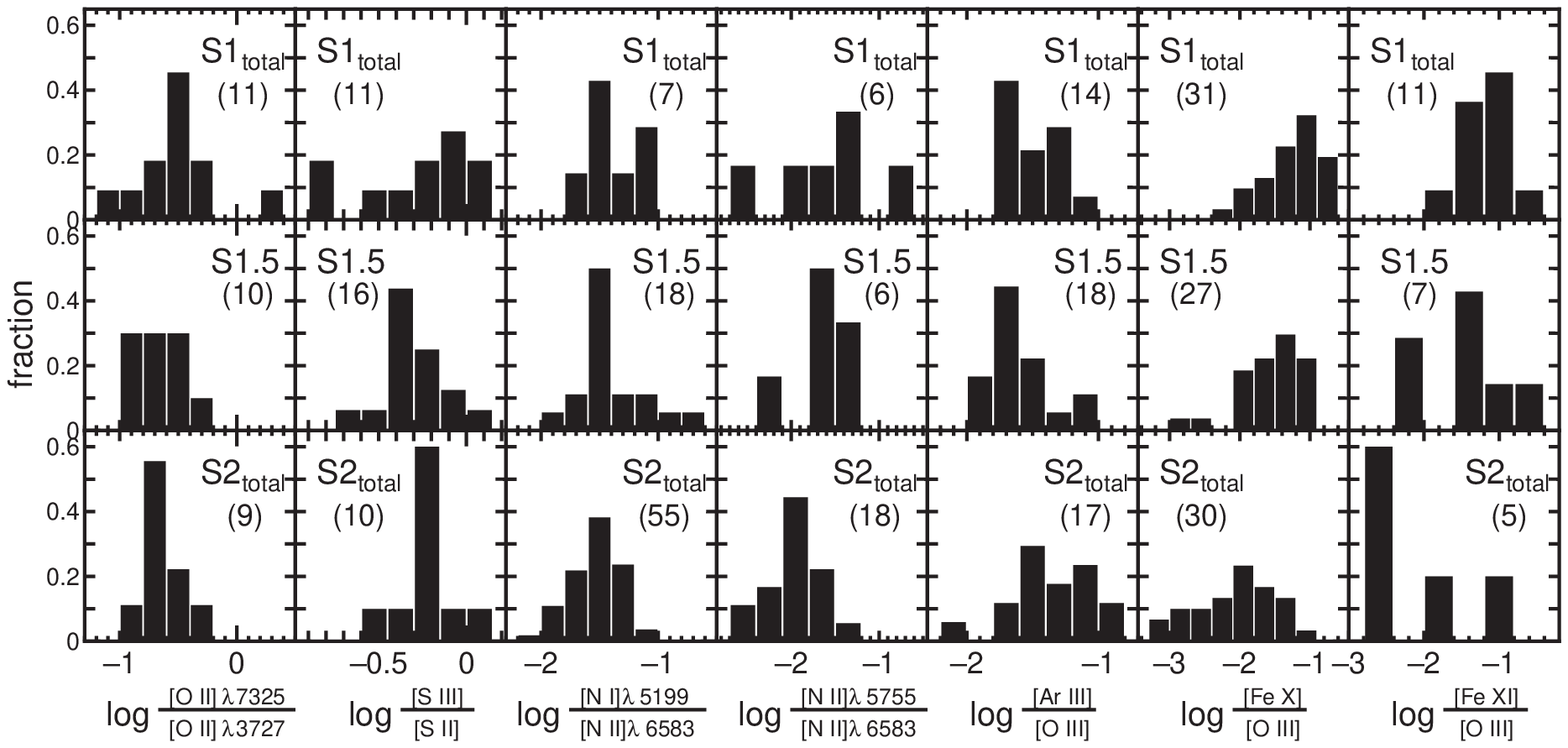}
\caption{Frequency distributions of the various emission-line flux ratios.
         Since the number of objects in each sample is small, the
         histograms are shown only for the S1$_{\rm total}$s, the S1.5s, 
         and the S2$_{\rm total}$.
         (a) Flux ratio of 
             [O {\sc ii}]$\lambda$7325/[O {\sc ii}]$\lambda$3727.
         (b) Flux ratio of 
             [S {\sc iii}]$\lambda$9069/[S {\sc ii}]$\lambda \lambda$6717,6731.
         (c) Flux ratio of 
             [N {\sc i}]$\lambda$5199/[N {\sc ii}]$\lambda$6583.
         (d) Flux ratio of 
             [N {\sc ii}]$\lambda$5755/[N {\sc ii}]$\lambda$6583.
         (e) Flux ratio of 
             [Ar {\sc iii}]$\lambda$7136/[O {\sc iii}]$\lambda$5007.
         (f) Flux ratio of 
             [Fe {\sc x}]$\lambda$6374/[O {\sc iii}]$\lambda$5007.
         (g) Flux ratio of 
             [Fe {\sc xi}]$\lambda$7892/[O {\sc iii}]$\lambda$5007.}
\end{figure*}

\begin{figure*}
\figurenum{8a}
\epsscale{0.6}
\plotone{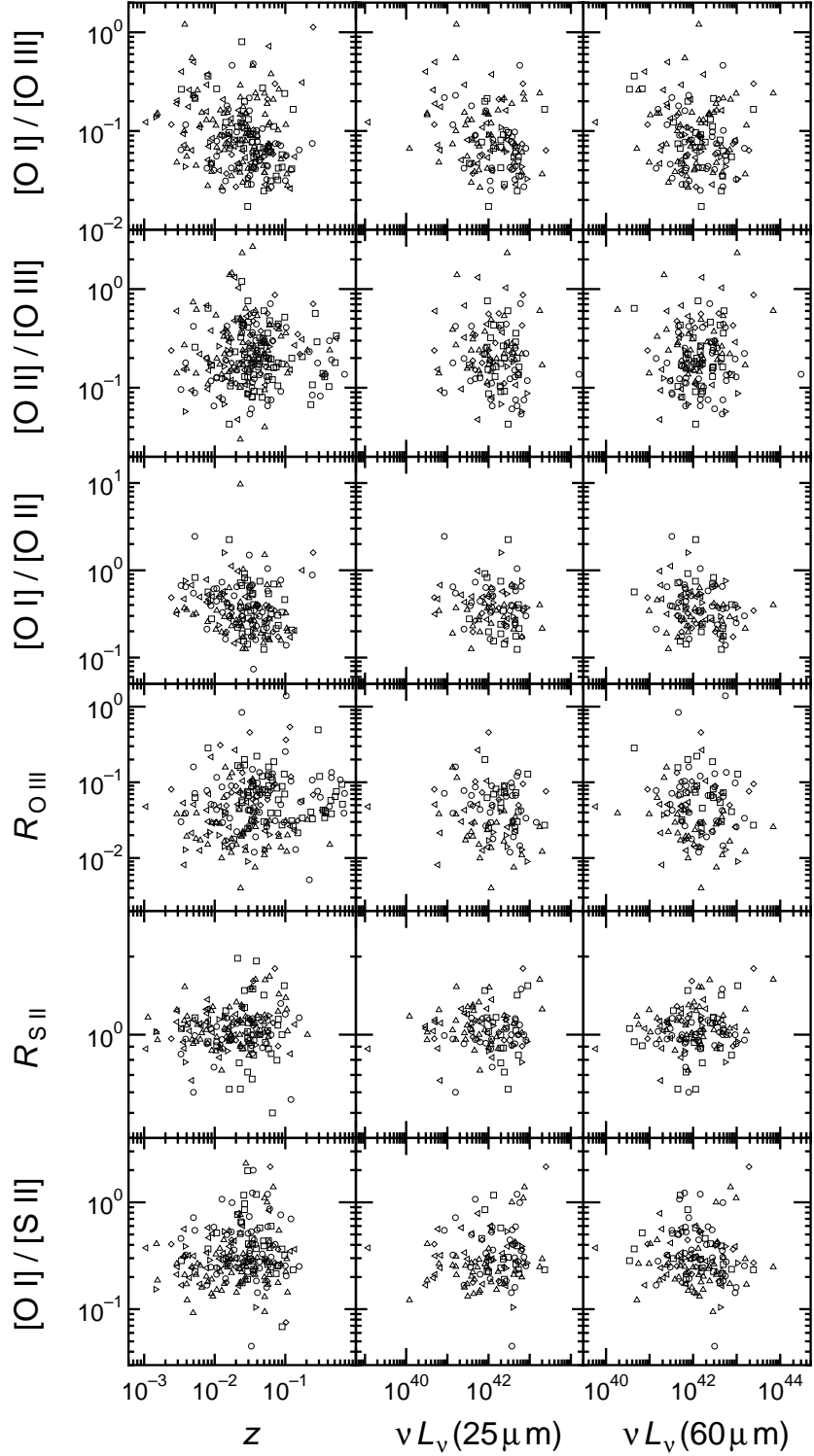}
\caption{Various emission-line flux ratios are plotted as functions of 
         redshift, IRAS 25 $\mu$m luminosity, and IRAS 60 $\mu$m luminosity.
         The symbols are the same as those in figure 4.
         (a) Flux ratios of
            [O {\sc i}]$\lambda$6300/[O {\sc iii}]$\lambda$5007,
            [O {\sc ii}]$\lambda$3727/[O {\sc iii}]$\lambda$5007,
            [O {\sc i}]$\lambda$6300/[O {\sc ii}]$\lambda$3727,
            [O {\sc iii}]$\lambda$4363/[O {\sc iii}]$\lambda$5007 
            (=$R_{\rm O III}$),
            [S {\sc ii}]$\lambda$6717/[S {\sc ii}]$\lambda$6731
            (=$R_{\rm S II}$), and
            [O {\sc i}]$\lambda$6300/[S {\sc ii}]$\lambda \lambda$6717,6731.
         (b) Flux ratios of
            [O {\sc ii}]$\lambda$3727/[S {\sc ii}]$\lambda \lambda$6717,6731
            [S {\sc ii}]$\lambda \lambda$6717,6731/[O {\sc iii}]$\lambda$5007,
            [O {\sc i}]$\lambda$6300/[N {\sc ii}]$\lambda$6583,
            [O {\sc ii}]$\lambda$3727/[N {\sc ii}]$\lambda$6583,
            [N {\sc ii}]$\lambda$6583/[O {\sc iii}]$\lambda$5007, and
            [S {\sc ii}]$\lambda \lambda$6717,6731/[N {\sc ii}]$\lambda$6583.
         (c) Flux ratios of
            [Ne {\sc iii}]$\lambda$3869/[O {\sc iii}]$\lambda$5007,
            [Ne {\sc iii}]$\lambda$3869/[O {\sc ii}]$\lambda$3727,
            [Ne {\sc v}]$\lambda$3426/[O {\sc ii}]$\lambda$3727,
            [Fe {\sc vii}]$\lambda$6087/[O {\sc iii}]$\lambda$5007,
            [O {\sc ii}]$\lambda$7325/[O {\sc ii}]$\lambda$3727
            (=$R_{\rm O II}$), and
            [S {\sc iii}]$\lambda$9069/[S {\sc ii}]$\lambda \lambda$6717,6731.
         (d) Flux ratios of
            [N {\sc i}]$\lambda$5199/[N {\sc ii}]$\lambda$6583,
            [N {\sc ii}]$\lambda$5755/[N {\sc ii}]$\lambda$6583 
            (=$R_{\rm N II}$), 
            [Ar {\sc iii}]$\lambda$7136/[O {\sc iii}]$\lambda$5007,
            [Fe {\sc x}]$\lambda$6374/[O {\sc iii}]$\lambda$5007, and
            [Fe {\sc xi}]$\lambda$7892/[O {\sc iii}]$\lambda$5007.}
\end{figure*}

\begin{figure*}
\figurenum{8b}
\epsscale{0.6}
\plotone{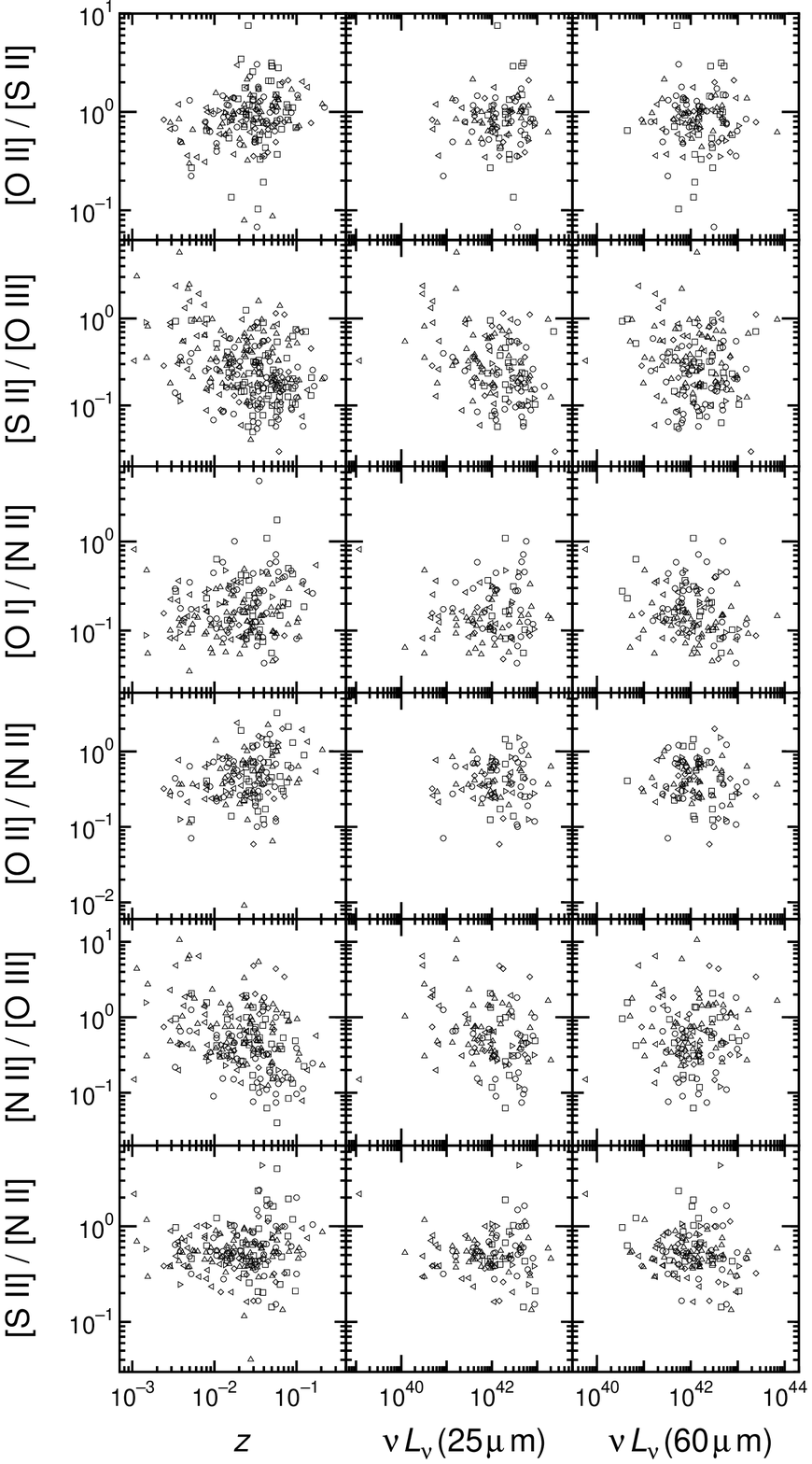}
\caption{continued.}
\end{figure*}

\begin{figure*}
\figurenum{8c}
\epsscale{0.6}
\plotone{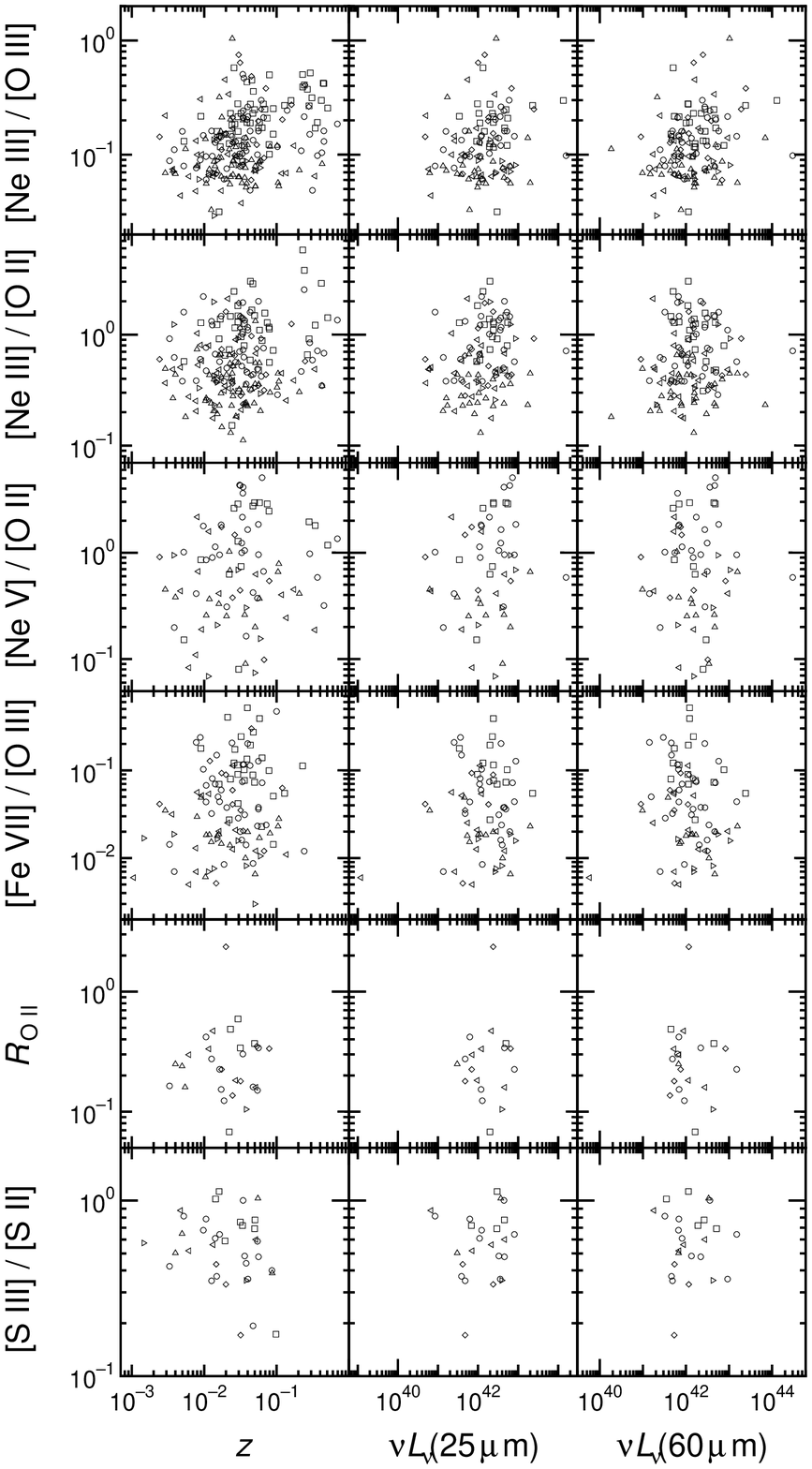}
\caption{continued.}
\end{figure*}

\begin{figure*}
\figurenum{8d}
\epsscale{0.6}
\plotone{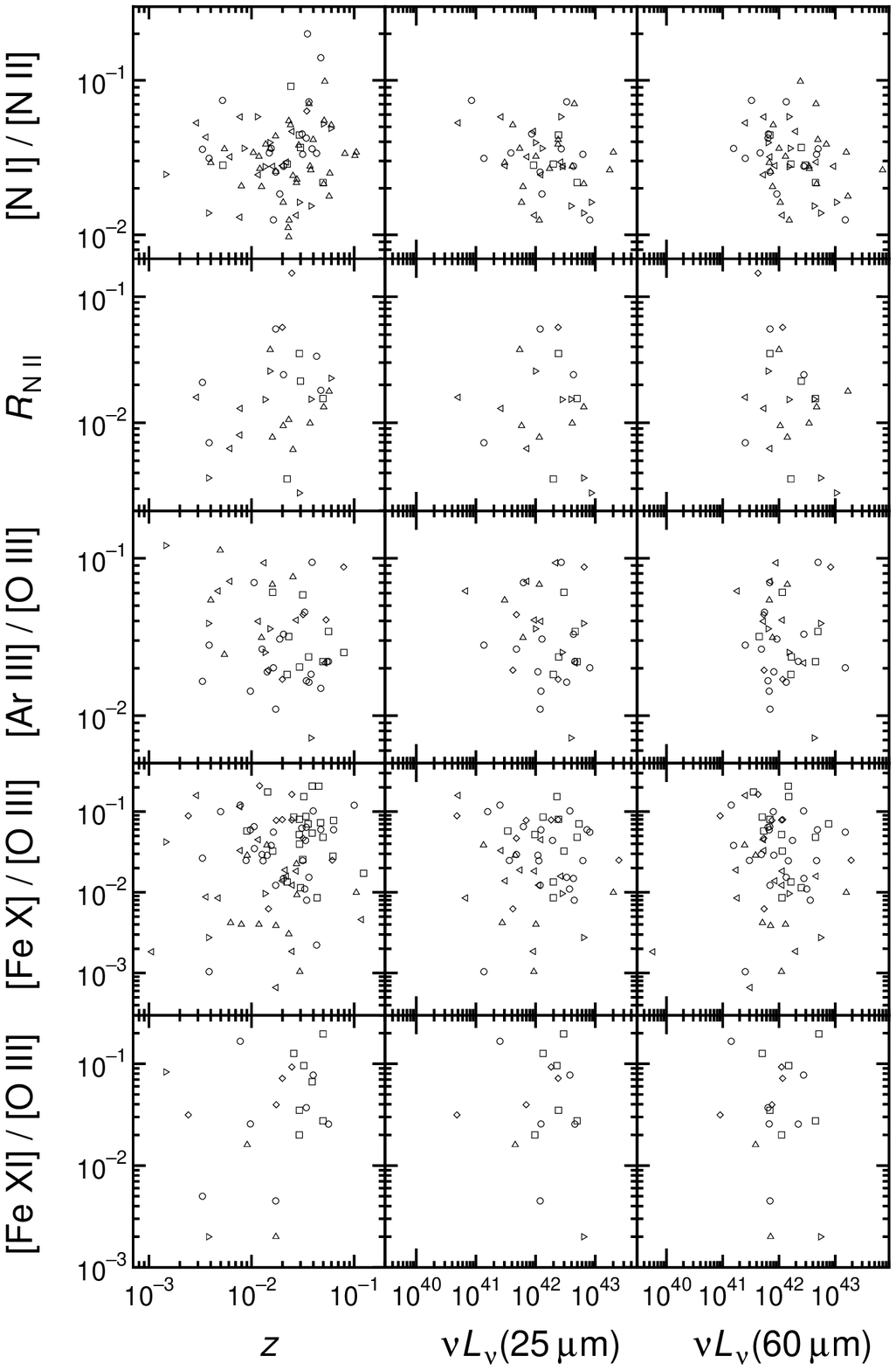}
\caption{continued.}
\end{figure*}

\begin{figure*}
\figurenum{9}
\epsscale{1.0}
\plotone{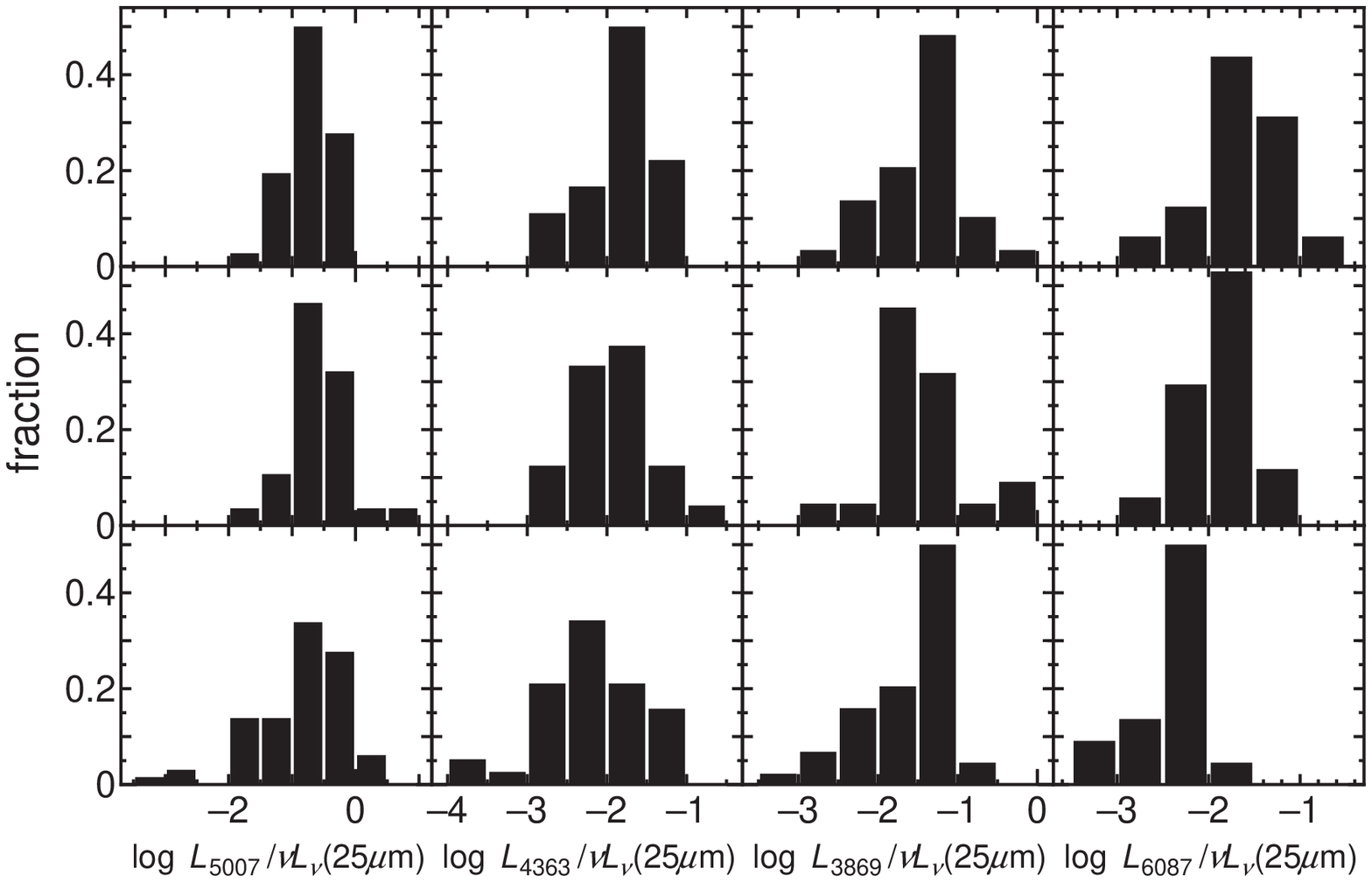}
\caption{Frequency distributions of the luminosities of
         the [O {\sc iii}]$\lambda$5007 emission,
         the [O {\sc iii}]$\lambda$4363 emission,
         the [Ne {\sc iii}]$\lambda$3869 emission and 
         the [Fe {\sc vii}]$\lambda$6087 emission for
         the S1$_{\rm total}$s, the S1.5s, and the S2$_{\rm total}$s.
         These luminosities are normalized by the IRAS 25 $\mu$m luminosity.}
\end{figure*}

\clearpage

\begin{figure*}
\figurenum{10}
\epsscale{0.5}
\plotone{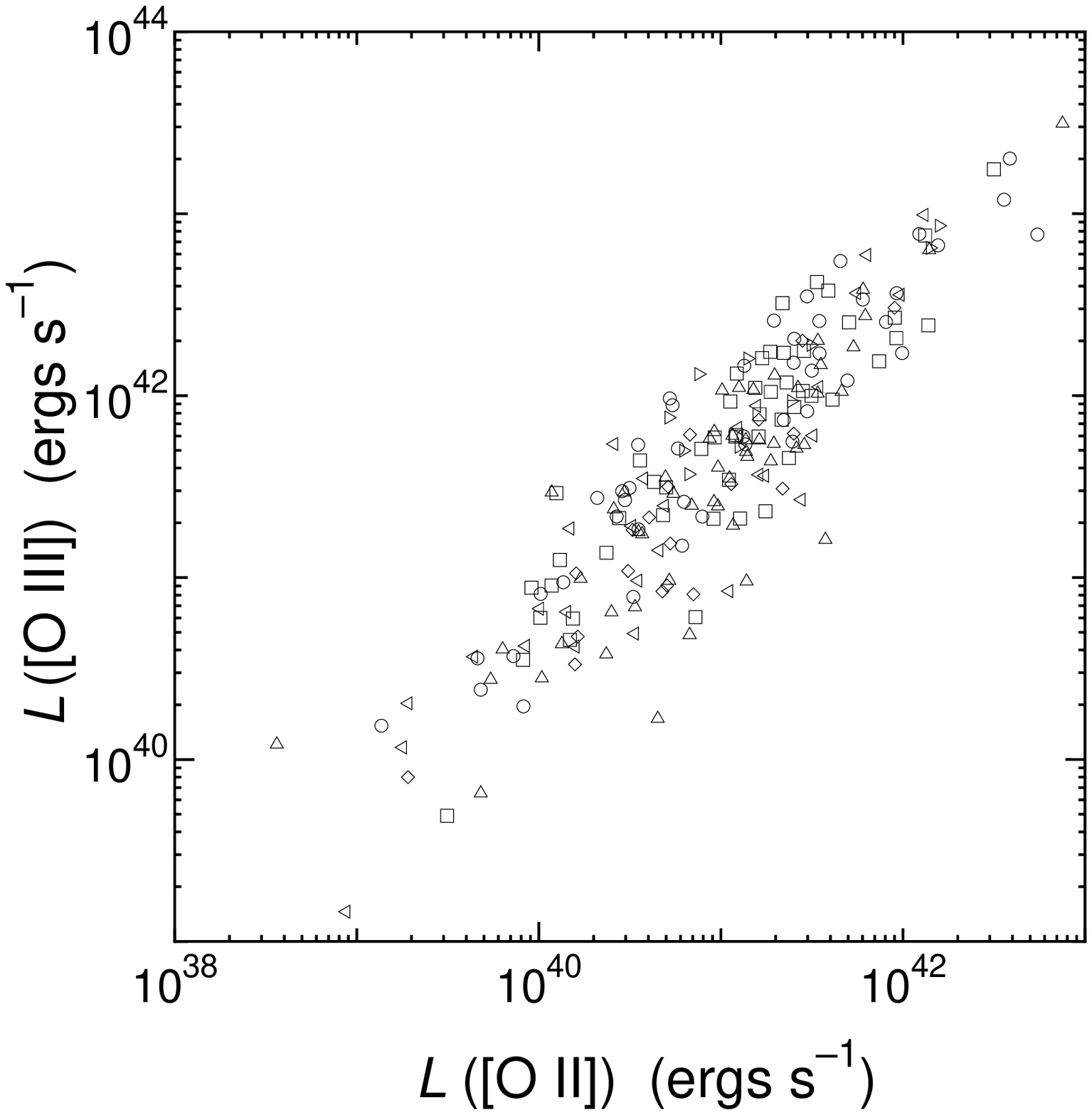}
\caption{Diagram of $L$([O {\sc iii}]$\lambda$5007) versus
         $L$([O {\sc ii}]$\lambda$3727). The symbols are the same as
         those in figure 4.}
\end{figure*}

\begin{figure*}
\figurenum{11}
\epsscale{0.6}
\plotone{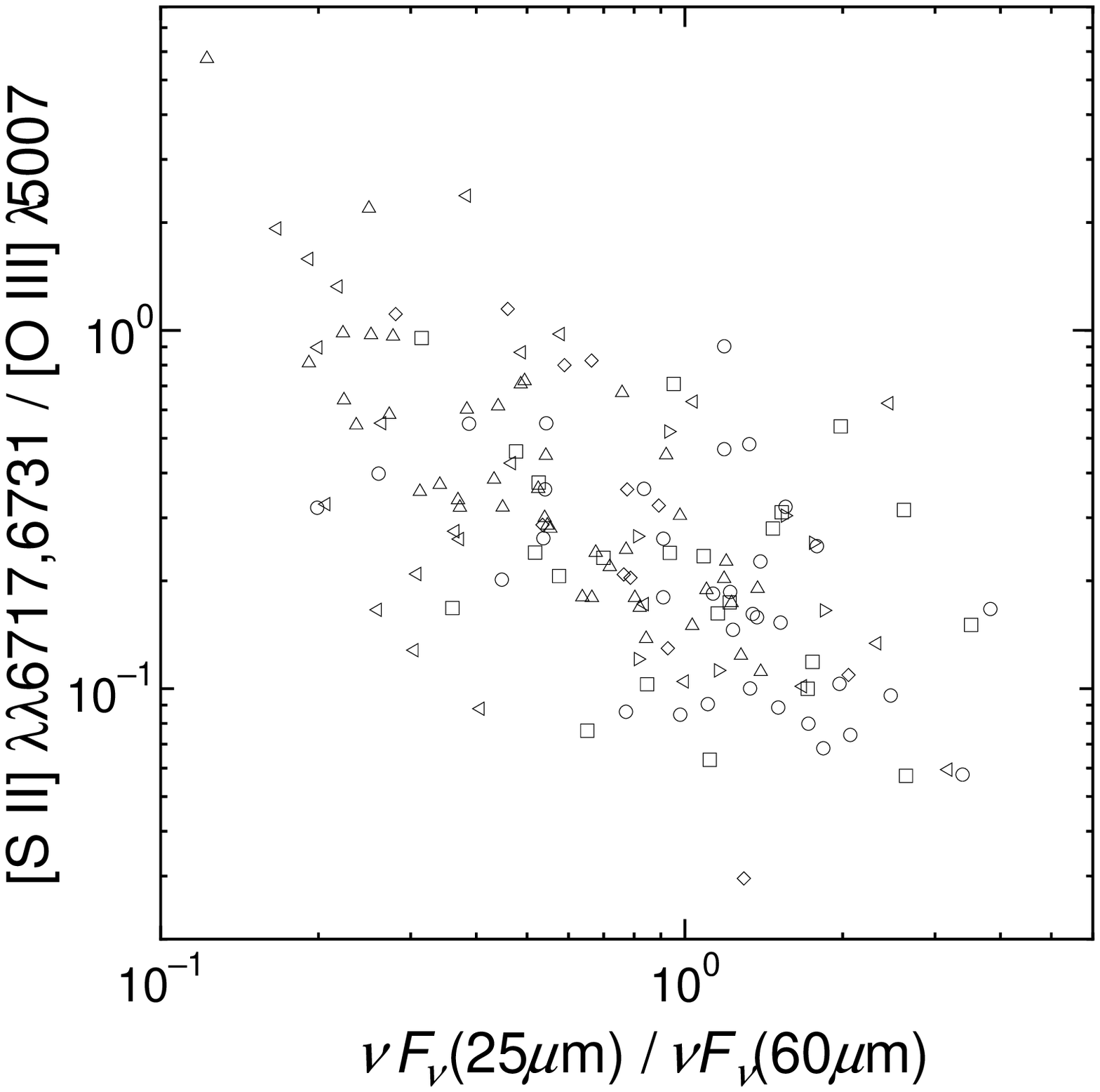}
\caption{Diagram of the emission-line flux ratio of
         [S {\sc ii}]$\lambda \lambda$6717,6731/[O {\sc iii}]$\lambda$5007
         versus that of $\nu F_{\nu}$(25 $\mu$m)/$\nu F_{\nu}$(60 $\mu$m).
         The symbols are the same as those in figure 4.}
\end{figure*}

\begin{figure*}
\figurenum{12}
\epsscale{0.6}
\plotone{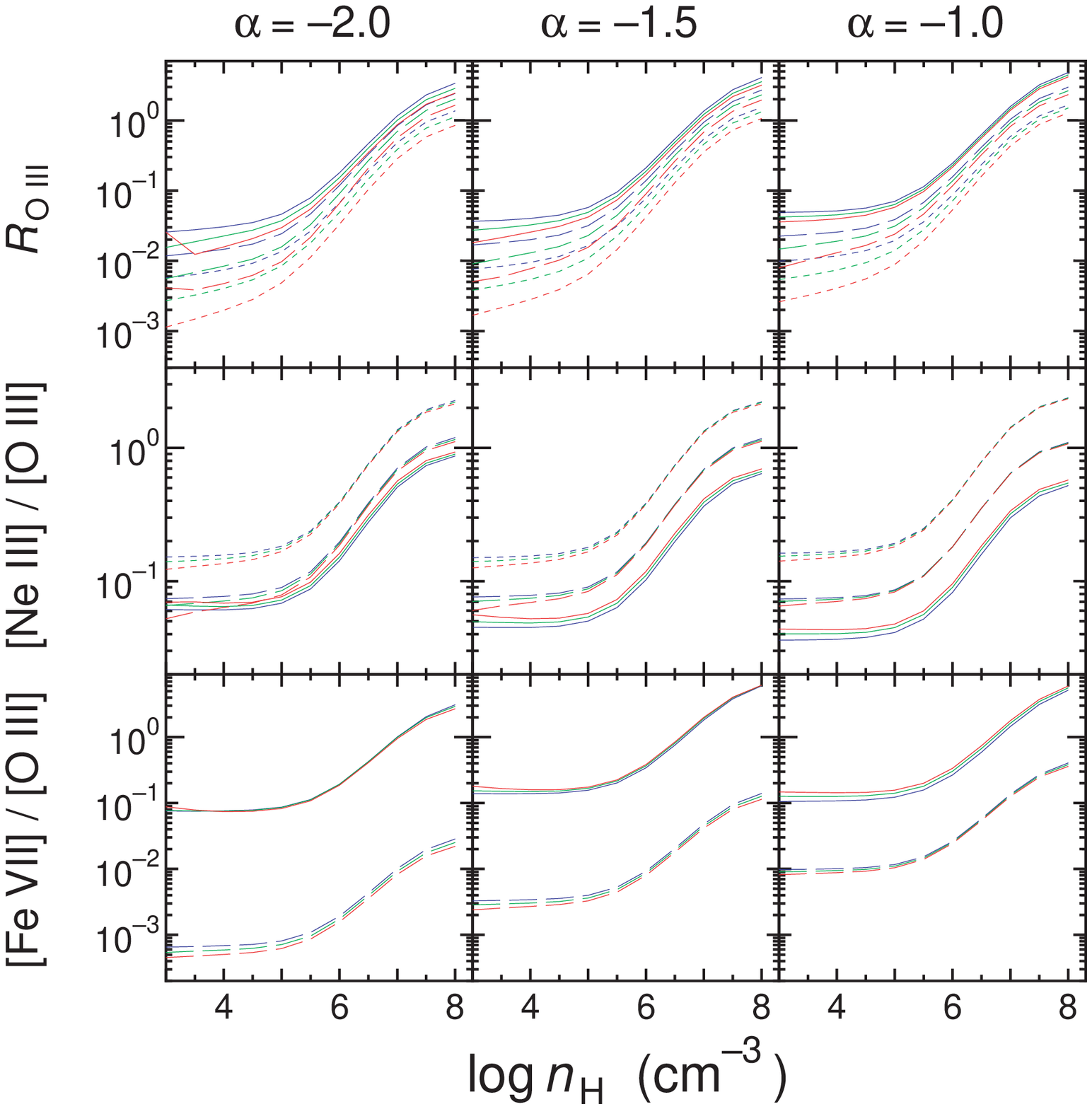}
\caption{Calculated emission-line flux ratios of
         [O {\sc iii}]$\lambda$4363/[O {\sc iii}]$\lambda$5007 
         ($R_{\rm O III}$),
         [Ne {\sc iii}]$\lambda$3869/[O {\sc iii}]$\lambda$5007 and
         [Fe {\sc vii}]$\lambda$6087/[O {\sc iii}]$\lambda$5007
         are shown as a function of the gas density.
         Here, the Lyman optical depth is assumed to be $\tau_{912} = 0.1$.
         The dotted lines, the dashed lines, and the solid lines denote
         the models adopting $U = 10^{-3.5}$, $10^{-2.5}$, and 10$^{-1.5}$,
         respectively.
         The blue lines, the green lines, and the red lines denote the
         models adopting the metallicity of half the solar, the solar,
         and twice the solar one, respectively.}
\end{figure*}

\begin{figure*}
\figurenum{13}
\epsscale{0.6}
\plotone{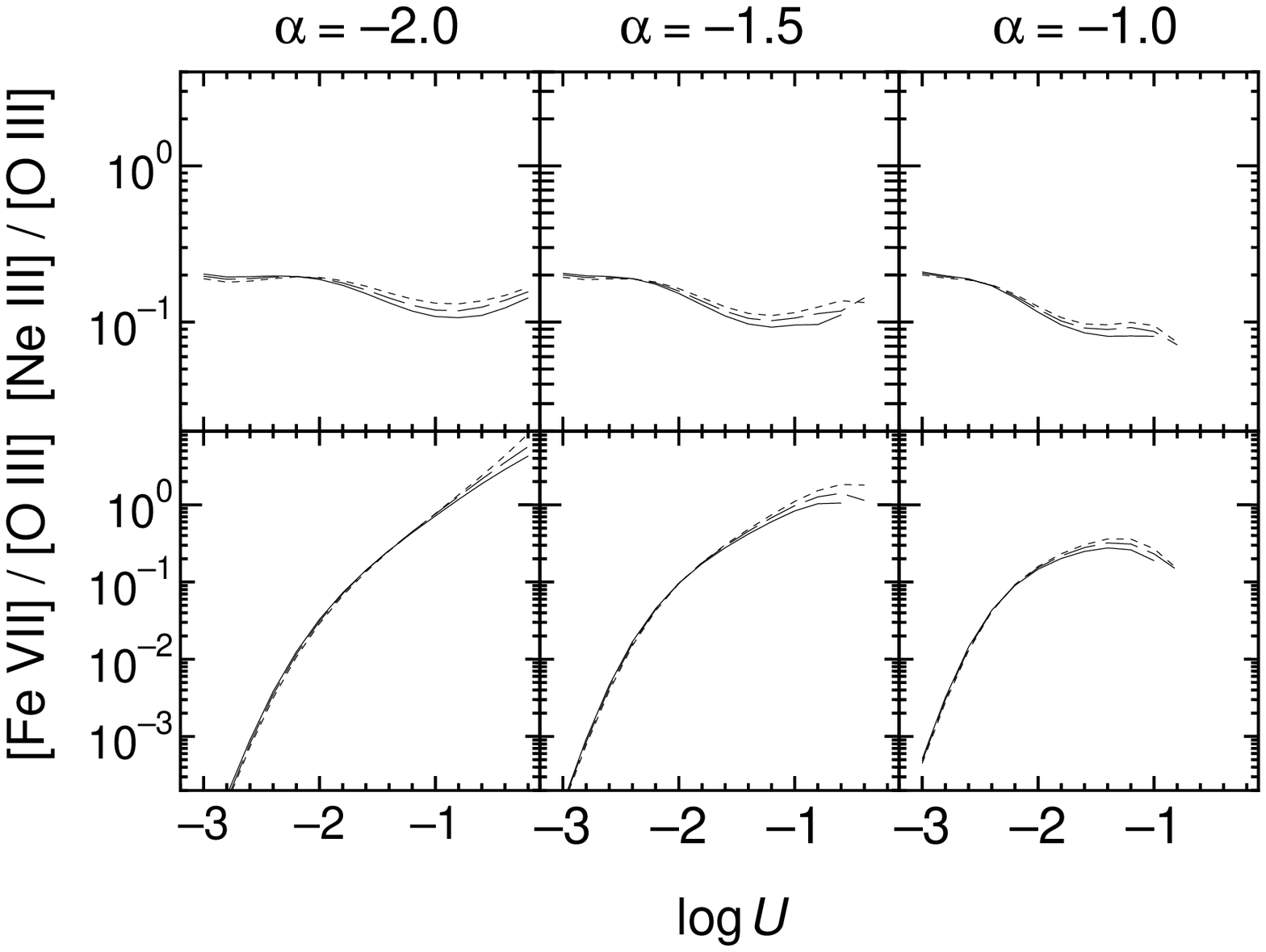}
\caption{Calculated emission-line flux ratios of
         [Ne {\sc iii}]$\lambda$3869/[O {\sc iii}]$\lambda$5007 and
         [Fe {\sc vii}]$\lambda$6087/[O {\sc iii}]$\lambda$5007
         shown as a function of the ionization parameter.
         In these models, $n_{\rm H} = 10^{6.0}$ cm$^{-3}$ is assumed.
         The solid lines denote the models with the metallicity of 
         half the solar value, the dashed lines denote the models with
         the solar metallicity, and the dotted lines denote the models
         with the metallicities of twice the solar value.}
\end{figure*}

\begin{figure*}
\figurenum{14}
\epsscale{0.55}
\plotone{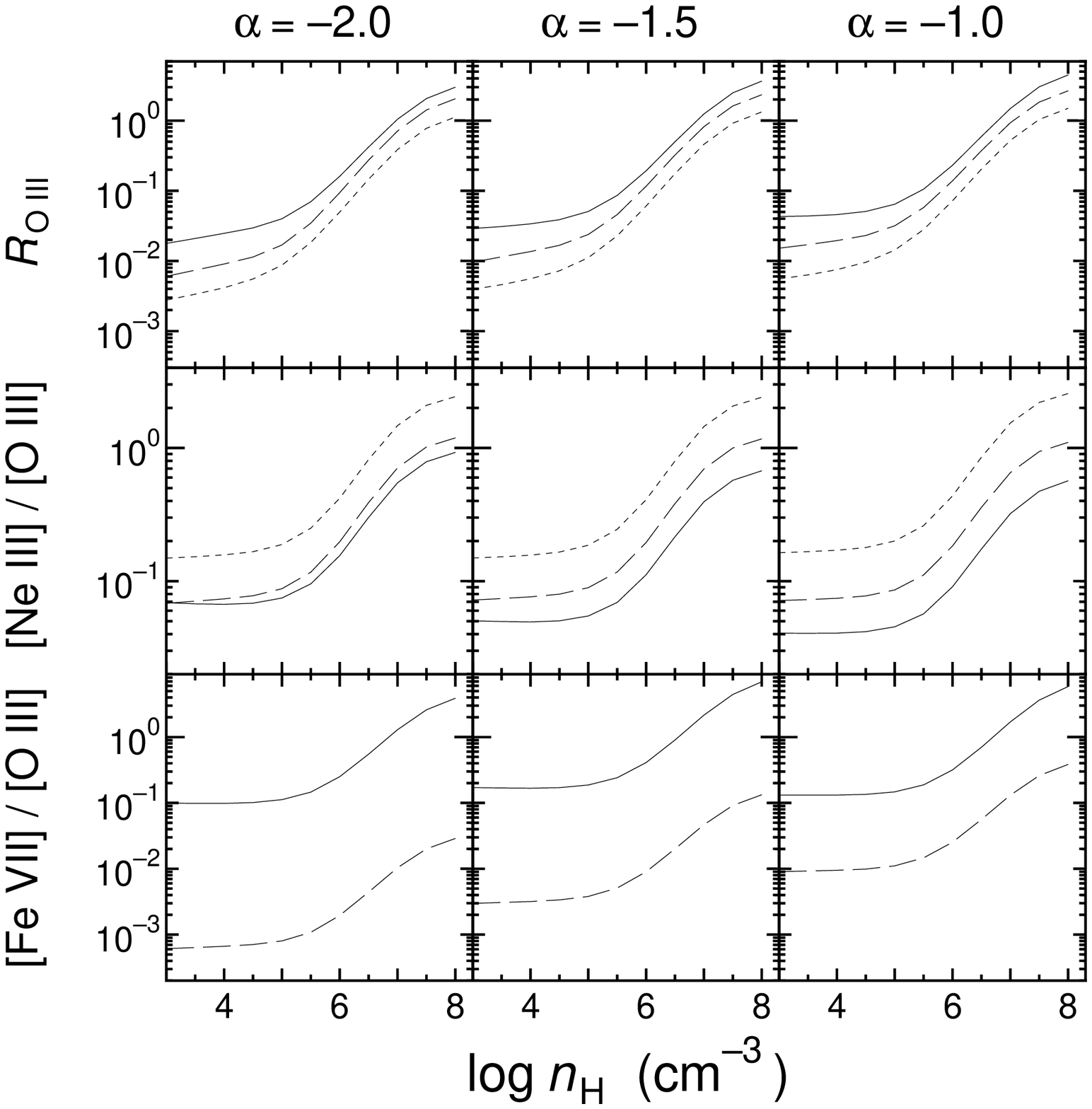}
\caption{Same as figure 12, but for $\tau_{912} = 0.01$.
         Only the models with solar metallicity are shown.}
\end{figure*}

\begin{figure*}
\figurenum{15}
\epsscale{0.55}
\plotone{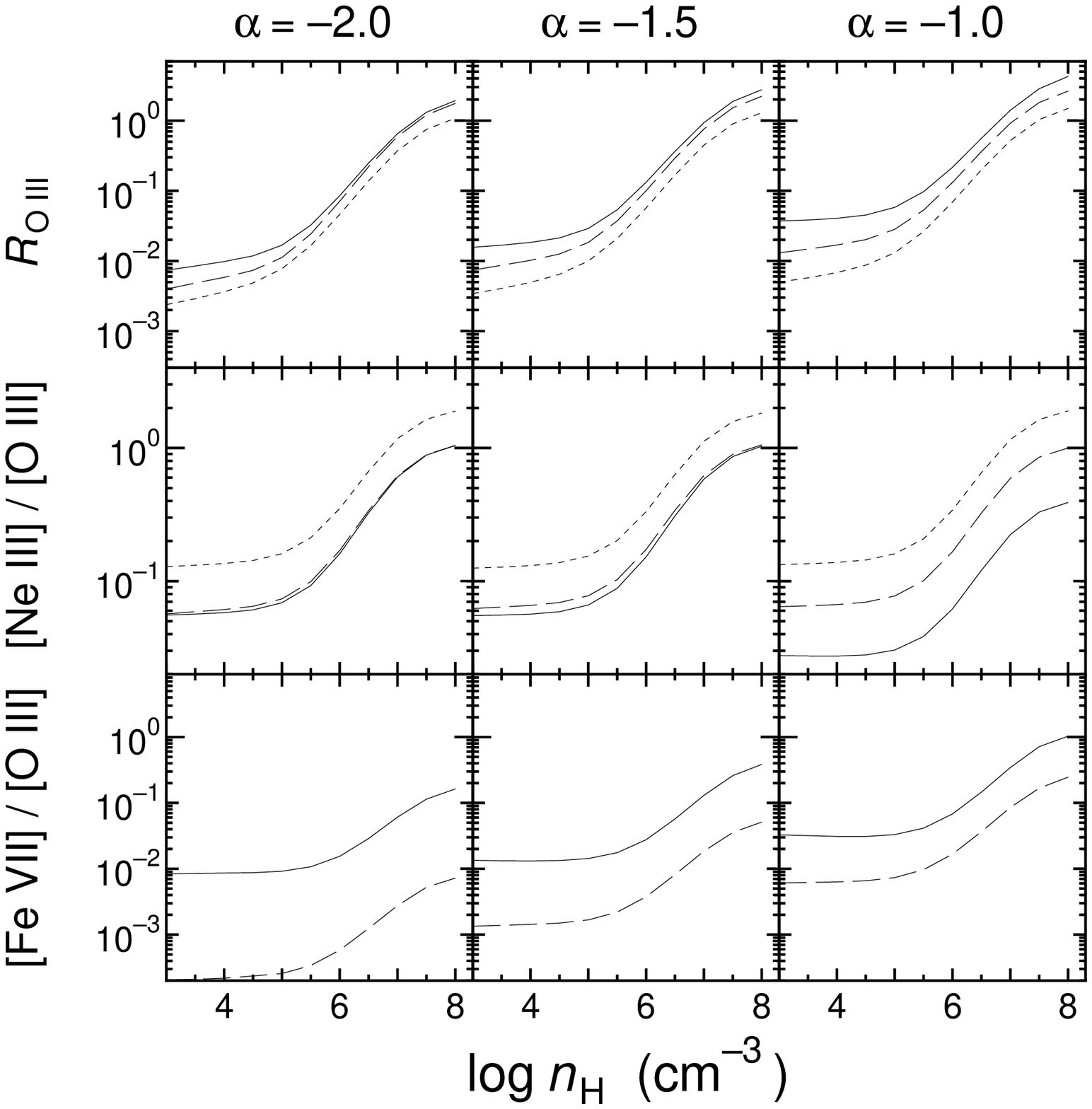}
\caption{Same as figure 12, but for $\tau_{912} = 1.0$.
         Only the models with solar metallicity are shown.}
\end{figure*}



\begin{references}
\reference{1}{	Antonucci, R.\ 1993, ARA\&A, 31, 473}
\reference{1}{	Baker, J. C.\ 1997, MNRAS, 286, 23}
\reference{1}{	Baker, J. C., \& Hunstead, R. W.\ 1995, ApJ, 452, L95}
\reference{1}{	Baldwin, J. A., Phillips, M. M., \& Terlevich, R.\ 1981, 
              PASP, 93, 5}
\reference{1}{	Barth, A. J., Tran, H. D., Brotherton, M. S., 
              Filippenko, A. V., Ho, L. C., van Breugel, W., Antonucci, R., \& 
              Goodrich, R. W.\ 1999, AJ, 118, 1609}
\reference{1}{	Barthel, P. D.\ 1989, ApJ, 336, 606}
\reference{1}{	Binette, L.\ 1985, A\&A, 143, 334}
\reference{1}{	Cardelli, J. A., Clayton, G. C., \& Mathis, J. S.\ 1989, 
              ApJ, 345, 245}
\reference{1}{	Cohen, R. D.\ 1983, ApJ, 273, 489}
\reference{1}{	Dahari, O., \& De Robertis, M. M.\ 1988, ApJS, 67, 249}
\reference{1}{	De Zotti, G., \& Gaskell, C. M.\ 1985, A\&A, 147, 1}
\reference{1}{	di Serego Alighieri, S., Cimatti, A., Fosbury, R. A. E., \&
              Hes, R.\ 1997, A\&A, 328, 510}
\reference{1}{	Efstathiou, A., \& Rowan-Robinson, M.\ 1995, MNRAS, 273, 649}
\reference{1}{	Fadda, D., Giuricin, G., Granato, G. L., \& Vecchies, D.\ 
              1998, ApJ, 496, 117}
\reference{1}{	Ferland, G. J.\ 1996, Hazy: A Brief Introduction to Cloudy 
              (Lexington: Univ. Kentucky Dept. Phys. Astron.)}
\reference{1}{	Ferland, G. J., \& Netzer, H.\ 1983, ApJ, 264, 105}
\reference{1}{	Grevesse, N., \& Anders, E.\ 1989, in AIP Conf. Proc. 183,
              Cosmic Abundance of Matter, ed. C. J. Waddington
              (New York: AIP), 1}
\reference{1}{	Grevesse, N., \& Noels, A.\ 1993, in Origin \& Evolution of
              the Elements, ed. N. Prantzos, E. Vangioni-Flam, \&
              M. Cass\'{e} (Cambridge: Cambridge Univ. Press), 15}
\reference{1}{	Halpern, J. P., \& Steiner, J. E.\ 1983, ApJ, 269, L37}
\reference{1}{	Heckman, T. M., \& Balick, B.\ 1979, A\&A, 79, 350}
\reference{1}{	Heckman, T. M., Gonzalez-Delgado, R., Leitherer, C.,
              Meurer, G. R., Krolik, J., Wilson, A. S., Koratkar, A., \&
              Kinney, A.\ 1997, ApJ, 482, 114}
\reference{1}{	Heckman, T. M., Krolik, J., Meurer, G., Calzetti, D., 
              Kinney, A., Koratkar, A., Leitherer, C., Robert, C., \&
              Wilson, A.\ 1995, ApJ, 452, 549}
\reference{1}{	Hes, R., Barthel, P. D., \& Fosbury, R. A. E.\ 1993, 
              Nature, 362, 326}
\reference{1}{	Izotov, Y. I., Thuan, T. X., \& Lipovetsky, V. A.\ 1994, 
              ApJ, 435, 647}
\reference{1}{	Jackson, N., \& Browne, I. W. A.\ 1990, Nature, 343, 43}
\reference{1}{	Khachikian, E. Y., \& Weedman, D. W.\ 1974, ApJ, 192, 581}
\reference{1}{	Kinney, A. L., Antonucci, R. R. J., Ward, M. J., 
              Wilson, A. S., \& Wittle, M.\ 1991, ApJ, 377, 100}
\reference{1}{	Koski, A. T.\ 1978, ApJ, 223, 56}
\reference{1}{	Kuraszkiewicz, J., Wilkes, B. J., Czerny, B., \&
              Mathur, S.\ 2000, ApJ, 542, 692}
\reference{1}{	Masegosa, J., Moles, M., \& Campos-Aguilar, A.\ 1994, 
              ApJ, 420, 576}
\reference{1}{	McCall, M. L., Rybski, P. M., \& Shields, G. A.\ 1985, 
              ApJS, 57, 1}
\reference{1}{	Moshir, M., Kopman, G., \& Conrow, T. A. O.\ 1992, 
              Explanatory Supplement to the
              $IRAS$ Faint Source Survey, version 2 (Pasadena: JPL)}
\reference{1}{	Mulchaey, J. S., Koratkar, A., Ward, M. J., Wilson, A. S.,
              Whittle, M., Antonucci, R. R. J., Kinney, A. L., \&
              Hurt, T.\ 1994, ApJ, 436, 586}
\reference{1}{	Murayama, T., \& Taniguchi, Y.\ 1998a, ApJ, 497, L9}
\reference{1}{	Murayama, T., \& Taniguchi, Y.\ 1998b, ApJ, 503, L115}
\reference{1}{	Murayama, T., Taniguchi, Y., \& Iwasawa, K.\ 1998, 
              AJ, 115, 460}
\reference{1}{	Nagao, T.\ 2001, Master's thesis, Tohoku University}
\reference{1}{	Nagao, T., Murayama, T., \& Taniguchi, Y.\ 2001a, 
              ApJ, 546, 744}
\reference{1}{	Nagao, T., Murayama, T., \& Taniguchi, Y.\ 2001b, 
              ApJ, 549, 155}
\reference{1}{	Nagao, T., Taniguchi, Y., \& Murayama, T.\ 2000, AJ, 119, 2605}
\reference{1}{	Ohyama, Y.\ 1996, Master's thesis, Tohoku University}
\reference{1}{	Osterbrock, D. E.\ 1978, Lick Obs. Bull, No. 775 }
\reference{1}{	Osterbrock, D. E.\ 1989, Astrophysics of Gaseous Nebulae and
              Active Galactic Nuclei (Mill Valley: University Science Books)}
\reference{1}{	Osterbrock, D. E., Koski, A. T., \& Phillips, M. M.\ 1976,
              ApJ, 206, 898}
\reference{1}{	Osterbrock, D. E., \& Pogge, R. W.\ 1985, ApJ, 297, 166}
\reference{1}{	Peterson, B. M.\ 1993,PASP, 105, 247}
\reference{1}{	Pier, E. A., \& Krolik, J. H.\ 1992, ApJ, 401, 99}
\reference{1}{	Pier, E. A., \& Voit, G. M.\ 1995, ApJ, 450, 628}
\reference{1}{	Press, W. H., Teukolsky, S. A., Vetterling, W. T., \&
              Flannery, B. P.\ 1988, Numerical Recipes in C 
              (Cambridge: Cambridge University Press)}
\reference{1}{	Schmitt, H. R.\ 1998, ApJ, 506, 647}
\reference{1}{	Schmitt, H. R., \& Kinney, A. L.\ 1996, ApJ, 463, 498}
\reference{1}{	Shuder, J. M., \& Osterbrock, D. E.\ 1981, ApJ, 250, 55}
\reference{1}{	Simpson, C.\ 1998, MNRAS, 297, L39}
\reference{1}{	Storchi-Bergmann, T., Mulchaey, J. S., \& Wilson A.S.\ 1992, 
              ApJ, 395, L73}
\reference{1}{	Storchi-Bergmann, T., \& Pastoriza, M. G.\ 1989, ApJ, 347, 195}
\reference{1}{	Storchi-Bergmann, T., \& Pastoriza, M. G.\ 1990, 
              PASP, 102, 1359}
\reference{1}{	Thuan, T. X.\ 1984, ApJ, 281, 126}
\reference{1}{	van Zee, L., Salzer, J. J., Haynes, M. P., \& 
              Balonek, T. J.\ 1998, AJ, 116, 2805}
\reference{1}{	Veilleux, S., \& Osterbrock, D. E.\ 1987, ApJS, 63, 295 (VO87)}
\reference{1}{	V\'{e}ron-Cetty, M. -P., \& V\'{e}ron, P.\ 2000,
              ESO Sci. Rept. 19 (European Southern Observatory)}
\reference{1}{	Wilkes, B. J., Schmidt, G. D., Smith, P. S., Mathur, S., \&
              McLeod, K. K.\ 1995, ApJ, 455, L13}

\end{references}
\end{document}